TERRANOVA CONSORTIUM

# Wireless Terahertz System Architectures for Networks Beyond 5G

Version 1.0, July 2018

# Contributors, Reviewers and Editors

| Name | Company/Institute/University | E-mail |
|---|---|---|
| Alexandros-Apostolos A. Boulogeorgos | University of Piraeus (UPRC), Greece | al.boulogeorgos@ieee.org |
| Angeliki Alexiou | University of Piraeus (UPRC), Greece | aalexiou@ieee.org |
| Dimitrios Kritharidis | Intracom (ICOM), Greece | dkri@intracom-telecom.com |
| Alexandros Katsiotis | Intracom (ICOM), Greece | alexkat@intracom-telecom.com |
| Georgia Ntouni | Intracom (ICOM), Greece | gntouni@intracom-telecom.com |
| Joonas Kokkoniemi | University of OULU, Finland | Joonas.Kokkoniemi@oulu.fi |
| Janne Lethtomaki | University of OULU, Finland | Janne.lehtomaki@oulu.fi |
| Markku Juntti | University of OULU, Finland | Markku.juntti@oulu.fi |
| Dessy Yankova | JCP, France | Dessy.yankova@jcp-connect.com |
| Ahmed Mokhtar | JCP, France | ahmed.mokhtar@jcp-connect.com |
| Jean-Charles Point | JCP, France | pointjc@jcp-connect.com |
| José Machado | Altice Labs SA, Portugal | Jose-e-machado@alticelabs.com |
| Robert Elschner | Fraunhofer HHI, Germany | Robert.elschner@hhi.fraunhofer.de |
| Colja Schubert | Fraunhofer HHI, Germany | Colja.schubert@hhi.fraunhofer.de |
| Thomas Merkle | Fraunhofer IAF, Germany | Thomas.merkle@iaf.fraunhofer.de |
| Ricardo Ferreira | PICadvanced, Portugal | Ricardo.ferreira@picadvanced.com |
| Francisco Rodrigues | PICadvanced, Portugal | francisco@picadvanced.com |
| Jose Lima | PICadvanced, Portugal | Jose.lima@picadvanced.com |

# Executive Summary

The present white paper focuses on the system requirements of TERRANOVA. Initially details the key use cases for the TERRANOVA technology and presents the description of the network architecture. In more detail, the use cases are classified into two categories, namely backhaul & fronthaul and access and small cell backhaul. The first category refers to fibre extender, point-to-point and redundancy applications, whereas the latter are designed to support backup connection for small and medium-sized enterprises (SMEs), internet of things (IoT) dense environments, data centres, indoor wireless access, ad hoc networks, and last mile access. Then, it provides the networks architecture for the TERRANOVA system as well as the network elements that need to be deployed. The use cases are matched to specific technical scenarios, namely outdoor fixed point-to-point (P2P), outdoor/indoor individual point-to-multipoint (P2MP), and outdoor/indoor "quasi"-omnidirection, and the key performance requirements of each scenario are identified. Likewise, we present the breakthrough novel technology concepts, including the joint design of baseband signal processing for the complete optical and wireless link, the development of broadband and spectrally efficient RF-frontends for frequencies >275 GHz, as well as channel modelling, waveforms, antenna array and multiple-access schemes design, which we are going to use in order to satisfy the presented requirements. Next, an overview of the required new functionalities in both physical (PHY) layer and medium access control (MAC) layers in the TERRANOVA system architecture will be given. Finally, the individual enablers of the TERRANOVA system are combined to develop particular candidate architectures for each of the three technical scenarios.

The main outcomes of the white paper are:

- The identification of a set of TERRANOVA use cases that are used to identify requirements, both in terms of functionality and performance, that the system must support.
- The description of the preliminary network architecture as well as the network elements of envisioned TERRANOVA system
- The determination of the technical scenarios that correspond to the application use cases as well as their key performance requirements.
- The illustration of candidate architectures of the optical link.
- The presentation of the technological enabler that will allow us to achieve these requirements.
- An overview of the required new functionalities in both PHY layer and MAC layer to support the identified use cases of the TERRANOVA system.
- The description of implementation options for the co-design of the optical/THz system, the THz RF frontend and the THz MAC/RRM protocols.
- The presentation of candidate architectures for the implementation of the three specific technical use-scenarios, namely outdoor fixed point-to-point (P2P), outdoor/indoor individual point-to-multipoint (P2MP), and outdoor/indoor "quasi"-omnidirectional.

# Table of Contents

# List of Figures

# List of Tables

# List of Acronyms and Abbreviations

| Acronym/Abbreviation | Description |
|---|---|
| 1D | One-Dimensional space |
| 2D | Two-Dimensional space |
| 3D | Three-Dimensional space |
| 2G | Second Generation |
| 3G | Third Generation |
| 3GPP | Third Generation Partnership Project |
| 4G | Forth Generation |
| 5G | Fifth Generation |
| A-BFT | Associate BeamForming Training |
| ACK | Acknowledgement |
| ACO | Analog Coherent Optics |
| ADC | Analog-to-Digital Converter |
| AE | Antenna Element |
| AFC | Automatic Frequency Correction |
| AFE | Analogue FrontEnd |
| AGC | Automatic Gain Control |
| AiP | Antenna-in-Package |
| AM | Amplitude Modulation |
| AMC | Adaptive Modulation and Coding |
| AP | Access Point |
| ASIC | Application-Specific Integrated Circuit |
| ATDE | Adaptive Time Domain Equalizer |
| ATI | Announcement Transmission Interval |
| AWG | Arrayed Waveguide Gratings |
| AWGN | Additive White Gaussian Noise |
| AWV | Antenna Weight Vector |
| BB | BaseBand |
| BC | Beam Combining |
| BER | Bit Error Rate |
| BF | BeamForming |
| BH | BackHaul |
| BHI | Beacon Header Interval |
| BI | Beacon Interval |
| BOC | BackOff Counter |
| BPSK | Binary Phase Shift Keying |
| BRP | Beam Refinement Protocol |
| BS | Base Station |
| BTI | Beacon Transmission Interval |
| BW | BeamWidth |
| CA | Consortium Agreement |
| CAP | Contention Access Period |
| CAUI | 100 gigabit Attachment Unit Interface |
| CBAP | Contention-Based Access Period |
| CapEx | Capital Expenditure |
| CC | Central Cloud |
| CCH | Control CHannel |
| CDF | Cumulative Distribution Function |
| CDR | Clock and Data Recovery |
| CFP | C-Form Factor Pluggable |
| CMOS | Complementary Metal–Oxide–Semiconductor |
| CoMP | Coordination Multi-Point |
| COTS | Commercial Off-The-Shelf |
| CPR | Carrier Phase Recovery |
| CRC | Cyclic Redundancy Code |
| CSI | Channel State Information |
| CSMA/CA | Carrier Sense Multiple Access with Collision Avoidance |
| CTA | Channel Time Allocation |
| CTAP | Channel Time Allocation Period |
| CTS | Clear-To-Send |
| CTS-NI | Clear-To-Send-Node-Information |
| CW | Continuous Wave |
| D2D | Device-to-Device |
| DAC | Digital to Analog Converter |
| DC | Direct Current |
| DCH | Data CHannel |
| DDC | Digital Down Conversion |
| DEMUX | DE-MUltipleXer |
| DL | DownLink |
| DMG | Directional Multi-Gigabit |
| DMT | Discrete Multi-Tone |
| DoA | Direction of Arrival |
| DoF | Degree of Freedom |
| DP | Detection Probability |
| DP-IQ | Dual Polarization In-phase and Quadrature |
| DPD | Digital PreDistortion |
| DSP | Digital Signal Processing |
| DTI | Data Transfer Interval |
| DUC | Digital Up Conversion |
| DWDM | Dense Wavelength Division Multiplexing |
| EC | European Commission |
| EDCA | Enhanced Distributed Channel Access |
| EDMG | Enhanced Directional Multi-Gigabit |
| EKF | Extended Kalman Filter |
| E/O | Electrical-Optical |
| ESE | Extended Schedule Element |
| ETSI | European Telecommunications Standards Institute |
| eWLB | embedded Wafer Level Ball grid array |
| FAN | Fixed Access Network |
| FAP | False-Alarm Probability |
| FCS | Frame Check Sequence |
| FD | Full Duplex |
| FDD | Frequency Division Duplexing |
| FDMA | Frequency Division Multiple Access |
| FEC | Forward Error Correction |
| FFE | Feed Forward Equilizer |
| FH | FrontHaul |
| FIFO | First In First Out |
| FM | Frequency Modulation |
| FPGA | Field-Programmable Gate Array |
| FrD | Frequency Domain |
| FSO | Free-Space Optics |
| FSPL | Free Space Path Loss |
| FTTH | Fiber To The Home |
| FWA | Fixed Wireless Access |
| G.fast | Transmission Technology for Telephone Lines up to 1 Gbit/s |
| GA | Grant Agreement |
| GaAs | Gallium Arsenide |
| HEMT | High Electron Mobility Transistor |
| HFT | High Frequency Trading |
| HSPA | High Speed Packet Access |
| HSPA+ | evolved High Speed Packet Access |
| I/Q | In-phase and Quadrature |
| $I^2C$ | Inter-Integrated Circuit |
| IA | Initial Access |
| ICF | Intermediate Carrier Frequency |
| IEEE | Institute of Electrical and Electronics Engineers |
| IF | Intermediate Frequency |
| IoT | Internet of Things |
| IM/DD | Intensity Modulation/Direct Detection |
| IP | Internet protocol layer |
| ISI | InterSymbol Interference |
| ISM | Industrial Scientific and Medical band |
| ITU | International Telecommunication Union |
| ITU-R | Radiocommunication sector of the International Telecommunication Union |
| IQ COMP. | In-phase and Quadrature impairments COMPensator |
| IQD | Indoor Quasi Directional |
| KPI | Key Performance Indicator |
| LDPC | Low-Density Parity-Check |
| LO | Local Oscillator |
| LoS | Line of Sight |
| LTE-A | Long Term Evolution Advanced |
| MAC | Medium Access Control |
| MAL | Molecular Absorption Loss |
| MCE | MAC Coordination Entity |
| MID | Multiple sector IDentifier |
| MIMO | Multiple Input Multiple Output |
| MMIC | Monolithic Microwave Integrated Circuit |
| mmWave | Millimetre Wave |
| MSA | Multi Source Agreement |
| MU | Multi User |
| MUE | Mobile User Equipment |
| MUX | MUltipleXer |
| MZI | Mach-Zehnder Interferometer |
| NAV | Network Allocation Vector |
| NETCONF | NETwork CONFiguration |
| NI | Node Information |
| NGPON2 | Next-Generation Passive Optical Network 2 |
| nLoS | Non-Line Of Sight |
| NR | New Radio |
| NRZ | Non-Return to Zero |
| OFDM | Orthogonal Frequency Division Modulation |
| OIF | Optical Internetworking Forum |
| OLT | Optical Line Terminal |
| ONUs | Optical Network Units |
| OOK | On-Off Keying |
| OpEx | Operating Expenses |
| P2MP | Point-to-Multi-Point |
| P2P | Point-to-Point |
| PA | Power Amplifier |
| PAM | Pulse Amplitude Modulation |
| PBSS | Personal Basic Service Set |
| PCB | Printed Circuit Board |
| PCP | Personal basic service set control point |
| PDM | Polarization-Division Multiplexing |
| PDM-QAM | Polarization Multiplexed Quadrature Amplitude Modulation |
| PER | Packet Error Rate |
| PFIS | Point coordination Function Inter-frame Space |
| PHY | PHYsical |
| PIN | Positive-Intrinsic-Negative |
| PL | PathLoss |
| PLL | Phased Locked Loop |
| PNC | Picocell Network Coordinator |
| PONs | Passive Optical Networks |
| PS | Phase Shifter |
| PSP | Pulse Shaping Filter |
| PSF | Primary Synchronization Signal |
| QAM | Quadrature Amplitude Modulation |
| QoE | Quality of Experience |
| QoS | Quality-of-Service |
| QSFP | Quad Small Form-Factor Pluggable |
| RA | Random Access |
| RAN | Radio Access Network |
| RAT | Radio Access Technology |
| RAR | Random Access Response |
| RAU | Remote Antenna Unit |
| RB | Resource Block |
| RF | Radio Frequency |
| RH | Radio Head |
| RoF | Radio over Fiber |
| RRM | Radio Resource Management |
| RSRP | Reference Signal Received Power |
| RSSI | Received Signal Strength Indicator |
| RTS | Request-To-Send |
| RTS-NI | Request-To-Send-Node Information |
| RX | Receiver |
| SC | Small Cell |
| SCM | Sub-Carrier Multiplexing |
| SD-FEC | Soft-Decision Forward-Error Correction |
| SDM | Space Division Multiplexing |
| SDMA | Space Division Multiple Access |
| SDN | Software Define Network |
| SFF | Small Form Factor |
| SFP | Small Form-Factor Pluggable |
| SiGe | Silicon-Germanium |
| SISO | Single Input Single Output |
| SLS | Sector Level Sweep |
| SM | Spatial Multiplexing |
| SME | Small and Medium-sized Enterprise |
| SMF | Single Mode Fiber |
| SNR | Signal to Noise Ratio |
| SOTA | State Of The Art |
| SP | Service Period |
| SPI | Serial Parallel Interface |
| SRC | Sample Rate Conversion |
| SSB | Single-SideBand |
| SSW | Sector SWeep |
| SSW-FBCK | Sector SWeep FeedBaCK |
| STA | STAtion |
| STM-1 | Synchronous Transport Module, level 1 |
| STS | Symbol Timing Synchronization |
| TAB-MAC | Terahertz Assisted Beamforming Medium Access Control |
| TD | Time Domain |
| TDD | Time Division Duplexing |
| TDM | Time Division Multiplexing |
| TDMA | Time Division Multiple Access |
| TERRANOVA | Terabit/s Wireless Connectivity by Terahertz innovative technologies to deliver Optical Network Quality of Experience in Systems beyond 5G |
| THz | Terahertz |
| TIA | TransImpedance Amplifier |
| TWDM | Time and Wavelength Division Multiplexed |
| Tx | Transmitter |
| TXOP | Transmission Opportunity |
| UL | Uplink |
| UE | User Equipment |
| VCO | Voltage Controlled Oscillator |
| VGA | Variable Gain Amplifier |
| VLC | Visible Light Communication |
| WLAN | Wireless Local Area Network |
| WDM | Wavelength Division Multiplexing |
| WiFi | Wireless Fidelity |
| WiGig | Wireless Gigabit alliance |
| WLBGA | Wafer Level Ball Grid Array |
| WM | Wireless Microwave |
| | Waveform Multiplexing |
| XG-PON | 10 Gbit/s Passive Optical Network |
| XPIC | Cross Polarization Interference Cancellation |
| YANG | Yet Another Next Generation |

# 1. Introduction

Wireless data traffic has drastically increased accompanied by an increasing demand for higher data rate transmission. In particular and according to the Edholm's law of bandwidth [1], wireless data rates have been doubled every 18 months over the last three decades and are quickly approaching the capacity of wired communication systems [2]. In order to address this tremendous capacity demands, the wireless word has moved towards the fifth generation (5G) era, by introducing several novel approaches, such as massive multiple-input multiple-output (MIMO) systems, full-duplexing, and millimetre wave (mmWave) communications. However, there is a lack of efficiency and flexibility in handling the huge amount of quality of service (QoS) and experience (QoE) oriented data services [3].

In view of the fact that the currently used frequency spectrum for 5G has limited capacity, THz wireless became an attractive complementing technology to the less flexible and more expensive optical-fibre connections as well as to the lower data-rate systems, such as visible light communications (VLCs), microwave links, and wireless fidelity (WiFi) [4], [5], [6], [7], [8], [9], [10]. As a consequence, by enabling wireless THz communications, we expect not only to address the spectrum scarcity and capacity limitations [11] of the current cellular systems, but also to boost a plethora of life-changing applications. Therefore, they will influence the main technology trends in wireless networks within the next 10 years and beyond.

The implementation of wireless THz systems will have to leverage breakthrough novel technological concepts. Indicative examples can be considered the joint-design of baseband digital signal processing (DSP) for the complete optical and wireless link, the development of broadband and highly spectral efficient radio frequency (RF) frontends operating at frequencies higher than 275 GHz, and new standardized electrical-optical (E/O) interfaces. Additionally, to address the extremely large bandwidth and the propagation properties of the THz regime, improved channel modelling and the design of appropriate waveforms, multiple access control (MAC) schemes and antenna array configurations are required.

## 1.1. Scope & contribution

Motivated by the potentials of the wireless THz systems, this white paper is devoted to present the beyond 5G applications, which are expected to be enabled by THz technologies, and identify their requirements. Moreover, after revealing the limitations of the wireless THz systems, which results from the channel particularities and the technology gaps, it illustrates the appropriate enables and functionalities that will allow us to deliver optical network QoE to the end-user. Finally, it discusses candidate architecture that are expected to catalyse the road beyond 5G.

## 1.2. Structure

The rest of the white paper is organized as follows. In Section 2, the applications that need to be supported by the beyond 5G networks as well as their requirements are summarized and classified into three technical scenarios. Likewise, Section 3 focuses on describing the system architecture, whereas, Section 4 discusses the optical link and TERRANOVA media converter design. Likewise, Section 5 reveals the physical limitations of the wireless systems operating in the THz band. Section 6 is devoted on presenting the required functionalities and enablers of the wireless THz system, while Section 7 provides the candidate architectures. Finally, Section 8 summarize our observations and conclude the white paper.

# 2. Beyond 5G applications & requirements

Section 2 presents a summary of the identified key use cases for the TERRANOVA technology. In more detail, the use cases are classified into two categories, namely backhaul & fronthaul, mobile and fixed wireless access. The first category refers to fibre extender, point-to-point and redundancy applications, whereas the latter is designed to support corporate backup connection for large and SMEs, internet of things (IoT) dense environments, data centres, indoor wireless access, ad hoc networks, sport and music events, and last mile access.

## 2.1. Use cases

This section is focused on presenting the key use cases of the TERRANOVA's technology. As illustrated in Figure 1, the use cases are classified into two categories, namely backhaul (BH) & fronthaul (FH) and access. The first category details the fibre extender, point-to-point (P2P) and redundancy applications, whereas the latter refers to business, last mile and open spaces events.

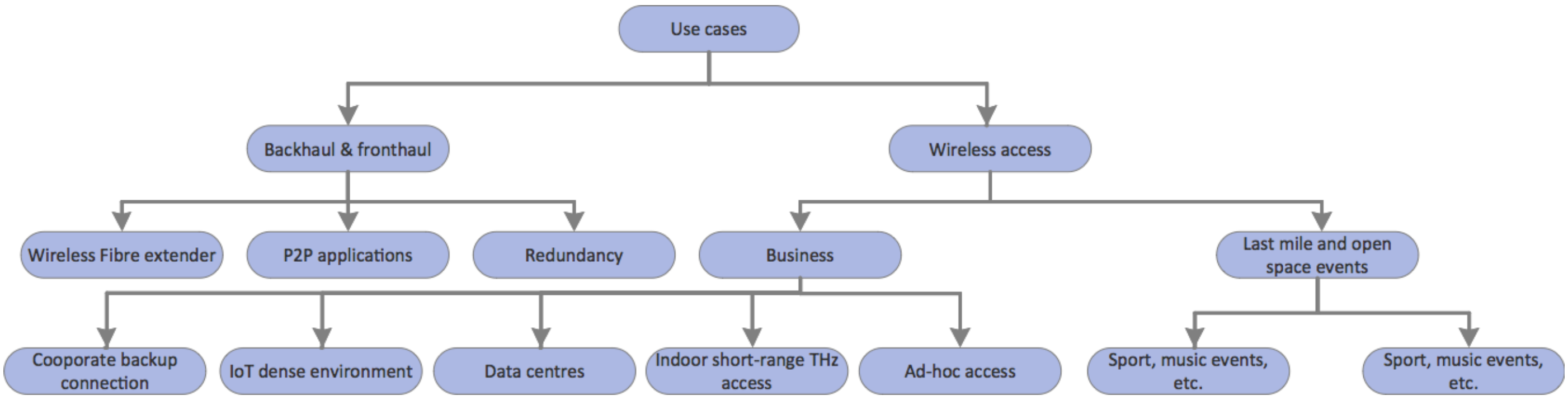

*Figure 1: Use cases classification.*

### 2.1.1. Backhaul & fronthaul.

The BH and FH use cases refer to the possible application of TERRANOVA outcomes in communication to or between cell towers (BH) or between remote radio heads located at the cell towers and centralized baseband units (FH). BH and FH applications can be classified in wireless fibre extender, point-to-point (P2P) applications in the absence of installed fibre and redundancy scenarios. Next, we focus in presenting the particulates of these applications.

#### 2.1.1.1. Wireless fibre extender

Data rates in both fibre-optic and wireless communications have been increasing exponentially over recent decades. For the upcoming decade this trend seems to be unbroken, at least as far as fibre-optic communications is concerned. On the other hand, in wireless communications, the spectral resources are extremely limited, because of the heavy use of today's conventional frequency range up to 60 GHz. Even with spectrally highly efficient quadrature amplitude modulation (QAM) and the spatial diversity achieved with multiple-input and multiple-output (MIMO) technology, a significant capacity enhancement to multi-gigabit or even terabit wireless transmission requires larger bandwidths, which are only available in the high millimetre-wave (mmWave) and terahertz (THz) region [12]. Between 200 and 300 GHz there is a transmission window with low atmospheric losses. In contrast to free-space optical (FSO) links, mmWave or THz transmission is much less affected by adverse weather conditions like rain and fog.

Motivated by this, the concept of wireless fibre extender in the THz band, which is demonstrated in Figure 2, is considered as a key application scenario for TERRANOVA. This application scenario is especially interesting when it is intended to provide reliable data communication with very high data throughputs (up to 1 Tbps) for small distances (up to 1 Km) on adverse geographies, such as lakes, rivers and dams, or even due to regulatory constraints, where it is not allowed to perform civil work construction for a certain period of time. Specifically, in the near future, the users in rural or remote regions, which nowadays suffer from low-connectivity, should enjoy 10 Gbps data-rates. In these adverse geographies, deploying optical fibre may be extremely complex, time-consuming, since one should wait to capitalize scheduled road-reconstructions, and often very costly. On the other hand, the installation of the TERRANOVA wireless fibre extender has low-complexity, higher flexibility and lower cost.

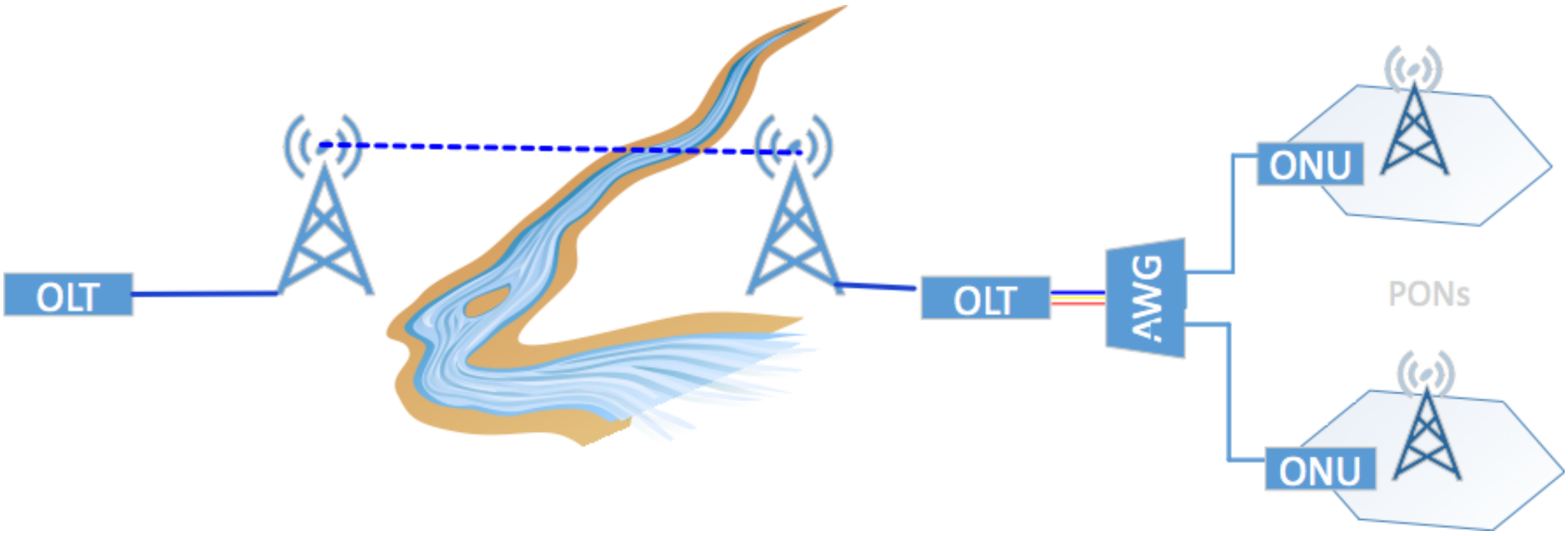

*Figure 2: The concept of wireless fibre extender.*

Currently, there are some commercial alternatives, which allow data communication with high throughput on adverse geographies, such as Free-Space Optics (FSO). However, FSO have significant performance limitations related to the weather conditions [13], [14], which typically limit its reliable use for long links (up to 1 Km in this use case).

#### 2.1.1.2. Short-range P2P applications

The P2P application scenario is especially interesting, when it is intended to have P2P data communication, with low latency and very high data throughputs (up to 1 Tbps) for small distances (up to 1 Km), and there is no access to fibre infrastructure, either due to technical or regulatory constrains that limit the fibre deployment, or even due to the cost of using existing fixed lines. In this scenario, there is no access to the fibre that could be extended using TERRANOVA technology (use case described above), so a possible approach here is to create a dedicated P2P link. This use case is also interesting in some key business scenarios, such as the stock exchanges applications, where very high throughput and extremely low latencies are critical for high frequency trading (HFT) [15]. HFT players trade on milliseconds of differences between bid and ask quotes on the same or different stock exchanges. To achieve this HFT players install equipment in or close to the stock exchange data centre, paying a fee for the data transfer in advance of other market players.

#### 2.1.1.3. Redundancy

The redundancy application scenario is also an important use case, in which critical services can benefit from the use of the wireless THz technology as backup for existing fixed lines. In this scenario, there is the possibility of enabling very high data rates (up to 0.1 Tbps) with QoS requirements for sensitive applications (availability ~99.999%) that should continue operate, even if the existing fixed lines get damaged, due to severe natural phenomena, or get loss of service, due to power outage. Moreover, in rural areas and in places where it is not viable to perform significant civil work constructions, it is very common to find exposed fixed lines. These are even more prone to failures, making it more important to have a redundancy scenario, possibly using TERRANOVA technology as a failsafe alternative.

### 2.1.2. Wireless access

This subsection details the access and small cell backhaul use cases of the TERRANOVA's technology. These are divided into two categories, namely the business use cases and last mile and open spaces use cases. The first category is designed to support corporate backup connection, IoT dense environments, data centres, indoor wireless access and ad hoc networks, while the second category refers to sport events, music events and last mile access.

#### 2.1.2.1. Business

##### 2.1.2.1.1. Corporate backup connection

The corporate backup connection is an exciting application scenario when it is intended to have a backup data communication for large and SMEs with very high data throughputs for small distances (up to 1 Km) and there is no access to a backup fibre access infrastructure, either due to regulatory constrains that limit the fibre deployment or even due to the cost of using/renting existing telecom links. Due to the very high data throughputs (up to 0.1 Tb/s), a single backup link using TERRANOVA technology could be used to serve several enterprise networks.

##### 2.1.2.1.2. IoT dense environment

Fully adopting digital networking in industry, commerce and public services, including traffic control and autonomous driving, remote health monitoring services, supply chain, security and safety procedures, automation of large production sites, places stringent requirements for Tbps class access subject to fast response constraints [16]. These applications scenarios describe the true colours of the well-known Internet of Things (IoT). The IoT dense environments is an interesting use case, in which TERRANOVA can be used to leverage industrial networking with very high data rates (up to 0.1 Tb/s), high reliability (application dependent) and low latency (less than 1 ms) in dense environments. Due to the huge available THz frequency range and the antenna directionality, TERRANOVA can provide a high level of immunity to environments that have high interference at the typical lower frequencies (usually up to the GHz frequency range). This scenario is particularly interesting in Industry 4.0 for short- to medium-range industrial networking (up to 500 m) where, due to technical, regulatory or economic constraints, it is not viable to use optical fibre links.

##### 2.1.2.1.3. Data centres

The use of TERRANOVA technology in data centres has the potential of allowing ultra-high speed wireless data distribution (up to 0.2 Tbps) per link for short range (up to 100 m) in noisy environments. In this sense, TERRANOVA could be used for establishing multiple links between processing and/or data storage racks with the potential advantages of simplifying the installation, reducing the amount of wired circuits interconnecting racks and possibly reducing the rack space utilization. This application scenario is also appealing considering that data centres also have high electromagnetic noise in the GHz frequency range, and TERRANOVA architecture can provide a high level of immunity and reliability of connectivity in a low mobility environment.

##### 2.1.2.1.4. Indoor short-range THz access

The increasing demand for higher indoor data rates makes an attractive application scenario for TERRANOVA. In this scenario, the indoors access network would experience a significant boost in terms of aggregated data rates (up to 0.3 Tbps), enabling new dimensions of interconnectivity and quality of experience (QoE) for very short range communications (up to 20 m), allowing wireless connectivity for specific applications, such as high-definition holographic video conferencing (i.e., virtual reality office).

##### 2.1.2.1.5. Ad-hoc access

Ad-hoc business access is an interesting scenario for sporadic events or in emergency situations, where it is important to have low latency, very high data throughputs (up to 0.1 Tbps) and reliable communication for small distances (up to 500 m). Music festivals in remote places, concerts or sports events with demanding special effects (like holographic images) could benefit from the temporary installation of TERRANOVA outcomes. Here, a key factor is the reduced setup time, where the TERRANOVA technology is expected to require limited installation time in the order of a few hours per link.

#### 2.1.2.2. Last mile and open-space events

##### 2.1.2.2.1. Sport, music events, etc.

Sporadic outdoors events, including sports and music events are very interesting use cases, due to the predictable short time duration, while requiring massive data throughputs [17]. Unlike the applications in 2.1.2.1.5 Ad-hoc access where the event is organised in a venue without suitable infrastructure and TERRANOVA based solutions must be installed and removed, in the present group we focus on permanent installation for stadiums, entertainment centres, etc. When the number of communication devices is significantly increased in a specific area, the communication traffic will also dramatically increase. As a consequence, the users' QoS and QoE will decrease. On the other hand, it is expected from the communication provider to guarantee the quality of the service. In such scenarios, due to technical or economic constraints, it may not be viable to use optical fibre links. On the other hand, TERRANOVA technology is a better alternative, due to the reduced setup time, the aggregated very high data throughputs (up to 0.2 Tbps) and reliable communication for the expected small average distances (up to 500 m).

##### 2.1.2.2.2. Last mile access

The last mile access is a key application scenario for TERRANOVA. This application scenario is especially interesting, when it is intended to have data communication with very high aggregated data throughputs (up to 0.1 Tbps) on the last mile access (up to 1 Km) and where, due to technical, regulatory or economic constraints, it is not viable to use optical fibre links. The easy installation is also a relevant point here, since it allows a faster deployment of the last mile access network, thus increasing the revenue margin from the service subscribers.

Nowadays, technologies such as G.fast that use low RF frequencies over twisted pair or coax lines are commonly used when deploying fibre in the last mile. However, this technology is limited in terms of bandwidth and will not map with applications requiring massive data throughputs. In this sense, TERRANOVA may become a natural evolution for extending higher bandwidth fibre access including NG-PON2 and future PON standards.

In all the above scenarios, TERRANOVA is an attractive complementing technology to the costlier and less flexible optical fibre connections and to the lower data rate wireless technologies including visible light communications (VLC), FSO, mmWave and Wi-Fi. It is widely known that FSO have significant performance limitations related to the weather conditions including rain and fog, which typically limit its reliable use for long links [13]. On the other hand, microwave and mmWave links are more robust to adverse weather conditions, but the state of the art (SOTA) throughputs is far below the required in the above application scenarios [18], [19].

## 2.2. Technical scenarios

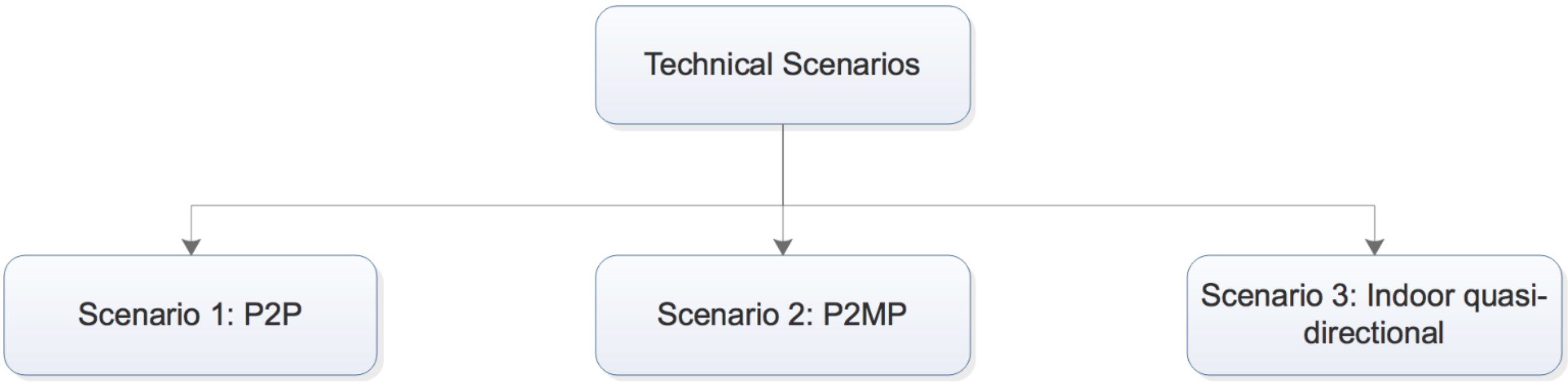

*Figure 3: Technical scenarios classification.*

As depicted in Figure 3, when classifying the identified uses cases presented in Section 2.1, we can identify the following relevant, physically different scenarios for the THz & optical network connections:

- Scenario 1: Point-to-point;
- Scenario 2: Point-to-multipoint; and
- Scenario 3: Indoor quasi-directional.

Next, we determine the characteristic of each technical scenario.

### 2.2.1. Scenario 1: Point-to-point

This scenario considers a THz point-to-point (P2P) connection with a single THz beam and is most relevant for stationary outdoor connections with large range and high capacity requirements, such as in the wireless fibre extender use case, presented in Section 2.1.1.1. In this case, the system's latency should not exceed the 1 ms, the link range should be in the order of 1 km, and support 1 Tbps data rate. Likewise, the P2P systems should be always available. Finally, the THz beam needs no or only limited steering capability.

### 2.2.2. Scenario 2: Point-to-multipoint

This scenario considers a THz point-to-multipoint (P2MP) connection with multiple THz beams, which are used to realize individual, independent communication links at the physical layer. The applications of such a scenario include stationary, nomadic or even mobile indoor/outdoor connectivity, with information broadcast or multiple independent data streams. For nomadic or mobile applications, the individual beams need to follow the receiver, therefore a significant and dynamic beam steering capability is required, depending on the solid angle to be served.

### 2.2.3. Scenario 3: Indoor quasi-directional

This scenario also considers a THz P2MP connection with multiple THz beams. However, the main difference from scenario 2 is that, the goal in this scenario is to cover a certain area, room or solid angle with uninterrupted THz connectivity. In this sense, the individual THz beams should complement each other to realize complete coverage for a nomadic or mobile "virtual" link rather than serving individual links. Each beam serves multiple "users" within its coverage with virtual links, sharing the physical link using multiplexing techniques and protocols. Beam steering (fixed or slow) may be only needed to generate multiple, slightly overlapping beams, while nomadic and mobile applications are served with handover techniques between the different beams.

## 2.3. Technical requirements

This section is devoted to derive the requirements on the link performance for the relevant use case scenarios for the co-designed THz and fibre-optical network. In order to evaluate the link and system performance the following key performance indicators are evaluated:

- Aggregate throughput of wireless access for any traffic load/pattern;
- Throughput of the point-to-point 'fibre optic - THz wireless' link;
- Link latency of the 'fibre optic - THz wireless';
- Range of the 'fibre optic - THz wireless' link;
- Reliable communications (probability of achieving a target bit error rate - BER and packet error rate – PER); and
- Availability ('Always' available connectivity of 'infinite' number of devices).

Table 1 illustrates the use cases presented in Section 2.1 matched with the technical scenarios defined in Section 2.2. From this table, it is evident that the key performance requirements depend on the nature of the application use case.

Table 1: Mapping between the use cases and the technical scenarios.

| Application Use Case | Scenario | Use case fundamental requirements |
|---|---|---|
| Fibre extender | 1 | Data Rate: 1 Tbps, Range: ~1 km |
| P2P | 1 | Data Rate: 1 Tbps, Distance < 1 km |
| Redundancy | 1 | Data Rate: ~0.1 Tbps, Availability:~99.999 % |
| Corporate backup connection | 1 | Data Rate: 0.1 Tb/s, Range: ~1 km |
| IoT dense environment | 2 and 3 | Data Rate: 0.1 Tbps, Latency: < 1 ms, Reliability (target BER): Application dependent |
| Data centres | 2 | Data Rate: 0.2 Tbps, Range < 100 m |
| Short range THz access indoors | 2 and 3 | Data Rate: up to 0.3 Tbps, Range < 20 m |
| Ad-hoc access | 2 and 3 | Data Rate: 0.1 Tbps, Range ~ 500 m, Target installation time < 1 hour |
| Sport events, music events, etc. | 2 | Data Rate: 0.2 Tbps, Range ~ 500 m |
| Last mile access | 2 | Data Rate: 0.1 Tb/s, Range: ~1 km |

Finally, the basic KPIs for the technical scenarios can be summarized in the Table 2, given at the top of the next page.

*Table 2: Basic KPIs for each technical scenario.*

| KPI | Scenario 1 | Scenario 2 | Scenario 3 |
|---|---|---|---|
| Max. THz link latency (ms) | 1 | 1 | 1 |
| Max. THz link range (m) | 1000 | 500 | 10 |
| Max. optical (wired) link range (km) | 50 | 10 | 1 |
| Number of connections per THz node | 1 | 10 | 100 |
| Max. THz link throughput x range (Gbps x m) | 1000 x 1000 | 100 x 1000 | 10 x 10 |
| Max. THz link throughput (Gbps) x connections (= aggregate throughput) | 1000 x 1 | 100 x 10 | 10 x 100 |
| Target BER | $10^{-12}$ | Application dependent | Application dependent |
| Availability | Critical | Critical | Application dependent |

# 3. System architecture

TERRANOVA envisions a heterogeneous, highly-flexible hybrid optical-wireless network architecture, which becomes an enabler of ultra-fast (in the order of 1 Tbps and more) beyond 5G systems. Therefore, it is critical to efficiently and flexibly handle the massive amount of QoS/QoE-oriented data that will be exchanged in a future big-data-driven society, along with the super high data rate and almost zero-latency requirements. Hence, wireless Tbps communications and the supporting BH network infrastructure are expected to become the main technology trend within the next 10 years and beyond.

TERRANOVA intends to provide unprecedented performance, not only by targeting data rates in the Tbps regime, but also by supporting novel usage scenarios and applications that combine these extreme data rates with agility, reliability and almost zero response time. Figure *4* depicts a high-level network architecture containing several different key scenarios where TERRANOVA can be used, including BH, FH and access.

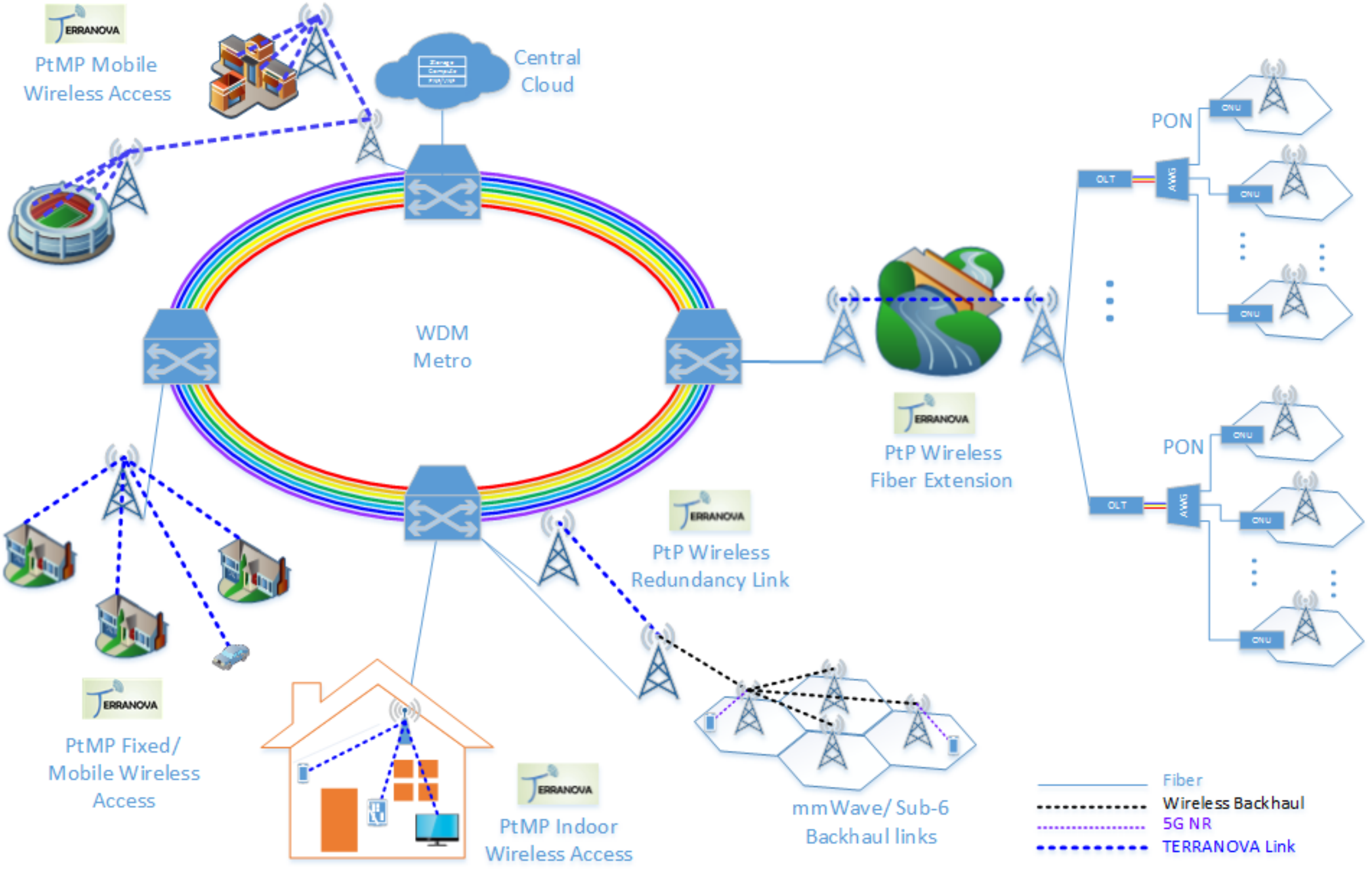

Figure 4: Physical network architecture.

The fixed access network (FAN) connects the THz link with the core network. As illustrated in Figure 2 and Figure *4*, radio heads (RHs) and base stations (BSs) can be connected either through dedicated optical fibres or via THz wireless links, which are called *wireless fibre extenders*. Given that in the near future, high-speed access should be made ubiquitously available to guarantee equal opportunities in the global competition, rural or remote regions that are difficult to access (e.g., mountains and islands), should be connected with high data rates up to 10 Gbit/s per user. This is either infeasible or prohibitively costly, when using solely optical fibre solutions. As a result, the use of wireless terahertz (THz) transmission as wireless backhaul extension of the optical fibre is an important building block to bridge the 'divide' between rural and dense urban areas and guarantee high-speed internet access everywhere in a cost-efficient manner, in the beyond 5G era.

Besides, since, the THz wireless link bandwidth can reach 1 THz, wireless THz links could be also considered as promising candidates to be integrated into the beyond 5G network as another *wireless access* branch for several bandwidth-hungry use cases (both indoor and outdoor, fixed and mobile users). Wireless THz access technology can be utilized complementary to the 5G new radio (NR) and offer connectivity between ultra-high-speed wired networks and personal wireless devices, achieving full transparency and rate convergence between the two links.

Indicative examples of bandwidth-hungry use cases are stadiums and concert halls during events. The unique characteristic of these cases is the extremely high concentration of mobile equipment for a short period of time. The need for ultra-high throughput can be covered by terahertz wireless access links. Furthermore, all this traffic can be aggregated and sent back to the network wirelessly. Since this extremely high traffic is not continuous in these cases, it can be forwarded to the network in an ad-hoc fashion, by employing equipment dedicated to other links. For example, as it can be seen in Figure *4*, a backhaul BS which is used for the traffic of a mall, can be utilized at night, where the mall has no traffic, to forward the traffic of a stadium.

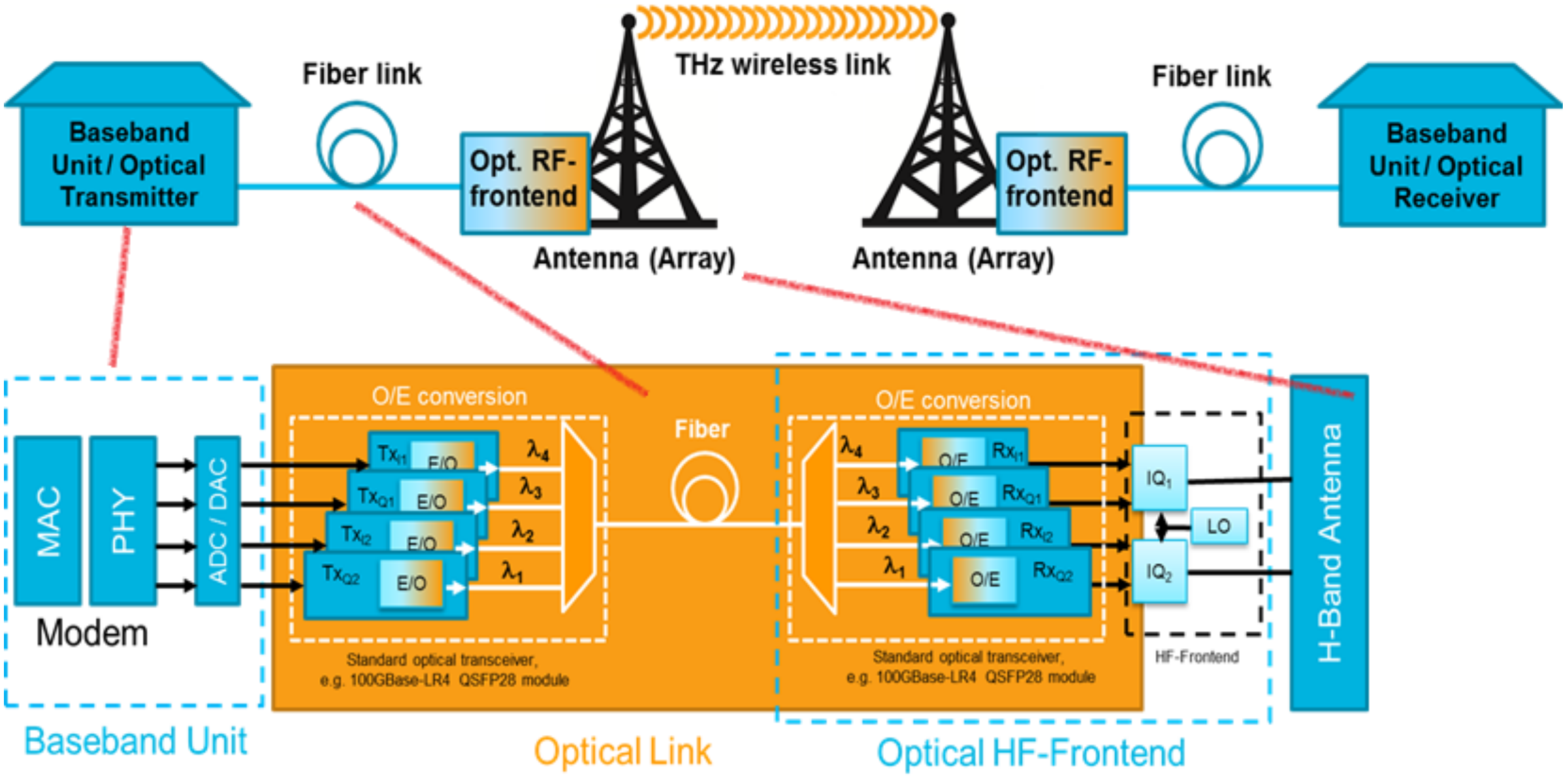

Figure 5: Schematic depiction of optical-wireless systems for replacement of fibre link by a wireless THz link (above) and an indicative example for an optical-wireless system with optical RF-frontend based on 100GBase-LR4 QSFP 288 transponder modules (below).

Figure *5* illustrates a schematic depiction of the TERRANOVA optical-wireless system for fibre replacement. Note that, in order to 'transparently' extend the capacity, range and reliability of the fibre optic link to wireless, novel THz transceiver designs should be presented.

For optical transportation, TERRANOVA employs passive optical networks (PONs). PONs have been considered as an effective solution for the access networks, because of the fact that PONs are able to provide huge bandwidth in a cost efficient manner [20]. A PON consists of an optical line terminal (OLT), which serves as the service provider endpoint, an optical link to a passive power splitter deployed at a remote site and individual fibres to optical network units (ONUs). In order to connect multiple ONUs in a single OLT an arrayed waveguide grating (AWG) unit is used. PONs share the optical bandwidth by a flexible assignment of transmission opportunities among the ONUs; thus, they enable statistical multiplexing[1]. Current PONs need further evolution in order to achieve the 1 Tb/s goal. Therefore, NG-PON2, which uses wavelength division multiplexing (WDM) to multiply capacity to Nx10 Gb/s, seems to be an attractive solution in order to deal with the intense telecommunication traffic, which is expected to be caused in the beyond 5G systems [21].

All access architectures end in the central cloud (CC), where the traffic from multiple BSs is aggregated. The CC is placed in a centrally located data centre. The centre hosts a large collection of processing, storage, networking, and other fundamental computing resources. In this node, tenants are allowed to deploy and run arbitrary software, such as operating systems and applications.

The metro network domain is responsible for providing reliable connectivity between the access branches and the regional/core data centres, as well as to the global internet. Metro solutions employ dense WDM (DWDM) in metropolitan areas [22]. As a consequence, up to 100 wavelength channels can be transported in parallel over the same fibre to satisfy the large aggregated traffic from the access branches. Besides advantages for longer distances typical in the metro domain (using coherent optical transport technologies), DWDM leads to aggregate capacity in the range of some Tbps over a single optical fibre [23].

[1] The number of ONUs in a PON is physically limited, because of the path loss at the splitting point.

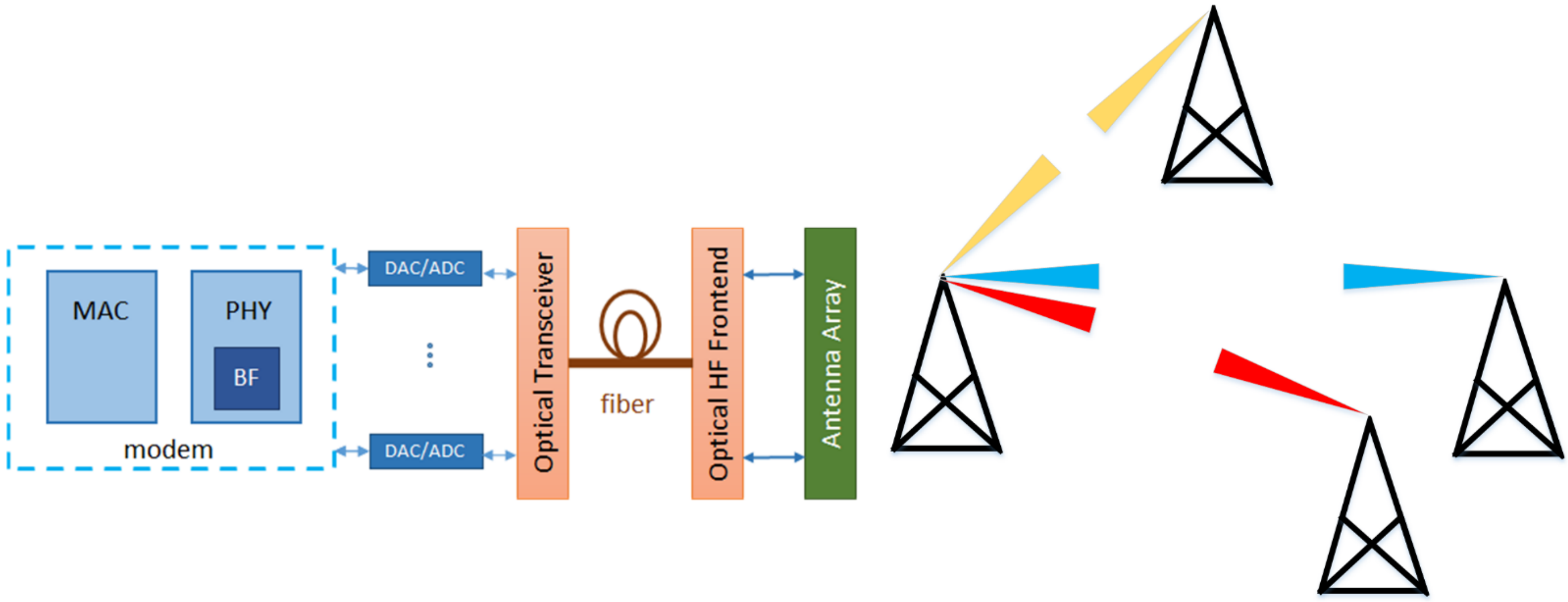

*Figure 6: P2MP Fixed Wireless Access (Backhaul)*

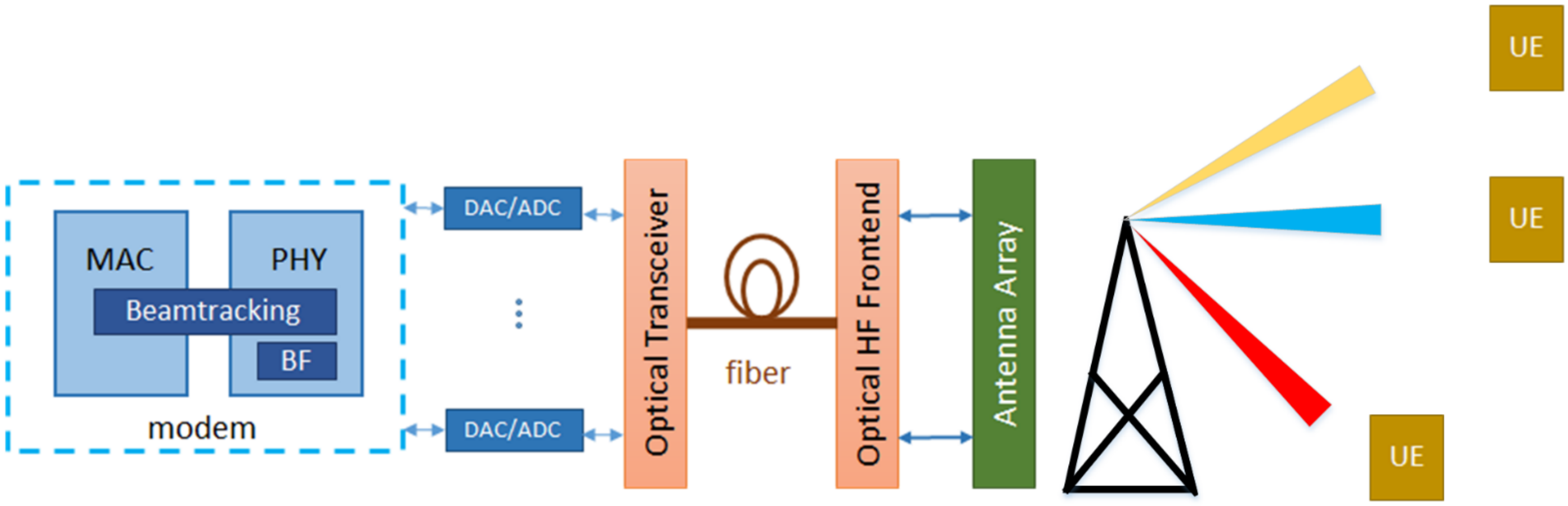

*Figure 7: P2MP Mobile Wireless Access*

Wireless THz links can be used for P2MP Fixed Wireless access (see Figure 6), thus, constituting a backup for the backhaul of BSs in several application scenarios, such as redundant infrastructure for disaster recovery or for coverage in emerging markets. Actually, this corresponds to the redundancy scenario where TERRANOVA technology is used to backup existing fixed lines in the case of critical services, which require a failsafe alternative.

Wireless access or indoor scenarios involve shorter distance links and, thus, relax the directivity demands, while, on the other hand, intensify the problem of discovering and tracking the users and controlling the interference in multi-user cases. As pencil beamforming is needed to realize both high available link SNR and narrow beamsteering angle, UE detection and tracking will be major challenges for point-to-multipoint scenarios, both at the physical and MAC layers. In the P2MP Mobile Wireless Access scenario, where independent THz pencil beams are potentially directed to individual users (see Figure 7), user equipment (UE) detection and tracking could be realized using (legacy) lower frequency regimes with omni-directional antennas, while only the high capacity connection is realized via the THz wireless connection. Interference issues between the beams will likely result in limitations with respect to how close the individual users can be located.

In the indoor scenario depicted in Figure 8, uninterrupted communication can be guaranteed by using beam tracking as well as coordination multi-point (CoMP). One option to solve the issue of UE tracking and interference is to realize a system with many fixed beams pointing in slightly different directions, achieving full coverage of the indoor environment. In all the P2MP scenarios, the key technology challenge is the realization of antenna arrays with sufficient number of antenna elements to implement the beam steering. The required array factors will depend on the use case, as well as the question for a one- or two-dimensional array. The same holds for the different beamforming options (local oscillator, RF or digital beam steering), which have different advantages and disadvantages.

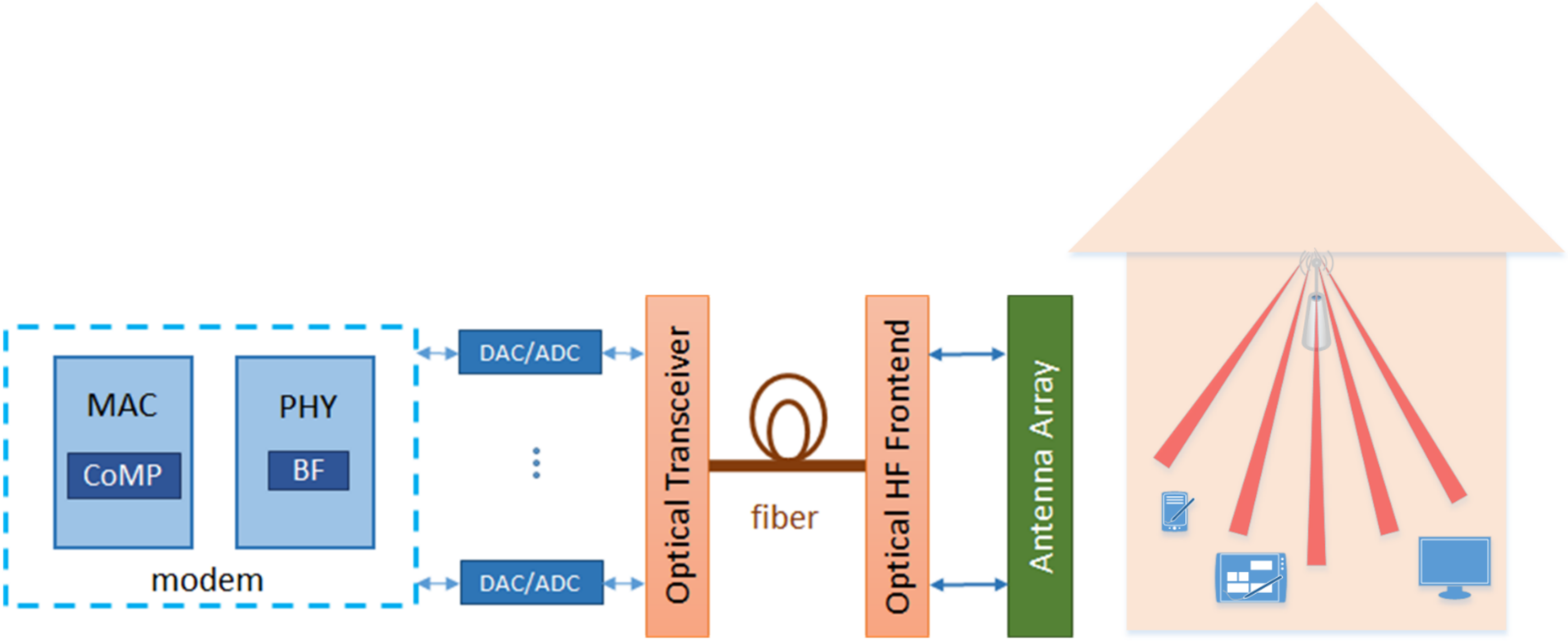

Figure 8: P2MP indoor wireless access.

# 4. OPTICAL LINK AND TERRANOVA MEDIA CONVERTER DESIGN

This section is focused on discussing the optical link options and the TERRANOVA candidate media converter design. In more detail, Section 4.1 reviews the standardized optical transceivers, and Section 4.2 discusses the candidate media converter designs.

## 4.1. Standardized Optical Transceivers

### 4.1.1. IM/ID Transceivers

Nowadays, two main detection schemes can be used to convert an optical signal to the electrical domain: direct or coherent detection. The first one is simply based on intensity modulation and direct detection (IM/DD), and so at the receiver side only the amplitude of the optical field is converted into an electrical signal by a simple photodiode. In the second, the received signal is mixed and boosted with an optical source signal provided by a local oscillator (LO), and all optical information of the signal, i.e., amplitude, phase and polarization, can be converted into the electrical domain [24].

Most applications in data-centres, metro and access networks employ the use of IM/DD pluggable transceivers as they are reliable for relatively long distances and are cost-effective for these applications compared to coherent solutions.

Most common form-factors of the IM/ID transceivers are the well-known SFP+, QSFP28, XFP and CFP2, which bit-rates vary from 1 Gbps to 100 Gbps, and distances from hundreds of meters to 80km and more.

*XFP transceivers:* began in P2P applications, but soon spread to PON applications. Since they are cost-effective solution, they can support the recent technologies, such as XG-PON, XGS-PON and NG-PON2. PICadvanced is specialized in XFP form-factor development for the new access network technology NGPON2, which supports 10G/10G and up to 64 clients for each of the 4 wavelengths enabling 40 Gbps aggregate traffic in downstream. XFP transceivers (10 Gigabit Small Form Factor Pluggable) follow the multi-source agreement (MSA), so despite of the different technologies such as 10G Ethernet, SONET/SDH, ITU-T 10G PON, XFP transceivers follow the same basic recommendations such as electrical interface, management interface and power dissipation limits.

*Interface:* As it is common to many other form factors transceivers, XFP transceivers use I2C management interface, in order to control and read internal diagnostics. XFP transceivers use two lanes of 10 Gbps, transmitter (TX) and receiver (RR) data paths – CML or LVPECL - to provide 10 Gbps symmetrical service. When possible, they employ the use of CDR and CTLE to improve even further the link budget and reach longer distances. A basic block diagram of an XFP can be found in Figure 9.

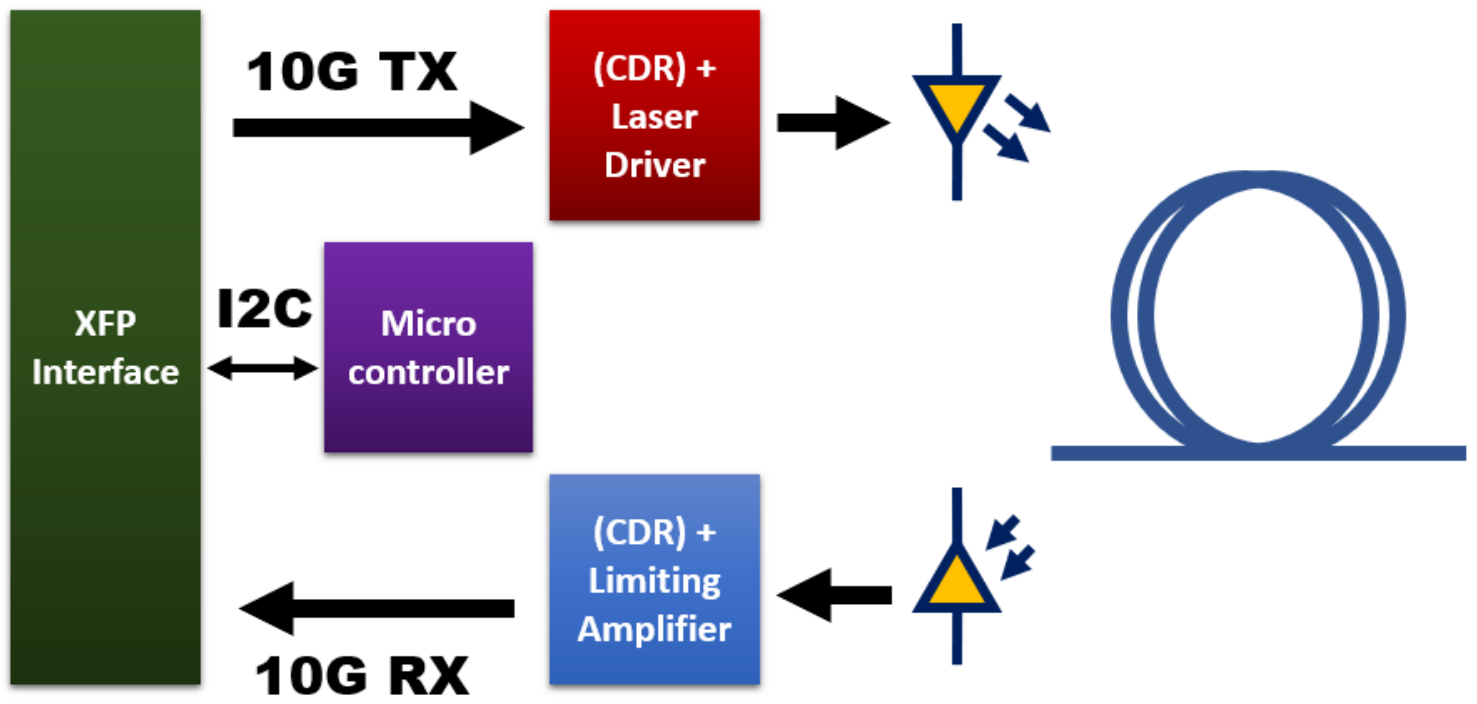

*Figure 9: XFP block diagram.*

### 4.1.2. Coherent transceivers

Backbone applications have the needs for higher capacity and longer fibre reach, in which IM/DD technology can no longer compete. Coherent transceivers have superior optical performance and can provide electronic equalization of fibre impairments, such as chromatic dispersion, more over as the phase from the signal is recovered higher order modulation schemes are allowed leading to the support of higher bitrates. Coherent solutions are more expensive when compared to common IM/DD transceivers as they integrate very high-speed analogue-to-digital converters (ADCs) and digital-to-analogue converters (DACs), along with digital signal processing (DSP) capabilities. Nonetheless such transceivers in the market can already achieve bitrate up to 200 Gbps.

Transceivers for coherent applications can be divided in analogue coherent optics (ACOs) and digital coherent optics (DCOs) subcategories. Both contain all necessary optical components for transmit and receive in a single package. ACO offer lower cost, and typically smaller form factor, however DCOs already comprises the use of ADCs/DACs and DSP inside the package, while ACO needs DSP capabilities in host board.

Typically, coherent optical transceivers have the CFP/CFP2 form factor and follow the Multi-source Agreement / OIF agreement for coherent optics and all the inherent guidelines. CFP/CFP2 uses MDIO management interface and can have different types of data path interfaces, digital (DCO) or analog/RF (ACO). A typical block diagram for a coherent RX can be found in Figure 10.

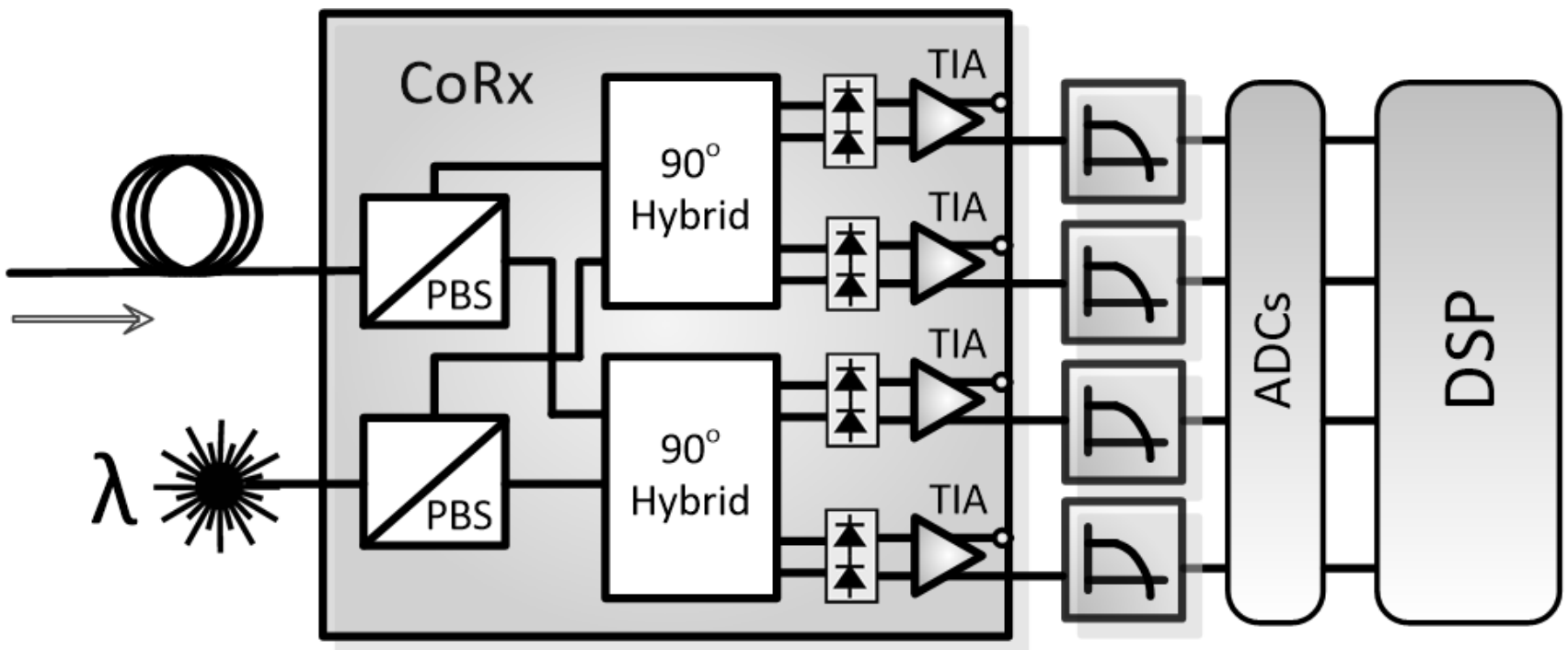

*Figure 10: Typical block diagram of a coherent RX.*

This detector employs polarization-diversity, where two phase-diversity configurations by using two 90° optical hybrid components are combined to detect the in-phase and quadrature signals of each polarization of the light. The polarization beam splitter (PBS) is used to split both signal polarizations, and after four balanced detectors, the in-phase and quadrature signals from each polarization can be sampled using ADCs. Therefore, a great advantage of these receivers is the ability to detect the full information of the optical field, enabling advanced modulation formats with the four dimensions available in the optical transmission. And finally, advanced DSP can be employed to manipulate the digitalized information towards to compensate for instance the distortions of the optical link.

## 4.2. TERRANOVA candidate media converter design

In this section we present several possible solutions for TERRANOVA Media Converter Design. TERRANOVA Media Converter enables the possibility to transmit from 100 Gbps up to 800 Gbps from TERRANOVA radio front-end through the fibre.

In order to transmit such high data rates, we have to consider state of the art IM/DD optical technologies and optical coherent transmission. Coherent solution requires conversion from analogue-to-digital domain and vice-versa, which can be very expensive due to the cost of ADCs/DACs and DSP. The IM/DD solution does not require the use of those components.

### 4.2.1. Proposal 1 – IM/DD using PICadvanced NG-PON2 technology

The first solution presented comprises the use of a time wavelength division multiplexing (TWDM) transceiver (included in PICadvanced portfolio) that allows to transmit multiple 10 Gbps links with different wavelengths in a single fibre. It can be viewed as a lower cost solution when compared to commercial IM/DD 100 Gbps solutions.

For this type of application the Figure 11 depicts the proposed optical architecture with the TERRANOVA Media Converter. Therefore, the idea behind the system is to multiplex a defined number of 10 Gbps modulated optical carrier in order to achieve the desired aggregated data rate over the optical link. For the XFP solution, the non-return-to-zero (NRZ) modulation format must be used due to the integrated subcomponents as limiting amplifiers and transimpedance amplifiers (TIAs), or the clock and data recovery (CDR) subsystem.

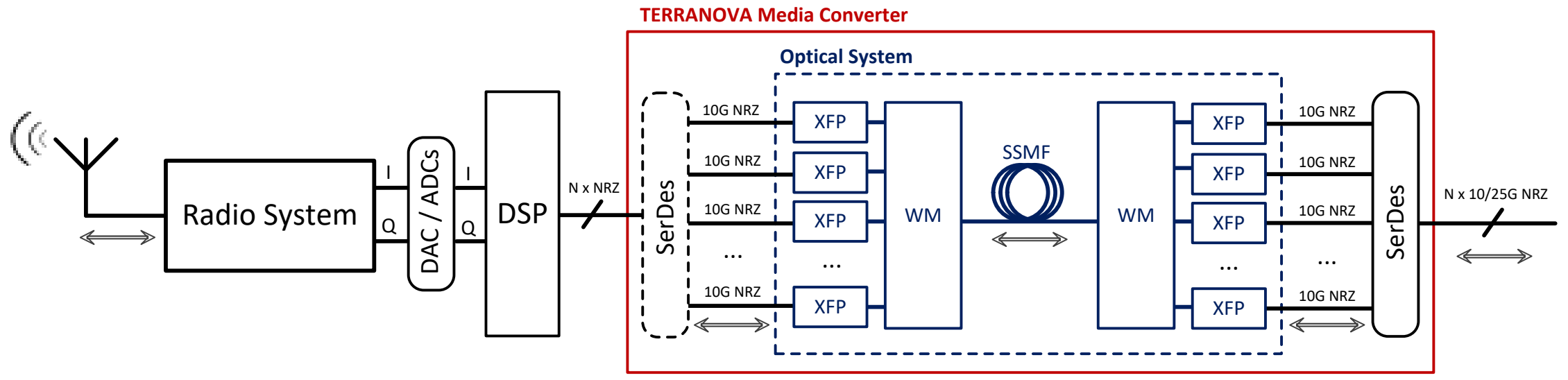

*Figure 11: Proposed optical architecture with TERRANOVA media converter.*

For the "central office-to-radio" direction, the N parallel bit stream lines reaches a serializer/de-serializer (SerDes) device, which converts the N parallel 10 Gbps / 25 Gbps data signals into 10 Gbps signals. The de-serialized 10 Gbps NRZ signal lanes will then feed the different XFP which will modulate the different optical carriers.

Each XFP transmits in a different optical wavelength, resulting in multiple wavelengths at 10 Gbps each (WDM network). The wavelength multiplexer (WM) is used to combine all wavelengths, and at the receiver side each one is then filtered in order to be direct detected at the optical transceiver level. After photo-detection, 10 Gbps electrical NRZ can then be connected to a DSP. Another option is to first convert to a different line rate using a SerDes device in order to match the connections. At the DSP, no digital equalization is required for the NRZ signals, therefore they can be directly linked to the DSP. The DSP is only used to generate the in-phase and quadrature electrical signals to be transmitted over the radio system.

For the "radio-to-central office" direction, the radio signal is affected with dispersion or phase noise, and therefore before being transmitted over the optical link, it must be compensated in the DSP. The ADCs are used to digitize the in-phase and quadrature electrical signals. After the DSP, the signal can then be sent to the optical link using multiple 10 Gbps electrical NRZ signals. The optical transceiver solution is also based on multiple XFPs, each one transmitting in an optical wavelength, and the WM is used to combine all wavelengths.

At the fibre, the "radio-to-central office" direction signal is composed by multiple 10 Gbps NRZ wavelengths. Note that the "central office-to-radio" direction wavelengths must be different from the "radio-to-central office" direction wavelengths in order to avoid non-linear crosstalk. At the receiver side, each wavelength is firstly filtered using an optical WM and then detected by the XFPs.

The advantages of this technology are the following:

- Low cost solution. Since the idea is to parallelize a high electrical bandwidth signal (100G/800G) into multiple 10 Gbps signals, the transceiver solution is cost-effective. In addition, the transceiver is fully based on IM/DD components;
- High flexible optical solution. The aggregated data rate can be easily extended by increase the number of parallel XFPs.

And the disadvantages are:

- Low spectral efficiency. Since the modulation format is NRZ, and a wavelength guard band must be used to separate the different wavelengths (e.g., 100 GHz as the used in the NG-PON2 technology), the optical spectral efficiency is low;

- Fibre length links up to 20/40 km. In case the NG-PON2 standard is used as baseline, maximum reach would be 40 km as the band of transmission presents chromatic dispersion.

### 4.2.2. Proposal 2 – IM/DD using COTS 100G

Commercial off the shelf transceivers for 100 Gbps applications with IM/DD already exists in the market - such as CFP to CFP4 form factors - and presents a concurrent solution for proposal 1. Although this solution is available, it can be more expensive than solution for Proposal 1, when applications for 10 km+ or 100 G+ are needed. A possible architecture of this proposal is depicted in Figure 12.

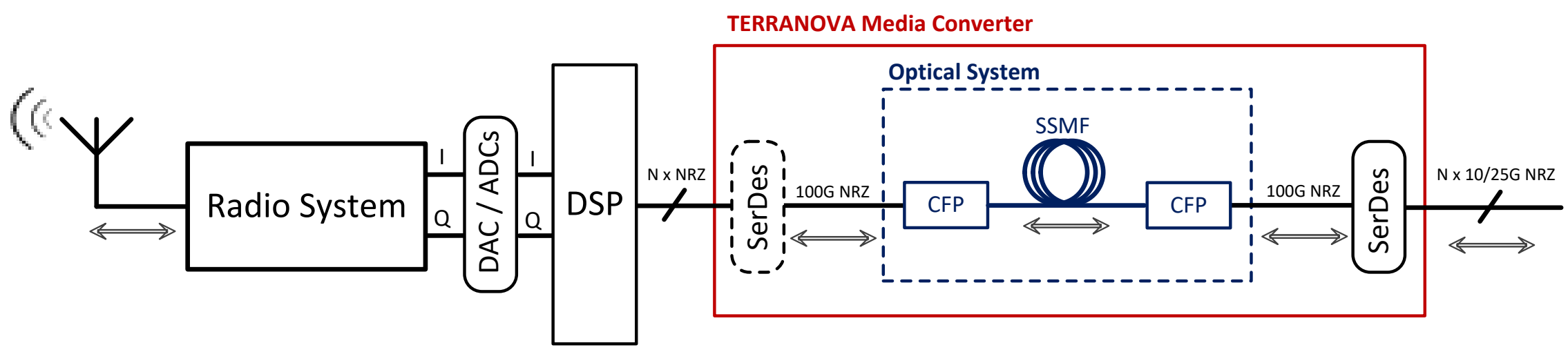

*Figure 12: Possible architecture of proposal 2.*

Although quite simple, the solution does not offer high flexibility. A commercial off the shelf transceiver is employed, and thus the solution is fixed at 100 Gbps, not allowing parallelization since the wavelength is fixed. The purpose of the DSP is to generate and demodulate the in-phase and quadrature signals from the radio system, since the optical interface is based on NRZ signals. This solution can be also limited at <20 km due to the chromatic dispersion.

### 4.2.3. Proposal 3 - IM/DD transceivers based on amplitude modulation

Proposal 3 comprise a similar scenario to Proposal 1. However, in this case the direct amplitude modulation is applied from the I/Q radio signals to the optical transceiver. The main advantage for this application is that the I/Q signals from the THz antenna can be directly mapped into the optical transceivers. Unfortunately, this advantage comes at the expense of the fibre range. The optical link would be reduced from the 20/40 km presented for Proposal 1.

Figure 13 shows the proposed optical architecture. The idea consists in to use two optical wavelengths per direction to transmit both in-phase and quadrature signals. Highly linear IM/DD transceivers must be used to achieve full amplitude modulation.

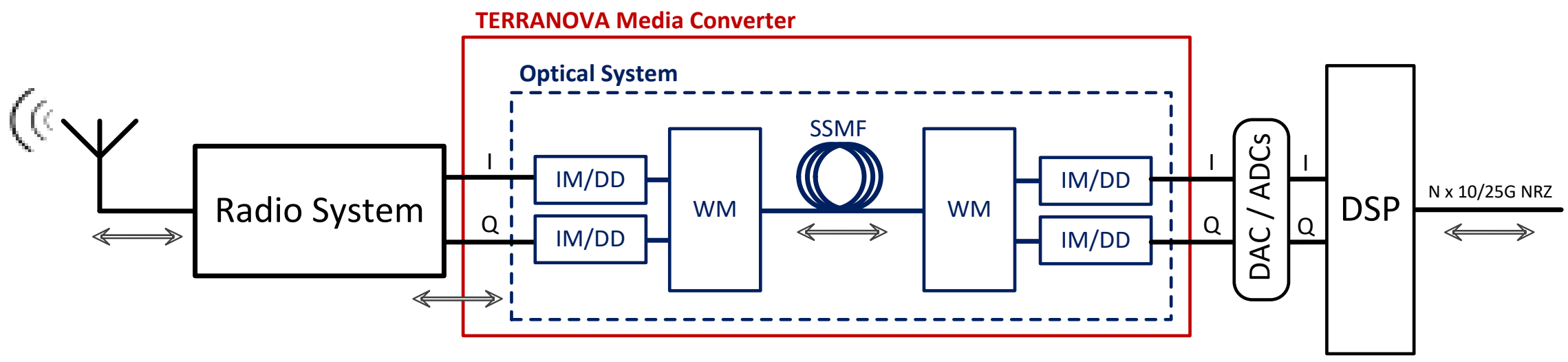

*Figure 13: Possible architecture of proposal 3.*

For the “central office-to-radio” direction, the DSP is used to generate the in-phase and quadrature electrical signals, and each signal is modulated in a different wavelength by using for example a Mach-Zehnder

Modulator (MZM). The WM is used to combine both wavelengths, and at the receiver side both are direct detected. After optical detection, the in-phase and quadrature signals are available to be transmitted to the radio system.

For the “radio-to-central office” direction, the concept is similar, and therefore other two wavelengths are used to transmit the in-phase and quadrature radio signals. Since both radio signals reach the optical system affected with dispersion or phase noise, at the receiver side both signals must be sampled and compensated in DSP. The DSP could be located also after the radio system IQ converter, but after the optical system it is also required to demodulate the in-phase and quadrature signals.

This technology is cost-effective since only four IM/DD optical transceivers are required and has a great advantage which is it’s easily scalable for higher baud rates. Modulation schemes of higher order, such as 256-QAM to 1024-QAM are transparent for this technology. On the other hand, since both in-phase and quadrature radio components are high bandwidth signals (>16 Gbaud), in the optical link they are affected by chromatic dispersion thus the typical optical reach for such scenario would be less than 10 km.

For the TERRANOVA project, and to the best of our knowledge, there is no commercial off the shelf solution available for this approach. Thus, this proposal requires further investigation, since most of the similar configurations in the literature are based on the radio over fibre (RoF) applications using for instance orthogonal frequency-division multiplexing (OFDM). In [25], experimental results of transmitting several digital 16-QAM RoF signals with high spectral efficiency and based on sub-carrier multiplexing (SCM) techniques were discussed. A high spectral efficiency is employed to drive a simple MZM and at the receiver, the optical signal is converted to electrical by a photoreceiver. Then a reference of the transmitted signal is available at the electrical domain, and no DSP subsystem at the receiver is applied for optical equalization, being only used filtering subsystems to improve the signal-to-noise ratio (SNR).

Since the idea of our proposed solution is based on transmitting two electrical baseband signals with amplitude modulation, i.e., both I/Q radio signals carrying multi-level modulation (which were previously slight distorted over the radio channel), an approach widely investigated that can be compared to the proposed is based on transmitting pulse amplitude modulation (PAM) signals using IM/DD transceivers. For instance, in [26], a single optical carrier with a single polarization 56 Gbaud PAM8 signal combined with raised cosine shaping and pre-emphasis, is propagated over 80 km fibre using simple IM/DD transceivers. However to mitigate the chromatic dispersion induced over the 80 km link or bandwidth limitations, a 3-tap feed-forward equalizer (FFE) is applied at the receiver. If the symbol rate is decreased and lower distances were used, the FFE may be avoided. In addition, in [27] and [28], Nyquist-shaped PAM signals are also employed using IM/DD transceivers to achieve single data rate lines of >100 Gbps for shorter-reach applications. Also, in [29], 56 Gbps PAM signals are demonstrated for inter-data centre connection optical networks over 100 km optical link. All these configurations are using amplitude equalizers (e.g., FFE) at the RX side towards to improve the eye diagram performance, and it could also be applied to this proposal since the radio signal should be restored (and equalized) in the DSP located at the CO.

#### 4.2.4. Proposal 4 – Optical Coherent transmission

In applications where fibre cannot be employed and it is needed an over the air fibre extension (TERRANOVA possible application), like the application below, analogue coherent transmission rise as an interesting solution. This solution could reach dozens or hundreds of kilometres and only requires, in a best case scenario, the DSP+ADC/DACs to be located in the central office to recover the signals and mitigate wireless and optical impairments. However, this kind of transceivers and DSP for coherent optical links are expensive which is a factor to take in consideration when one thinks of future implementation of such THz systems in the field.

Figure 14 depicts details of this technology. Both single or dual polarization can be used for the transmitter side in both traffic sides, but to use a standard coherent DSP, the coherent receiver must include a phase- and polarization-diversity configuration. The schematic represents the single polarization approach for the transmitter side, with the in-phase and quadrature signals directly connected to the IQ modulators of the coherent transceivers.

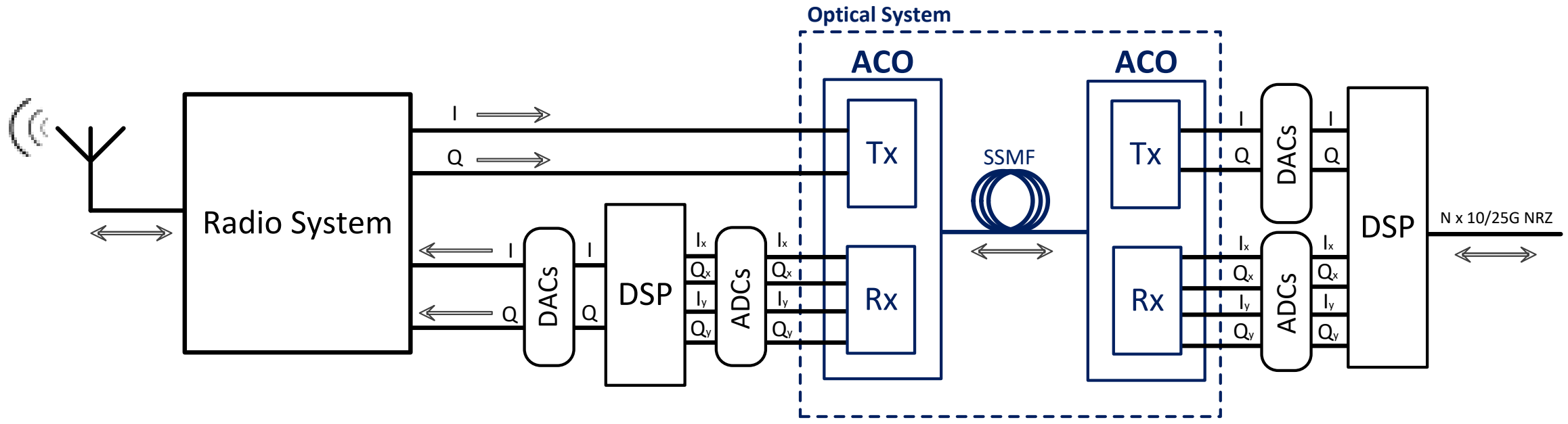

*Figure 14: Possible architecture of proposal 4.*

During the propagation over the optical link, the signal is affected with polarization rotation or chromatic dispersion. Therefore, from the "central office-to-radio" direction, since the signal generally reaches the coherent receiver destroyed, before being transmitted to the radio channel, it should be ideally recovered, otherwise the peak-to-average power ratio (PAPR) of the signal is high. Traditionally, the coherent DSP includes the following subsystems [30]:

- Pre-processing subsystems with a matched or a low-pass filter in order to mitigate noise interferences;
- Amplitude normalization subsystem to improve the dynamic range of the DSP;
- Clock recovery algorithm towards to compensate the ideal sampling instant of the ADCs;
- Static equalization to compensate the chromatic dispersion;
- Adaptive equalization to compensate the state-of-polarization (SOP) of the optical signal;
- Carrier frequency and phase recovery respectively to compensate the frequency and phase noise between the transmitted and the received local oscillators lasers.

For the "radio-to-central office" direction, the DSP located on the central office side is expected to compensate both radio and optical interferences.

# 5. THz challenges & physical system limitations

In this section, we focus on the fundamental characteristics of THz systems that will affect the design of the wireless THz system. These characteristics can be summarized as follows:

- In the THz region, because of the small wavelength, we are able to design high directional transmit antennas and receive antennas with low acceptance angle. These antennas are employed to countermeasure the high channel attenuation. However, at the same time, they require an extremely accurate alignment between the communication nodes.
- The high material absorption in the THz band makes doubtable the use of the non-line of sight (NLOS) communications. As a consequence, beam tracking schemes as well as coordination multi-point (CoMP) needs to be used in order to guarantee uninterrupted communication.

- Molecular absorption in the THz frequencies causes frequency and distance dependent pathloss (PL), which makes specific frequency windows unsuitable for establishing a communication link. Therefore, although the high bandwidth availability in the THz region, windowed transmission with time varying loss and per-window adaptive bandwidth usage is expected to be employed.
- In order to increase the links capacity, suitable multiple-input and multiple-output (MIMO) techniques in combination with beamforming (BF) need to be employed.
- In order to countermeasure the impact of transmitter (TX) and RX misalignment and support tracking of mobile or moving user equipment (UE), adaptive beamsteering is expected to be utilized. Adaptive beamsteering enables the low-complexity link installation and guarantees that the TX and RX antennas are aligned.
- In all cases, due to small wavelengths, there are high requirement on intra and inter beam coherence.
- Furthermore, due to small wavelengths, multi-path fading in case of NLOS link will be quickly changing already at small spatial movements leading to highly time-variable non-flat channel characteristics in nomadic applications.
- Finally, AMC schemes will be employed in order to increase both the range and the throughput of the THz system, while, at the same time, guarantee a pre-defined degree of reliability.

Due to the fundamental characteristics of the THz systems, it is evident that the propagation environment suffers from sparse-scattering. This causes to the majority of the channel direction of arrivals (DoAs) to be below the noise floor. As a consequence, a channel in a wireless THz system can be established in a specific direction with a range that varies according to the directionality level. However, the directionality of wireless THz channels result in two consequences, namely:

- *Blockage*, which refers to the high penetration loss, due to obstacles and cannot be solved by just increasing the transmission power; and
- *Deafness,* which refers to the situation, in which the main beams of the transmitter and the receiver are not aligned to each other. This prevents the establishment of the communication link.

In order to overcome *blockage*, the wireless THz system is required to search for and identify alternatives directed spatial channels, which are not blocked. However, this search entails a new BF overhead of significant amount and hence it introduces a new type of latency, in which we will refer to as BF latency. As a consequence, the medium access control (MAC) design for cellular networks is more complicated than the one of the conventional wireless local area networks (WLANs), in which short range communications can be also established through nLOS components. Additionally, the conventional notion of cell boundary becomes questionable in these systems, due to the randomly located obstacles. As a result, a redefinition of the notion of the “traditional cell” into “dynamic cell” is required.

On the other hand, *deafness* has a detrimental effect on the complexity of establishing the link and causes a synchronization overhead increase. This indicates the importance of redesigning the initial access (IA) procedures.

Likewise, in order to countermeasure the physical limitations of wireless THz systems, the MAC mechanisms may simultaneously exploit both the microwave and THz bands [31]. This was initially presented in [31], where the authors provided a MAC protocol that employs microwave frequencies for establishing the control channels (CCHs), and THz frequencies for the data channels (DCH). Additionally, MAC mechanisms may need to facilitate the co-existence of several communication technologies with different coverage. As a consequence, two different types of heterogeneity are observed in wireless THz networks, namely:

- Spectrum heterogeneity; and
- Deployment heterogeneity.

By *spectrum heterogeneity*, we refer to the scenarios in which wireless THz UEs use both high (THz) and lower frequencies (e.g. in the microwave band). On the one hand, THz frequencies provide a massive amount of bandwidth for high data rate communications. On the other hand, the microwave frequencies are used for control message exchange, which demands much lower data rates, but higher reliability than data communications. This facilitates the deployment of wireless THz networks, due to possible omnidirectional transmission/reception of control messages, as well as higher link stability, at lower frequencies. However, the use of both microwave and THz bands in UEs increases the fabrication cost and may result to an important reduction of the mobile UE (MUE) energy autonomy. Moreover, due to the blockage and deafness effects, the establishment of a microwave CCH might not result in establishing the corresponding THz data transmission channel.

The *deployment heterogeneity* introduces two scenarios for THz cellular network, namely:

- stand-alone networks; and
- integrated networks.

In the *stand-alone* scenario, a complete THz network (from macro to pico levels) will be deployed, whereas the *integrated network* solution is an amendment to existing microwave networks for performance enhancement, and includes wireless THz small cells (SCs) and/or THz hotspots [32].

# 6. Required functionalities & enablers

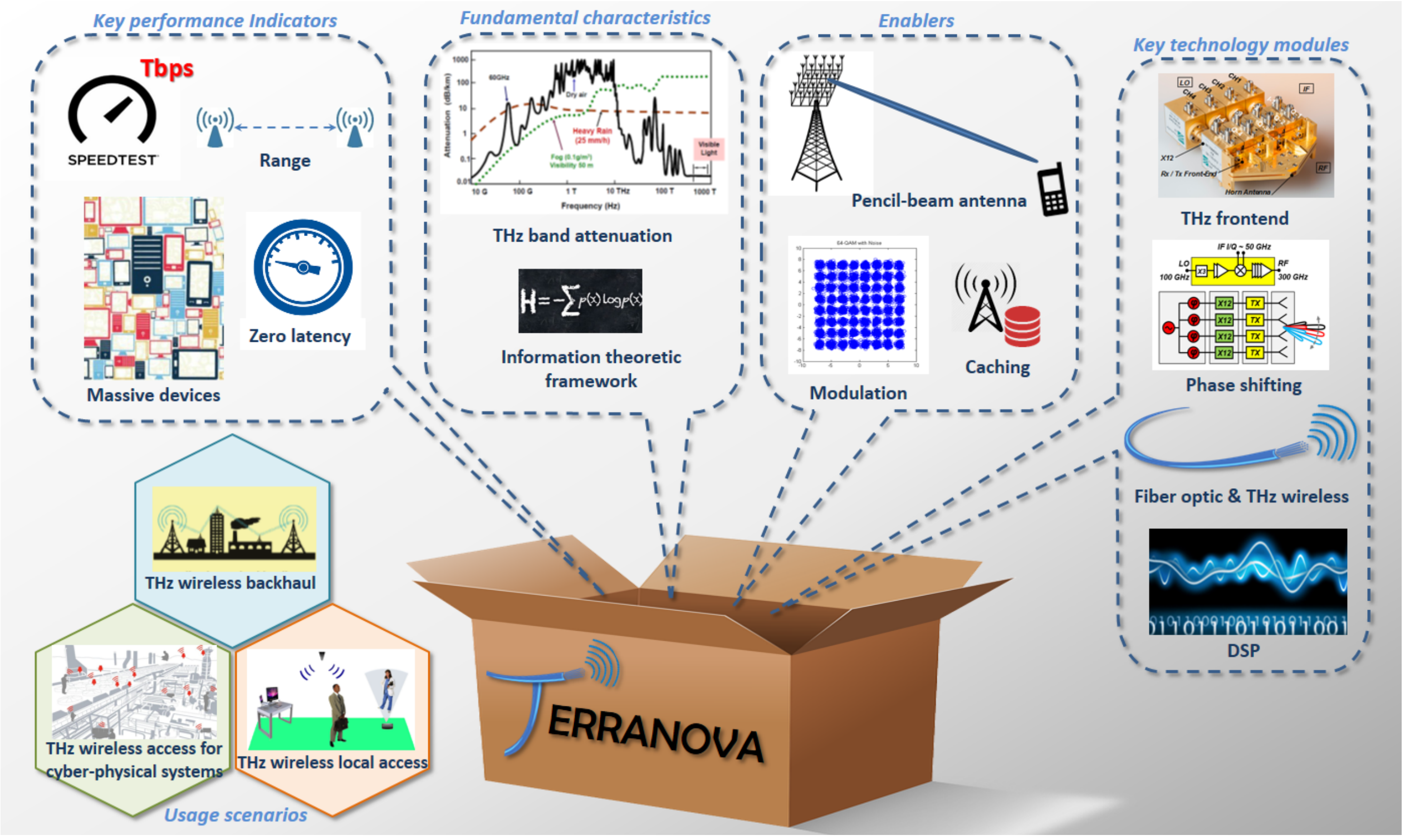

Figure 15: TERRANOVA system concept and enablers for the beyond 5G network.

As illustrated in Figure 15, in order to address the TERRANOVA use cases requirements, which are provided in Section 2, in the radio access network (RAN), a combination and integration of new concepts with excising technology enablers as well as key technology modules is anticipated. The use of THz communications is expected to reduce further the size of the small cells and make them denser than in the current setups. Additionally, the use of massive multiple-input multiple-output (MIMO) systems, which enable the utilization of pencil BF schemes, will allow the deployment of more efficient interference coordination and management schemes, basically due to the use of coordination multi-point – CoMP techniques [33]. Furthermore, it is necessary to jointly design baseband signal processing algorithms for the complete optical and wireless link, as well as to develop broadband and highly spectral efficient radio frequency (RF) frontends operating at frequencies higher than 275 GHz, and novel standardized electrical-optical (E/O) interfaces. Finally, to address the extremely large bandwidth, the propagation properties of the THz regime, and to achieve the almost zero latency requirement of several application scenarios, improved channel modelling, the design of appropriate waveforms, multiple access control (MAC) protocols, caching schemes and antenna array configurations are required.

## 6.1. Channel modelling

The THz links suffer from several path loss mechanisms, including the free space path-loss (FSPL) and the molecular absorption loss (MAL). The latter is a distinguishing feature of the millimetre and sub-millimetre bands. The main difference between the mmWave and the THz band is the progressively increasing MAL. At longer link distances, below 300 GHz, links are dominantly attenuated by the FSPL, whereas in the THz regime, the molecular absorption becomes more important, due to its exponentially increasing impact as a function of distance. At short distances (tens of meters), the FSPL remains the dominant loss below one THz frequencies. The existence of the molecular absorption loss highly depends on the level of the water vapor in the atmosphere, since, in this band, water vapor efficiently absorbs energy. This indicates that there are global variations in the level of absorption. Therefore, for standardization, it is important to take into account the regional propagation properties.

Although, the THz frequencies are often considered in full line-of-sight (LOS) conditions, it has been shown that communications via non-LOS (NLOS) paths is also possible, e.g., by reflections [34] and penetrations [35]. The THz signals are rather easily blocked by thicker objects, such as humans. Thus, the possibility of utilizing the NLOS paths as primary communications channels is plausible in the case the LOS path is not available. If the THz links are established outdoors, one will have to take into account that, e.g., rain, fog, and clouds also attenuate THz signals [36] [37]. These are random phenomena, but for which the probability of occurrence is also dependent on the time of the year, location, and altitude on Earth.

In order to accommodate with the THz channel particularities, several models have been presented in the open technical literature. For instance, in [38] is shown a simple channel model for 275 – 400 GHz frequency band. In [39], the authors presented an initial path-loss model for nano-sensor networks, that describes the THz propagation behavior through plant foliage, whereas, in [40], a channel model for intra-body nanoscale communications was provided. Additionally, in [41], a model, which evaluates the total absorption loss assuming that the propagation medium is the air, natural gas and/or water, was proposed. In [42], the authors presented a multi-ray THz model for THz communications. All the published results agree on the fact that THz communication channel has a strong dependence on both the molecular composition of the medium and the transmission distance.

## 6.2. THz frontend

The available THz frontends and their package integration still differ from mmWave systems drastically. When discussing the SOTA of THz frontends, we need to be aware that, nowadays, there is no commercial wireless THz frontend product available. The existing research prototypes need to still adopt many functionalities and features that are considered to be standard functions in microwave or mmWave wireless systems, for example automatic gain control (AGC) or frontend calibration and impairment correction schemes. The functional integration density is little advanced and it is the great THz frontend development challenge of the next coming years to alter this. Conceptually, mere frequency scaling of proven mmWave frontend architectures is one possible strategy. However, the wireless baseband modem technologies have not evolved in the same way so far. For this reason, there is a significant gap between available modem technology and the possible capacities of the wireless THz channel. There are also technological limitations. With every transition to smaller semiconductor transistor nodes, whatever specific technology one favours, the chip manufacturing process complexity increases and the yield drops.

The development of THz packaging and antenna integration technologies needs to evolve as well. At the THz band, quasi-optical antenna solutions may actually shrink to an acceptable size in comparison to implementing them at low mmWave band. This offers for MIMO, P2MP and BF systems new options that need to be taken into account. Solutions that minimize the required number of channels or antenna elements are considered to be disruptive solutions at THz frequencies. The reason is partly the integration density problem and the associated limitations of implementing larger active antenna arrays. Another important aspect is the energy efficiency and saving requirements; especially for mobile battery driven UE.

TERRANOVA follows a strategy other than mere frequency scaling though not in a dogmatic sense. The motivation is rather the pragmatic idea to exploit the progress of optical fibre transceiver technologies and the fibre optical infrastructure, technically and economically. The technical benefits are the available tremendous capacity, circumventing today's wireless modem performance bottleneck and complementing fibre optical networks. The economic charm is the access to available inexpensive and very advanced baseband (BB) ASICs and the ease to introduce modified DSP algorithms.

The electrical "baseband" interface of most common optical communication significantly influences the frontend architectures. For instance, the BB signal format determines the bandwidth or frequency plan, the number of channels and possible duplexing solutions of the wireless link. Vice versa, the available frontend capabilities partly determine if an optical DSP is required at the interface between the optical and THz wireless link and in that sense the design process attains the nature of a co-design. There are numerous frontend options suddenly to choose among and part of this chapter aims to summarize the major ones. The problem is also strongly related to multiplexing methods and in some way time domain and frequency domain multiplexing plays an important role in the interface.

Overall, it is one of the major challenges and targets of WP5 to analyse and develop the building blocks of the TERRANOVA media converter. This media converter has to perform a mapping between the different optical transmission standards and the THz frontend including the THz antenna. Having an optical DSP available in that media converter offers the most flexibility to modify signal formats (modulation, symbol rates, bandwidth efficiency) but increases also the complexity.

## 6.3. Hybrid optical/THz system co-design

With the vision to provide reliable and scalable connectivity of extremely high data rates in the Tbps regime at almost 'zero-latency', TERRANOVA proposes to extend the fibre optic systems' QoS and QoE as well as performance reliability into the wireless domain, by exploiting THz frequencies predominantly above 275 GHz for access and backhaul links.

This will be achieved by co-designing hybrid optical/THz systems supporting the identified main technical scenarios with P2P, P2MP and quasi-omnidirectional links. The co-designed hybrid optical/THz systems should allow seamless integration into legacy networks on PHY and MAC layers. In order to allow this, the THz link should be designed to provide data rate compatibility with standard optical transmission systems and transceivers and should use common baseband interfaces. In a step beyond this, it could be also possible to provide fully transparent optical/THz links with common baseband signal processing.

### 6.3.1. Architecture of the wireless transmission system

A typical architecture for the wireless transmission system is presented in Figure *16*, and is composed of the following main units:

- Digital baseband processing unit;
- ADC/DAC; and
- Analogue frontend (AFE).

Next, we analyse the functionalities of the different units.

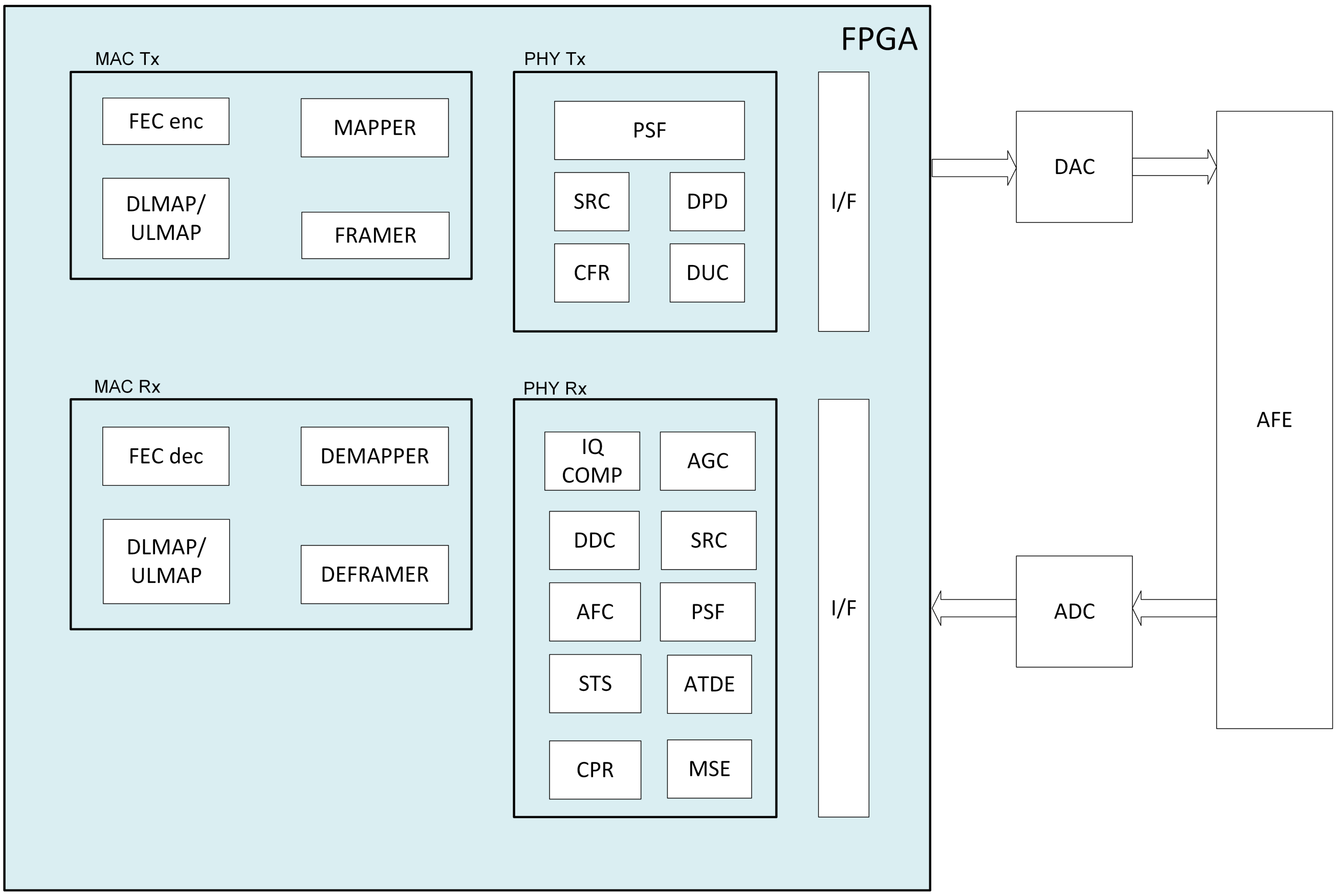

ADC: Analog to Digital Converter
AFC: Automatic Frequency Correction
AFE: Analog Front End
AGC: Automatic Gain Control
ATDE: Adaptive Time Domain Equalizer
CPR: Carrier Phase Recovery
DAC: Digital to Analog Converter
CFR: Crest Factor Reduction
DDC: Digital Down Conversion
DPD: Digital Predistortion
DUC: Digital Up Conversion
IQ COMP: I/Q impairments Compensator
FEC: Forward Error Correction
PSF: Pulse Shaping Filter
SRC: Sample Rate Conversion
STS: Symbol Timing Synchronizer

Figure 16: Architecture of the wireless transmission system.

Digital baseband processing unit: As illustrated in Figure *16*, the main parts of the BB unit of a transceiver are the MAC and PHY layer blocks. Within the MAC layer, DL/UP mapping is performed, i.e., the proper timeslots are created for the traffic. Forward error correction (FEC) coding is employed in order to increase the transmission quality, typically based on LDPC, Turbo and RS algorithms. Furthermore, constellation mapping/de-mapping is performed. Within the PHY layer, advanced signal processing algorithms are implemented for the optimum exploitation of the available spectrum (i.e. transmit more bits/s/Hz), synchronization, highest possible link availability, , etc. Finally, the PHY sub-unit is responsible for mitigating the impact of the RF chain hardware imperfections. In more detail, RF imperfections cause non-linearities, phase noise, in-phase and quadrature imbalances (IQI) at the transceivers. In general, their impact were studied in several works, including [43], [44], [45], [46], [47], [48], [49], [50], [51], [52], [53], [54], [55], [56], [57], [58], [59], whereas baseband signal processing approaches to mitigate their effect were presented in [60], [61], [62], [63].

ADC/DAC: Digital signals output from the baseband unit are converted to analogue and fed to the AFE in the TX direction. Similarly, analogue signals received at the AFE are sampled, then converted to digital, and finally read in the baseband processing unit, in the RX direction. The ADC/DAC unit places limits on the sampling rate that can be used, and as the system bandwidth depends on the utilized sampling rate, the ADC/DAC subsystem influences the achievable throughput.

AFE: Up-conversion of the analogue signal takes place to either an intermediate frequency (IF) or directly to the final RF frequency, in the TX direction. Respectively, down-conversion from IF/RF to baseband frequency is performed in the Rx direction. The AFE can feed the antenna through a duplexer in order to use a single antenna for both transmission and reception.

## 6.4. PHY layer

In this section, we present the PHY layer functionalities that are required in the THz systems. In particular, in Section 5.4.1, we discuss the concept of beamforming and analyse the different architectures that can be implemented. Additionally, Section 5.4.2 illustrates the concept of adaptive modulation and coding (AMC), which is expected to contribute to the increase of the link data rate, while, at the same time, it satisfies reliability requirements. Finally, in Section 5.4.3, the concept of simultaneous multi-window transmission is revisited, whereas, in Section 5.4.4, the use of multiple input multiple output (MIMO) schemes is discussed in order to achieve 1 Tb/s over 1 km line of sight (LOS) link.

### 6.4.1. Beamforming

Due to their high carrier frequencies, THz signals experience increased attenuation. This is due to PL as well as MAL. Particularly, FSPL grows with the square of the carrier frequency, if isotropic (RX and TX) antennas are employed [64], [65]. Furthermore, atmospheric absorption, mainly from water vapour molecules, becomes significant in the THz band [65]. Finally, the large bandwidth that can be used by the wireless THz systems leads to a significant signal-to-noise ratio (SNR) reduction, due to the wideband noise [64].

The use of antenna arrays is considered as an efficient approach in order to countermeasure the aforementioned phenomena [64], [65], [66], [67]. Fortunately, the small wavelengths, and hence the small-sized antenna elements (AE), enable the use of a large number of antenna elements, which leads to the development of powerful “beamformers”. Of note, BF can be employed either at the TX, or/and at the RX.

An antenna array is a set of antennas arranged in a pattern and constitutes one of the most prominent techniques to achieve electronic steerable beam, i.e., control the width and direction of the main beam. In general, antenna arrays can be classified in three main categories according to the relative locations of the AEs:

1. Linear antenna arrays (1D): the AEs are placed in a line;
2. Planar antenna arrays (2D): the AEs are placed in a plane; and
3. Volumetric arrays (3D): The AEs are placed in 3-D space.

As a consequence, the resulting pattern of the antenna array depends on the geometrical configuration of the array, the type and the number of the individual antenna elements, the excitation amplitude and phase of the elements, the operating frequency and the desired direction of the main beam [68]. Therefore, the design of a proper antenna array is critical, as it is expected to provide features in order to meet the requirements of the specific application. More precisely, a 2D rectangular antenna array is required, when BF has to be implemented both in azimuthal and elevation angles.

BF is the spatial equivalent of frequency filtering and requires multiple antenna elements cooperating to create constructive and destructive interference among them in order to produce a desired beam pattern [64]. Beamformers can be grouped into two classes:

- Conventional; and
- Adaptive.

*Conventional BF* utilizes fixed and predefined weights which are chosen independent of any signal received by the antenna array. On the other hand, *adaptive BF* utilizes signal-processing algorithms that choose the weights in real time based on information derived by the received signal [69]. Therefore, the adaptive concept outperforms the conventional one in terms of performance, because it provides extra degrees of freedom (DoF) and the ability to adapt in real time to the RF signal environment.

It should be emphasized that BF can be implemented either in the time or the frequency domain (FrD). In *time domain (TD) BF*, the desired weights can be either applied as time delays or as phase shifting in the digital or the analogue stage. On the other hand, *FrD BF* is implemented when the signal is processed in the FrD and the corresponding transformation tools (i.e., the fast Fourier transform (FFT) method) are utilized [70].

Apart from the structure of the RX and the TX in terms of hardware (i.e., number of AEs in the array, phase shifters, FPGA design) and the RF environment characteristics, the BF is affected by the bandwidth of the transmitted signal. Depending on the signal bandwidth, it can be divided into two categories, namely

- narrowband BF; and
- wideband BF.

In the case of narrow bandwidth, the signals transmitted/received by the most distant AEs of the array are correlated with each other. Thus, a delay in the timing of a narrowband signal can be approximated by a phase shift and *narrowband BF* can be achieved by an instantaneous linear combination of the received signals in the AEs. In contrast to the frequency-independent weights of the narrowband BF, wideband signals require the value of the weights to vary according to the different frequency components of which these signals consist. As a result, *wideband BF* is achieved by applying discrete FIR/IIR filters, tapped delay-lines, sensor delay-lines or by dividing the wideband signal into consecutive narrow sub-bands and adaptively processing each of them [70].

The *digital BF* architecture used in the microwave systems cannot be directly applied in the THz systems, due to the radio frequency (RF) hardware constraints. Usually phase and amplitude are the variables used for BF. In the digital BF architecture, the signal's phase and amplitude are controlled digitally at BB. The weights are applied directly on the digital signal by the DSP unit, before the digital-to-analogue converter. Thereby, as illustrated in Figure 17, the digital processing requires a dedicated BB and RF chain for each AE. Unfortunately, as large number of antennas are employed in the THz systems, having a dedicated baseband and RF chain for each AE is prohibitive in terms of complexity, cost and power consumption. As a result, alternative architectures have been considered to reduce the implementation complexity.

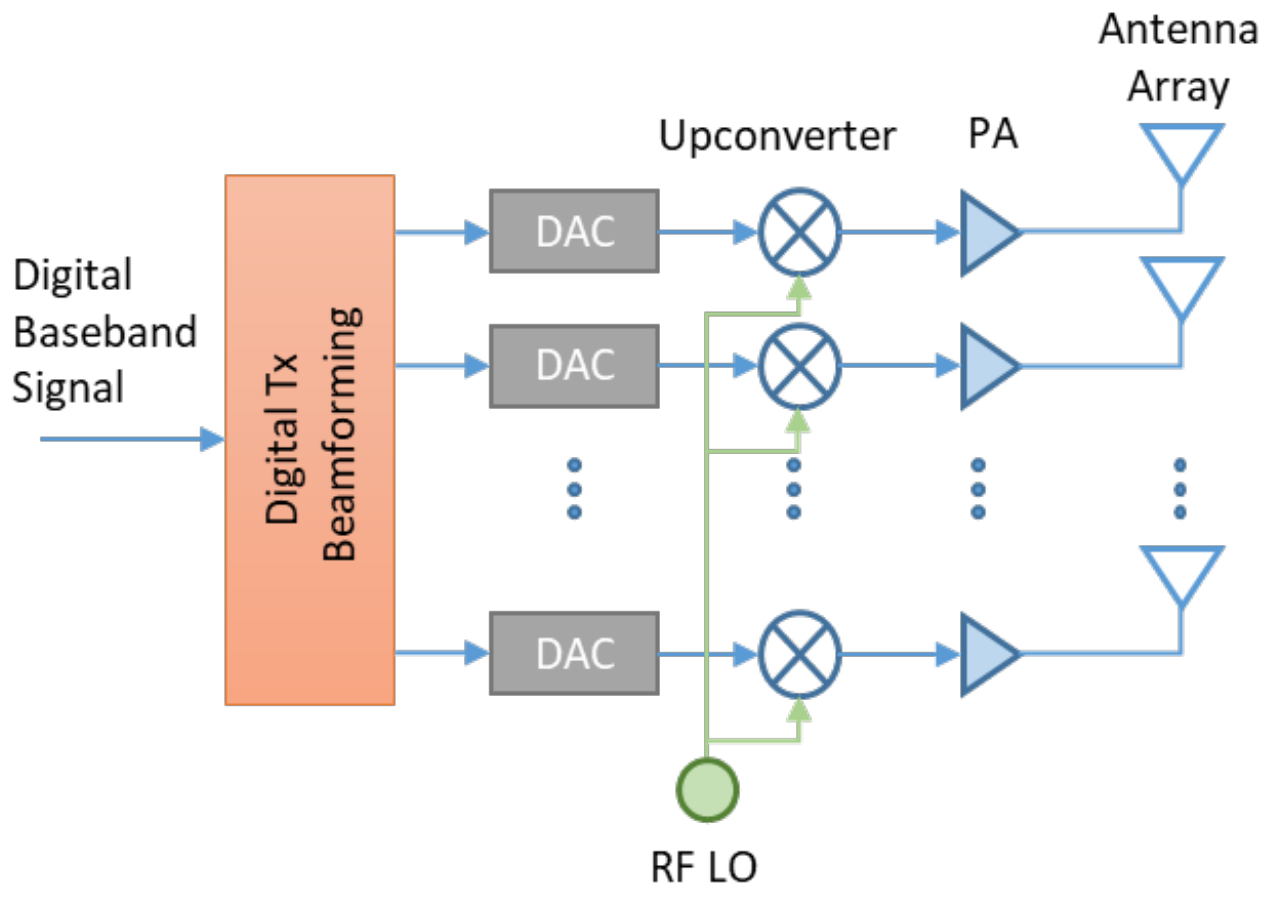

*Figure 17: TX with digital BB BF.*

The *analogue RF BF* architecture, which employs digital or analogue RF phase shifter (PS) has been considered in the technical literature for both mmWave and THz systems [71], [72], [73], [74], [75], [76]. Although this BF technique predates its digital counterpart, the approach is used in current mmWave systems, especially in short-range communication systems like the IEEE 802.11ad [71], [75]. In the analogue RF BF architecture, each RF chain is connected to many antennas through RF phase shifters, which are placed after the RF up-converters in the transmit mode and before them in the receive mode. The RF phase shifters can be implemented with digital or analogue control, while the first type is more common. The weights represented by the phase shifters are designed to steer the transmit and receive beams towards the dominant directions. In addition to RF, analogue BF can also be implemented at IF, BB, or LO.

As illustrated in Figure 18, in *analogue BB BF*, the PSs and PAs are placed before the frequency up-converters in the transmit mode and after them in the receive mode. On the other hand, in *analogue IF BF*, the PSs are placed after the IF up-converter in the transmit mode and before it in the receive mode (see Figure 19). Finally, in *analogue BF with LO phase shift*, the desired phase shifting is performed in the LO path, as depicted in Figure 20, before the up-conversion of the signal at each antenna element.

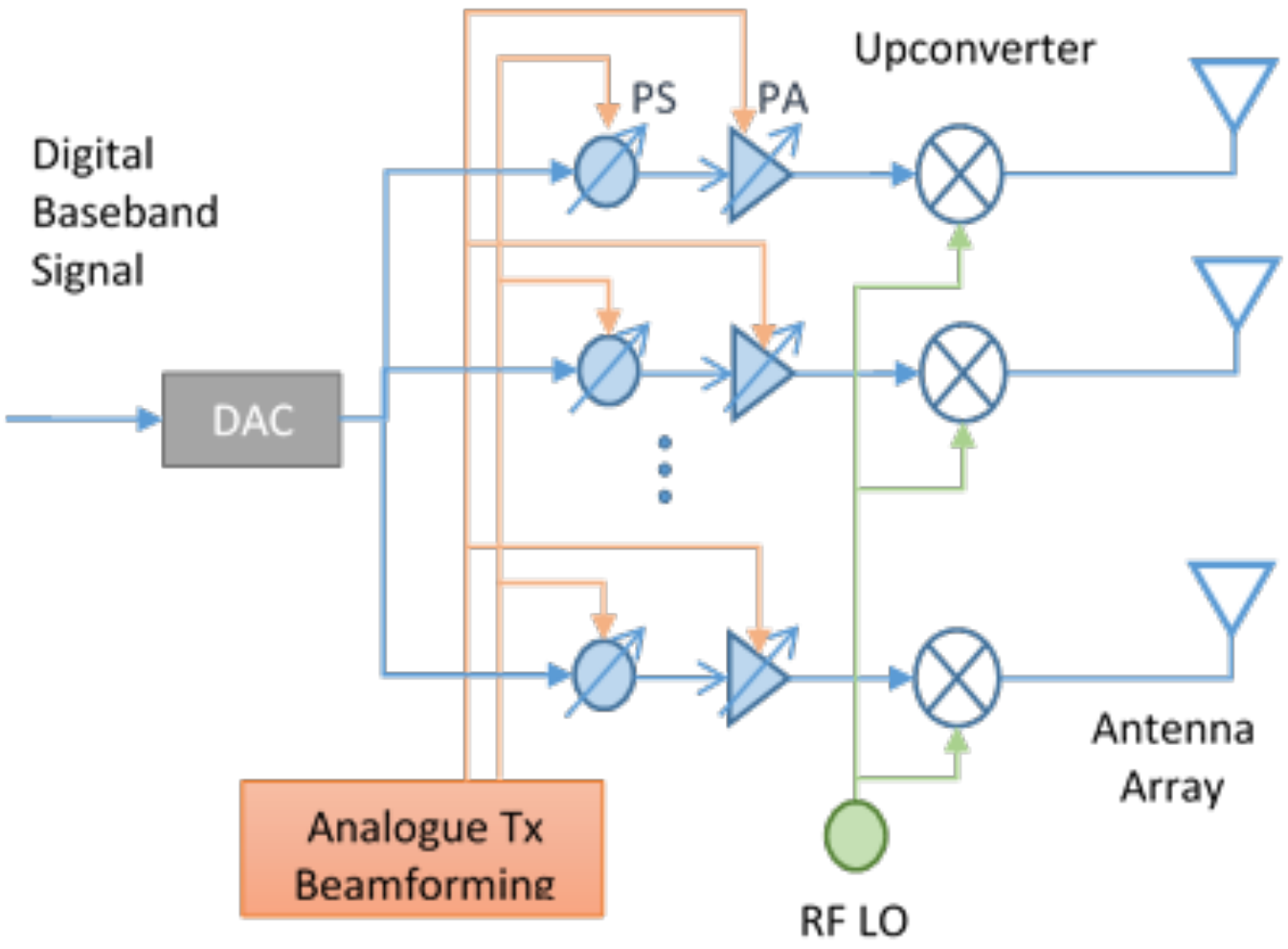

*Figure 18: TX with analogue BB BF.*

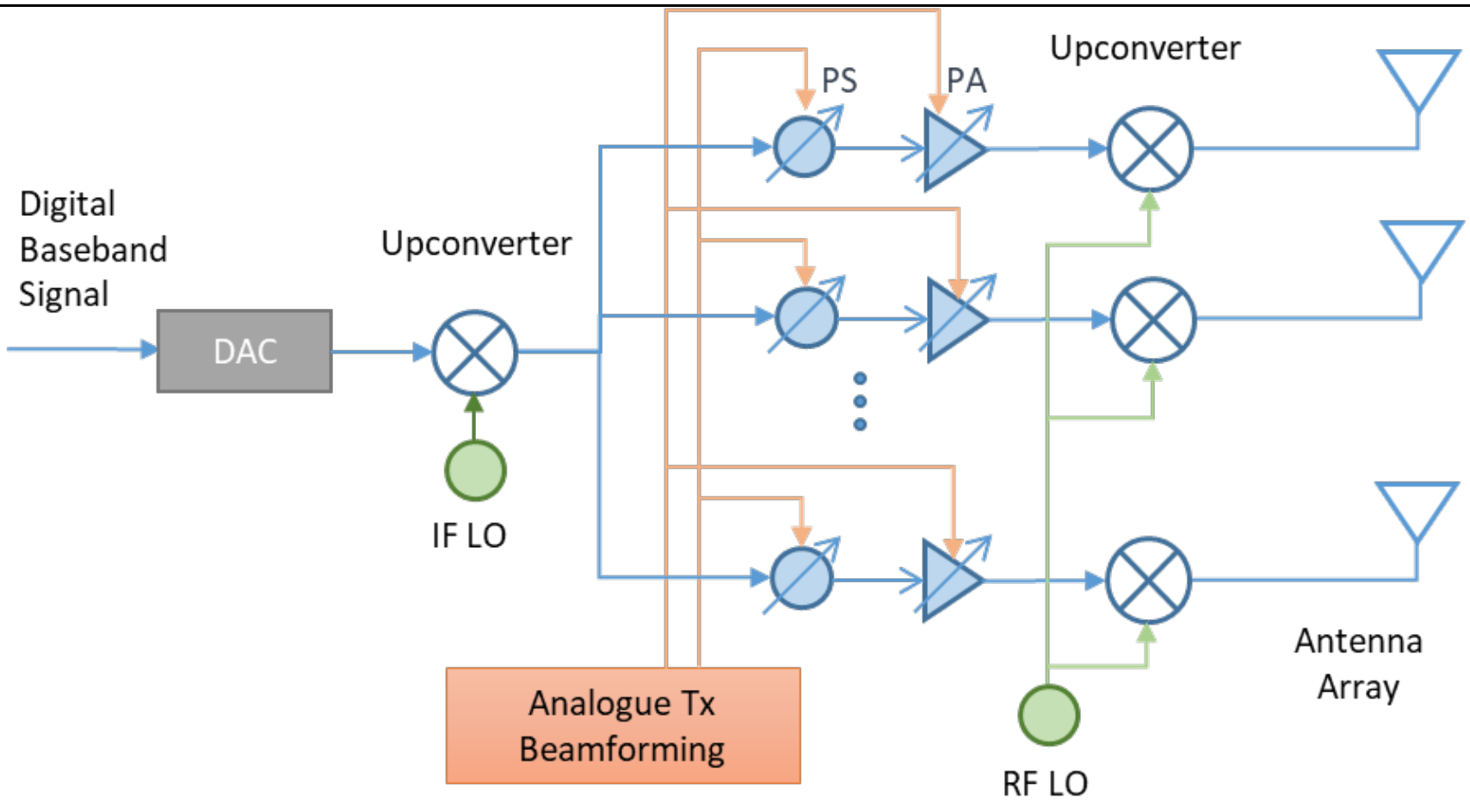

*Figure 19: TX with analogue IF BF.*

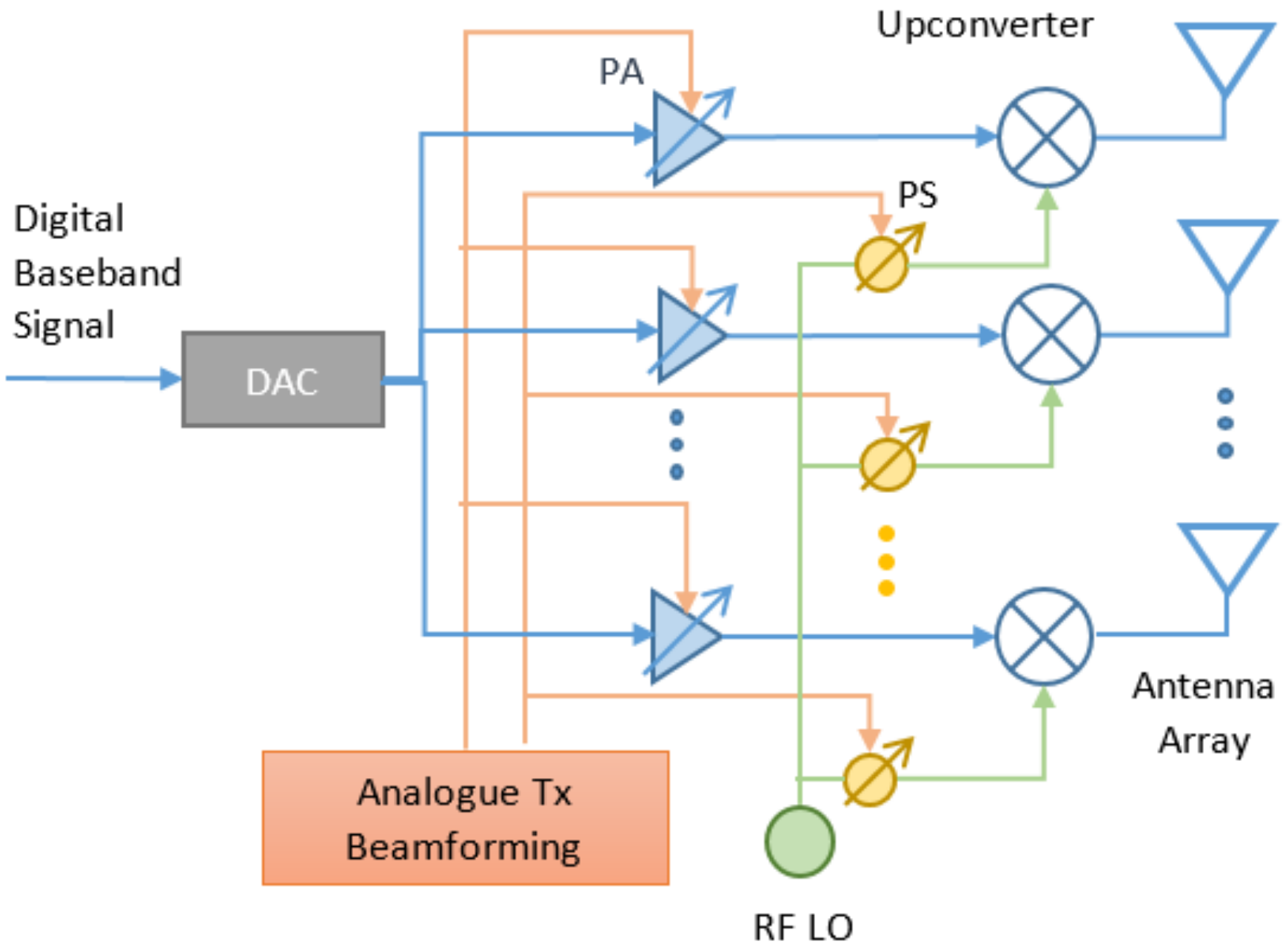

*Figure 20: Analogue BF with LO PS.*

Despite its simple utilization and energy efficiency, analogue RF BF has some drawbacks. One of them is the non-idealities of the required PS compared to digital BF. Conventional analogue and digital RF PSs show phase-dependent insertion-loss and frequency-dependent PSs. Likewise, digital PSs suffer from limited resolution, whereas analogue PSs require the extraction and interpolation of look-up tables. Another disadvantage is that realizing multiple beam transmissions using analogue BF is highly complex, since it would typically require signal splitting and multiple sets of phase adjustment hardware [75]. Table 3 compares analogue and digital BF.

*Table 3: Comparison of analogue and digital BF.*

| | **Analogue BF** | **Digital BF** |
|---|---|---|
| **Transmission BW** | It is mainly utilized for narrowband signals. Implementing wideband BF is difficult, due to the need for precise time delay. | Works well for narrowband signals. However, its performance for wideband signals is exceptional especially when FFT method (i.e., FrD BF) is utilized. |
| **Radiation pattern** | Only one beam is formed at a time, unless the antennas are partitioned. | Simultaneous beams at different angles and frequencies can be produced. |
| **Power Consumption** | RF phase shifters have high insertion losses and power consumption, while IF PSs have less circuit losses.<br>For large arrays, AFEs consume considerable power. | The FPGA and other processors require significant power.<br>Recently, there has been progress in the reduction of the digital system energy consumption. |
| **Band of application** | Can be realized in IF or BB, but is more frequently applied in the RF frequencies. | Performed at BB frequencies. |
| **Delay** | There exists no delay issue. However, analogue BF is slower, apart from the case where the digital BF FPGA is overloaded, due to a large antenna array. | The processing speed of the FPGA becomes a limiting factor in case of a very large number of AEs. |
| **Implementation** | One transceiver unit (RF chain) connected through a dedicated PS with each AE. | Requires a dedicated BB and RF chain per AE. This is prohibitive for large antenna arrays, due to cost and complexity. |
| **Cost** | The basic cost is relatively low and currently cheaper than the digital BF. For a large number of AEs equivalent PSs are required, increasing considerably the cost. | The main cost concerns the appropriate FPGA to implement the design. Its cost reduces over time, while the cost for adding further performance is relatively low. |
| **Flexibility** | Analogue BF is only to some extent flexible. | Digital BF is essentially flexible, as it is mainly implemented in the FPGA. |
| **Calibration** | Analogue beamformers provide limited calibration capabilities and a low resolution phase control. | Digital beamformers provide both frequency and direction calibration, which may be of high resolution. |

To strike a balance between complexity and performance gain, researchers have proposed a hybrid BF architecture [75], [77], [66], [78], [67], [79], [80], [81], [82], [83], [84]. The hybrid BF architectures have been under discussion also in radio astronomy at microwave frequencies. The hybrid architecture is a two-stage digital-analogue RF BF procedure, allowing the use of a large number of AEs, while limiting the number of RF chains as shown in Figure 15. The fewer number of RF chains compared to the number of AEs in the hybrid structure makes it more cost and power efficient compared to a fully digital structure. Furthermore, the additional RF chains (as compared to the fully analogue RF BF) allows the hybrid BF to perform more complicated processing and support multiple transmissions with different data streams. By exploiting the sparse nature of the mmWave channels, the results in [77] and [81] reveal that with proper BF design in a single user MIMO setup, hybrid BF can achieve near optimal data rates compared to fully digital BF. For a multi-user MIMO setup, the authors in [67] investigated the optimal design of hybrid BF structures, with the focus on an $N$ (the number of transceivers) by $M$ (the number of active antennas per transceiver) hybrid BF structure. Their study further discussed some guidelines on how to achieve an optimal energy/spectrum efficiency trade-off.

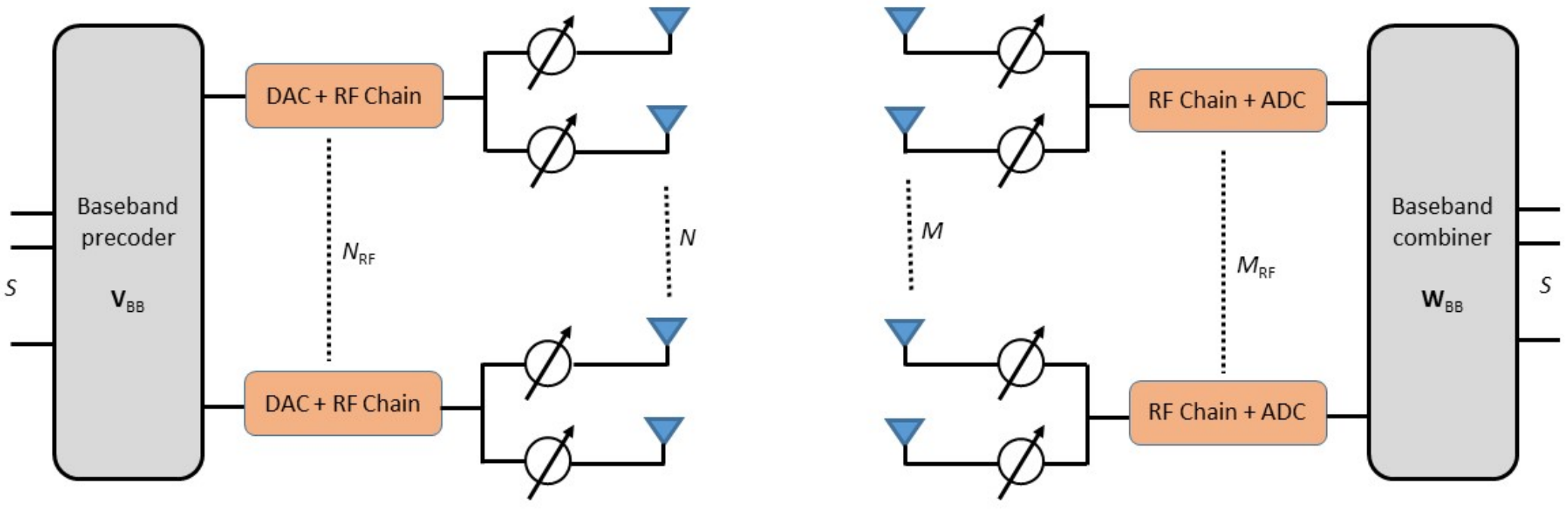

*Figure 21: Hybrid analog/digital MIMO system.*

### 6.4.2. Adaptive modulation and coding

It is widely known that the wireless channel experiences a time-varying behaviour [85]. As a consequence, a wireless system that is not able to adapt to the environment and is always working under the worst channel conditions, results in an inefficient management of the available spectrum. Therefore, modern wireless systems have to adapt their transmission to the channel conditions in order to improve their efficiency and performance. In particular, they should adapt the modulation order, as well as the channel code rate, in a hitless fashion, i.e., without relaxing a reliability predefined requirement, such as the BER, during changes.

As shown in [86], the bandwidth of the LOS THz transmission windows drastically changes even with small variations in the transmission distance; thus, the development of distance-aware AMC schemes that take into account the distance-dependent bandwidth is required. Distance-adaptive AMC should be developed that employ either: (i) the entire transmission bandwidth (for ultra-high data rate short-distance communications), (ii) the central part of the transmission window (to serve close and far nodes), or (iii) the sides of the transmission window (for lower data rate short-distance links) [3]. Within an individual transmission window, both single-carrier (such as quadrature amplitude modulation - QAM) as well as multi-carrier modulation schemes (such as orthogonal frequency division multiplexing – OFDM, and/or discrete multi-tone - DMT) should be considered to account for the required adaption to the time varying channel conditions.

Additionally, due to small wavelengths, multi-path fading in the case of NLOS links will be quickly changing already at small spatial movements leading to highly time-variable non-flat channel characteristics in nomadic applications. This phenomenon is expected to play an important role especially in indoor communications, i.e., Scenario 3.

### 6.4.3. Transmission through multiple frequency windows

By utilizing multiple carriers across the different non-contiguous transmission windows, a higher aggregate capacity could be achieved [87]. On the other hand, this also leads to an increased complexity of the system, as parallel individually optimized THz frontends would have to be used, due to the large differences in carrier frequencies. In [86], initial estimations for the theoretical PHY layer data rates in the different windows were presented by utilizing higher order modulations to increase the capacity at the expense of higher SNR requirements. Under the assumption of considering frequency windows above 275 GHz, no individual frequency band is able to support high data rates and achieve 1 Tbps at 1 km link distance. One option would be to consider at least parallel transmission in the first three windows including frequencies up to 450 GHz which would be sufficient to aggregate most of the available capacity. The other option would be to extend the first frequency window to frequencies below 275 GHz in order to achieve higher bandwidth and capacity. An additional advantage of the multiple frequency windows transmission would be that by selecting the available transmission windows that have the lower PL, the antenna gain requirements can be relaxed or the available SNR can be increased.

### 6.4.4. Spatially-multiplexed transmission

Another option, in order to achieve the 1 Tbps over 1 km LOS link, would be the use of spatial multiplexing (SM) schemes together with MIMO techniques in digital signal processing (DSP). SM can be achieved by utilizing the following approaches:

- two polarizations (factor 2);

- several antennas (factor N).

Due to the small wavelength of the THz wave, compact antenna arrays can be realized still satisfying the requirements of MIMO transmission. Employing a 2x2 MIMO transmission with polarization would already allow to reduce the required transmission bandwidth by a factor of 2, i.e., from 200 GHz to 100 GHz, as well as the RX noise floor by 3 dB. Since each MIMO channel can use its own power amplifier (PA), this also results in a decrease of the required per-antenna directivity of 1.5 dB. Using higher NxN MIMO with antenna arrays would result in higher savings in the required bandwidth and directivity; however, at the same time, the energy consumption might significantly increase.

## 6.5. Medium access control and radio resource management

This section is focussed on the MAC and RRM . In particular, in Section 5.5.1, the radio resource block (RB) is defined, while, in Section 5.5.2, an appropriate user association strategy for ultra-dense THz network is provided. Likewise, Section 5.5.3 discusses the initial access (IA) procedure, whereas Section 5.5.4 is focussed on UE tracking. Interference management and hand over schemes are presented in Sections 5.5.5 and 5.5.6, respectively, while the concept of a new MAC protocol that satisfies the requirements and take into accounts the particularities of the THz systems is provided in Section 5.5.7. Finally, the role of caching in TERRANOVA system is presented in Section 5.5.8.

### 6.5.1. Channelization

An important decision in the design of the MAC and RRM layers is the definition of the resource blocks (RBs), i.e., the unit of the physical resources. In LTE, a RB is defined as a portion of the time-frequency domain. On the other hand, in wireless THz systems, the directional transmission enables to also use the spatial dimension. As a result, the RB can be defined in the time-frequency-space domain. As illustrated in Figure 16, in order to proper utilize this type of RB, the BS needs to group a set of UEs together, which are non-distinguishable in the transmitted beam, and serves each group with one analogue BF vector [88]. This indicates that an analogue or hybrid BF should be employed. The analogue BF partially utilizes the spatial part of the RB, whereas the digital BF increase the multiplexing gain within each group.

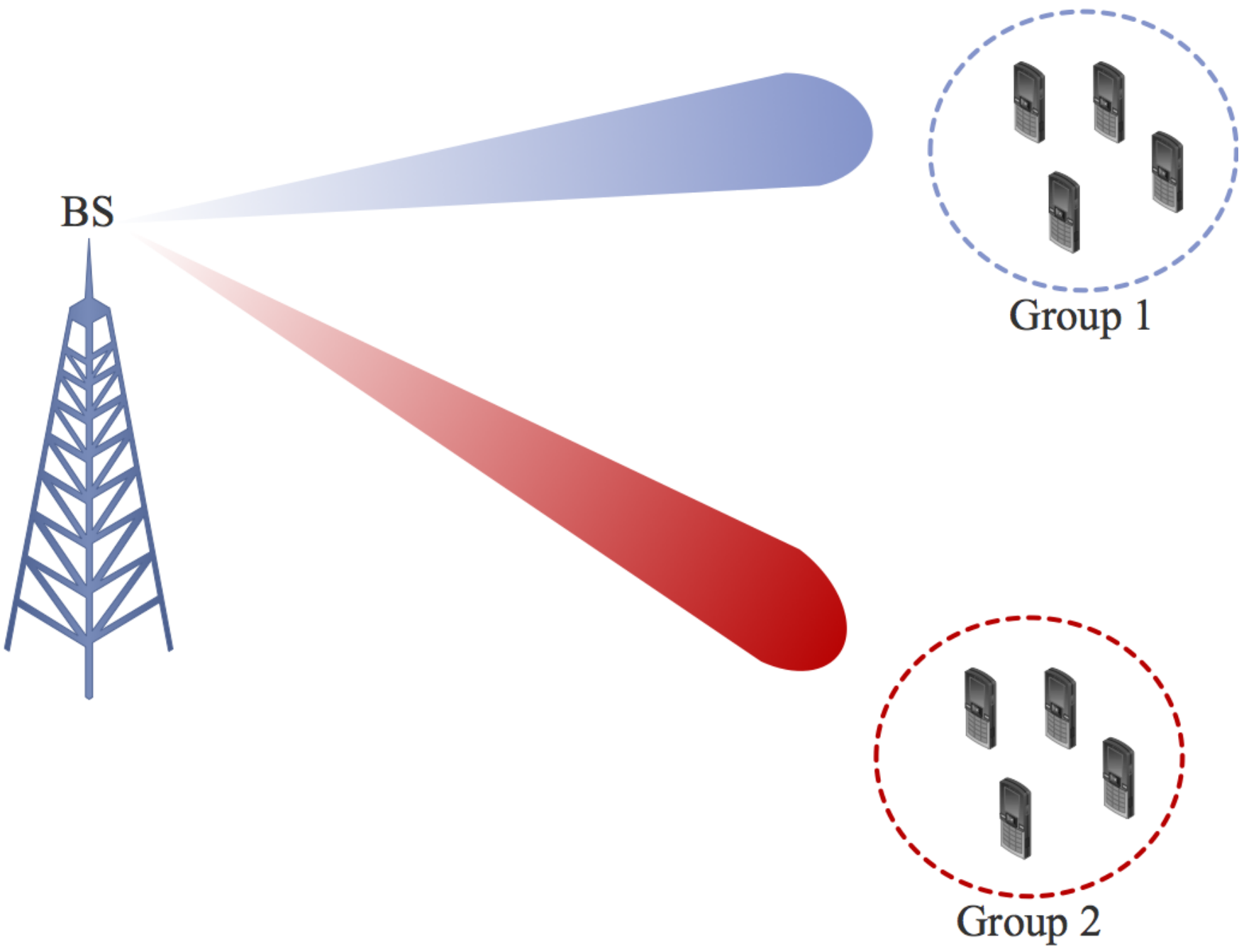

*Figure 22: Space-dependent UE grouping.*

### 6.5.2. Users association

Most of the current standards define a cell by the set of UEs that are associated using a minimum-distance rule, which leads to non-overlapping Voronoi tessellation of the serving area of every BS, exemplified by the well-known hexagonal cells [89], [90]. The minimum-distance rule leads to a simple association metric based on the reference signal received power (RSRP) and RSSI. However, the traditional RSRP/RSSI-based association may become significantly inefficient in the presence of non-uniform UE spatial distribution, and of heterogeneous BSs with a different number of antenna elements and different transmission powers [91]. This association entails an unbalanced number of UEs per cell, which limits the available resources per UE in highly populated cells, irrespective of individual signal strengths, while wasting them in sparse ones. The main disadvantage of the current static definition of a cell, as a predetermined geographical area covered by a BS, is that the static cell formation is independent of the cell load as well as of the UEs' capabilities. Therefore, the parameters that should affect the cell formation are:

- UE traffic demand;
- channel between UE and BSs; and
- BSs loads.

By taking into account, the massive number of degrees of freedom that fully-directional communication offers and possible MAC layer analogue BF, we can define a dynamic cell as a set of not necessarily collocated UEs that are served by the same analogue beamformer of the BS and dynamically selected to improve some objective function. Upon any significant fluctuations of the above three parameters, dynamic cell redefinition may be required. In this context, a full database in the macro-cell BS is needed, which records the following attributes:

- Dynamic cell formations;
- UEs' traffic demands;
- UEs' QoS levels; and
- UEs' connectivity to the neighbouring BSs.

Depending on the UEs' demands, microcell BSs dynamically group UEs together and form new cells so that:

- individual UE's demands are met (QoS provisioning);
- the trade-off between macro-level fairness and spectral efficiency is improved - e.g., through proportional fair resource allocation (network utility maximization); and
- every UE is categorized in at least two groups, to guarantee robustness to blockage (connection robustness).

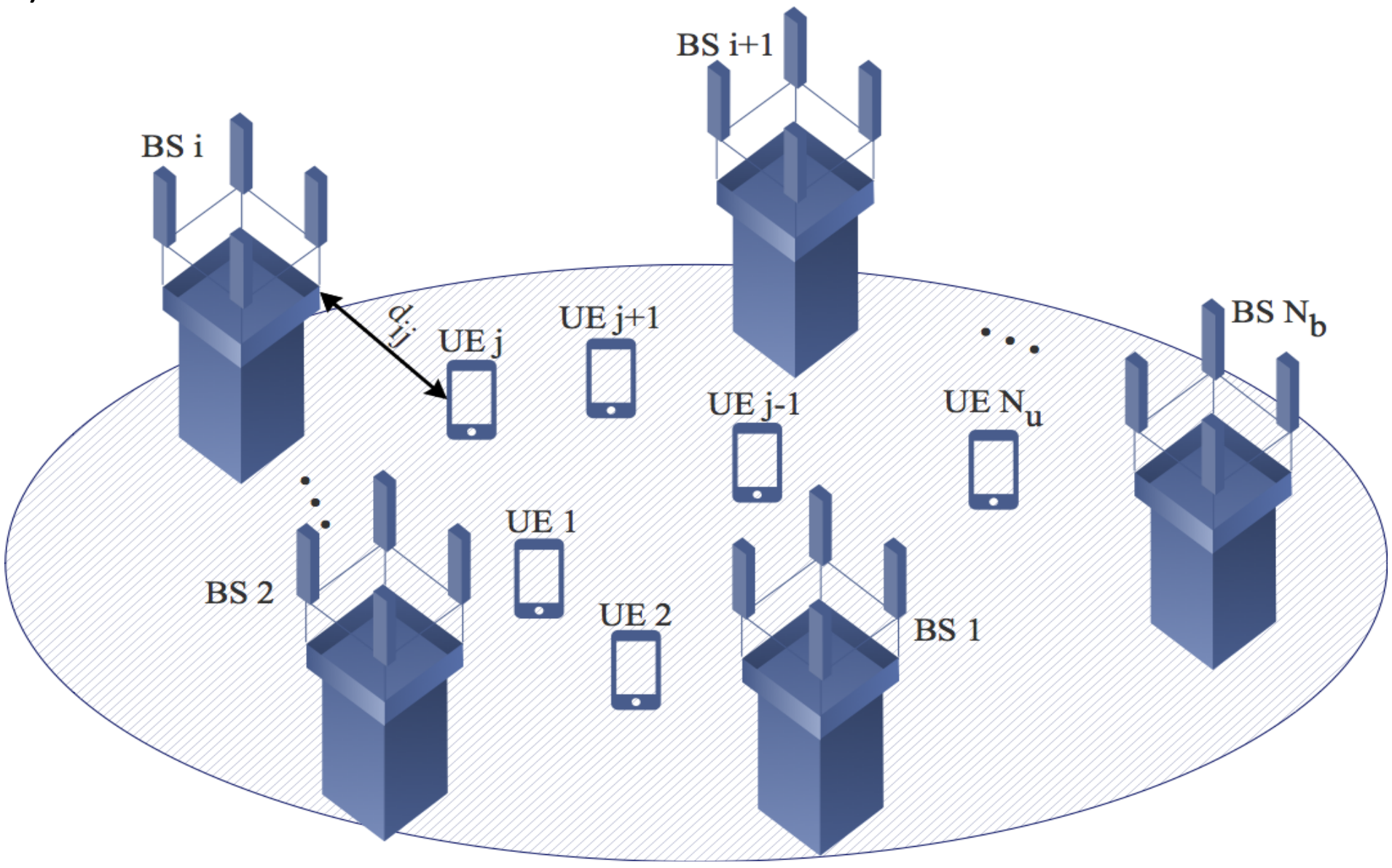

Figure 23: Dynamic cell network topology.

Based on the above criteria, we formulate the corresponding optimization problem for the BS-UE association, assuming a scenario, in which $N_b$ BS can serve $N_u$ UE (see Figure 17), in a full-directional mode. In more detail and as reported in [92], for a fixed network topology, which is assumed to be prior known to every BS i and UE j, the optimal cell formation should satisfy the following optimization problem:

$$\max_{\boldsymbol{\Psi},\boldsymbol{C}} \quad \sum_{j\in U, i\in B} r_{ij}c_{ij},$$

s. t.
- $C_1$: $\sum_{j\in U} c_{ij} \leq 1$, for each $i \in B$,
- $C_2$: $\sum_{i\in B} \psi_{ij} = 1$, for each $j \in U$,
- $C_3$: $0 \leq c_{ij} \leq \psi_{ij}, \psi_{ij} \in \{0, 1\}$, for each $i \in B$,
- $C_4$: $\psi_{ij} = 0$, if $R_{ij} \leq R_{j,min}$,
- $C_5$: $0 \leq \varphi_i^b \leq 2\pi$, for each $i \in B$,
- $C_6$: $0 \leq \varphi_j^u \leq 2\pi$, for each $j \in U$,

where $c_{ij}$ is the fraction of resources that the BS i employs to serve the UE j. Likewise, $\varphi_i^b$ and $\varphi_j^u$ respectively stand for the boresight angles of the BS I and UE j, whereas $R_{j,min}$ is the minimum required rate for the UE j. Finally, **Ψ** and **C** are matrices that collects all the user association variables, $\psi_{ij}$, and fraction of resources, $c_{ij}$, used by the BS i to serve the UE j, respectively.

In the above optimization problem, the constraint $C_2$ guarantees the association of the UE j to exactly one BS, whereas the constraints $C_3$ and $C_4$ ensures that the minimum acceptable QoS for every UE. Additionally, C3 ensures that hat every BS i will provide a positive resource share only to its associated UEs. Note that the solution of the optimization problem provides a long-term association policy along with proper orientation and operating beam widths for fully-directional wireless THz communications. This solution guarantees the optimal UE-BS association as long as the inputs of the optimization problem, namely network topology and UE demands, are unchanged. If a UE requires more resources or loses its connection, the optimization problem has to be re-executed.

In order to find the optimal solution for the association problem, we present Algorithm 1, which is based on the swarm intelligence meta-heuristic algorithm called grey wolf optimizer (GWO) [93] and returns the optimal user association matrix. According to the GWO, the solutions are grouped into four groups, namely α, β, δ, and ω, based on their optimality. In more detail, $x^\alpha$ vector represents the best solution, while $x^\beta$ and $x^\delta$ are respectively the second and third optimal solution vectors. Finally, ω is the set of the rest feasible solutions.

***Algorithm 1:** GWO algorithm for UE association*

**Input:** Population size: $N_p$
Maximum iterations: $G_{max}$
**Output:** x
**Initialize:** $\mathbf{x}^\alpha$ the best solution
f($\mathbf{x}^\alpha$) the objective function corresponding value.

```
Calculate the objective function of each vector.
Use Algorithm 2.
Sort vectors according to the objective function value.
Select the best solution x^α, the second best x^β, and the third best x^δ.
set G=1.
while G ≤ G_max do
    for k=1 to N_p do
        Update vector positions according to (1)
```

**end for**
Update vectors **α**, **A**, and **K**.
Calculate objective function value of each vector **k** according to **Algorithm 2**.
Select the new $\mathbf{x}^{\alpha}$, $\mathbf{x}^{\beta}$, and $\mathbf{x}^{\delta}$ vectors.
G=G+1
**end while**
**return** the best objective function vector $\mathbf{x}^{\alpha}$ and best objective function value f($\mathbf{x}^{\alpha}$).

A brief description of the GWO algorithm for the user association problem is given in Algorithm 1. In Algorithm 1, NP denotes the population size, and $G_{max}$ is the maximum number of generations. We notice that these are the only inputs for a given problem, since all the other parameters are randomly generated. Thus, GWO does not require additional control parameter setting like other algorithms. The population of NP vectors is initialized randomly from a uniform distribution. After the objective function value of each vector is calculated the algorithm sorts the vectors descending according to objective function value. The best solution is selected as the $x^{\alpha}$ vector, while the second best is the $x^{\beta}$, and the third best is the $x^{\delta}$ vector respectively. Then the algorithm main loop starts.

In the G+1 generation, the position of the k-th vector in the n-th dimension, is evaluated as

$$x_{nk}(G+1) = \frac{x_{nk}^{\alpha 1}(G)+x_{nk}^{\beta 2}(G)+x_{nk}^{\delta 3}(G)}{3} \quad (1)$$

where

$$x_{nk}^{ij} = x_{nk}^{i}(G) - A_{nk}^{j}(G)\, D_{nk}^{i}(G), \text{for } (i,j) \in \{(\alpha,1),(\beta,2),(\delta,3)\}$$

and

$$D_{nk}^{i}(G) = |K_{nk}^{j}(G)\, x_{nk}^{i}(G) - x_{nk}(G)|$$

with

$$A_{nk}^{j}(G) = a\,(2\, f_{nk}^{1}(G) - 1)$$

and

$$K_{nk}^{j}(G) = 2\, f_{nk}^{2}(G)$$

Of note, $f_{nk}^{1}(G)$ and $f_{nk}^{2}(G)$ are uniformly distributed randomly selected numbers in the range of [0, 1].

The algorithm updates the α, **A** and **K** vectors according to the above relations and evaluates the objective function value of the new vectors. Then again a new set of $\mathbf{x}^{\alpha}$, $\mathbf{x}^{\beta}$, $\mathbf{x}^{\delta}$ vectors is selected, if it improves the objective function value, provided by algorithm 2. Finally, note that the stopping criterion is the maximum number of generations.

***Algorithm 2:** Evaluation of the objective function.*

**Input:** A possible solution **y**
**for** i=1 to $N_u$ **do**
Calculate rate $r_{ij}$ for the i-th UE connecting to the j-th BS
**if** $R_{ij} \geq R_{j,min}$ **then**
Calculate $c_{ij}$
**end if**
**end for**
Calculate objective function

$$\sum_{j\in U, i\in B} r_{ij} c_{ij}$$

9: **Return** objective function value

Next, we present simulation results that evaluate the performance of the solution framework. We consider a scenario in which 120 UEs are randomly deployed and served by 6 BSs. Each UE requires a minimum data rate that is randomly selected within [1, 10 Gbps] with bandwidth equals 1 GHz. Moreover, we assume that the UE and the BS are placed in a uniform random manner within a circle of radius equals 50 m, while $10\ \log_{10}\left(\frac{P\,G_b\,G_u}{N_0}\right) = 120\ dB$, where P, $G_b$, $G_u$ and $N_0$ respectively stand for the transmission power, the BS and UE antenna gains, and the noise power. Note that for simplicity and without loss of generality, we assume that the BS and UE corresponding beams can be perfectly align. Moreover, it is assumed that all the BSs/UEs antennas can provide the same gain. Moreover, standard atmospheric conditions are assumed, i.e., the temperature, relative humidity and pressure are set to 25°C, 50% and 101325 Pa, respectively. The population size is 200 and the number of maximum generations is set to 150. Finally, in order to evaluate the effectiveness of the proposed approach, we compare it with the corresponding particle swarm optimizer (PSO) method. Both algorithms are executed for 20 independent topologies.

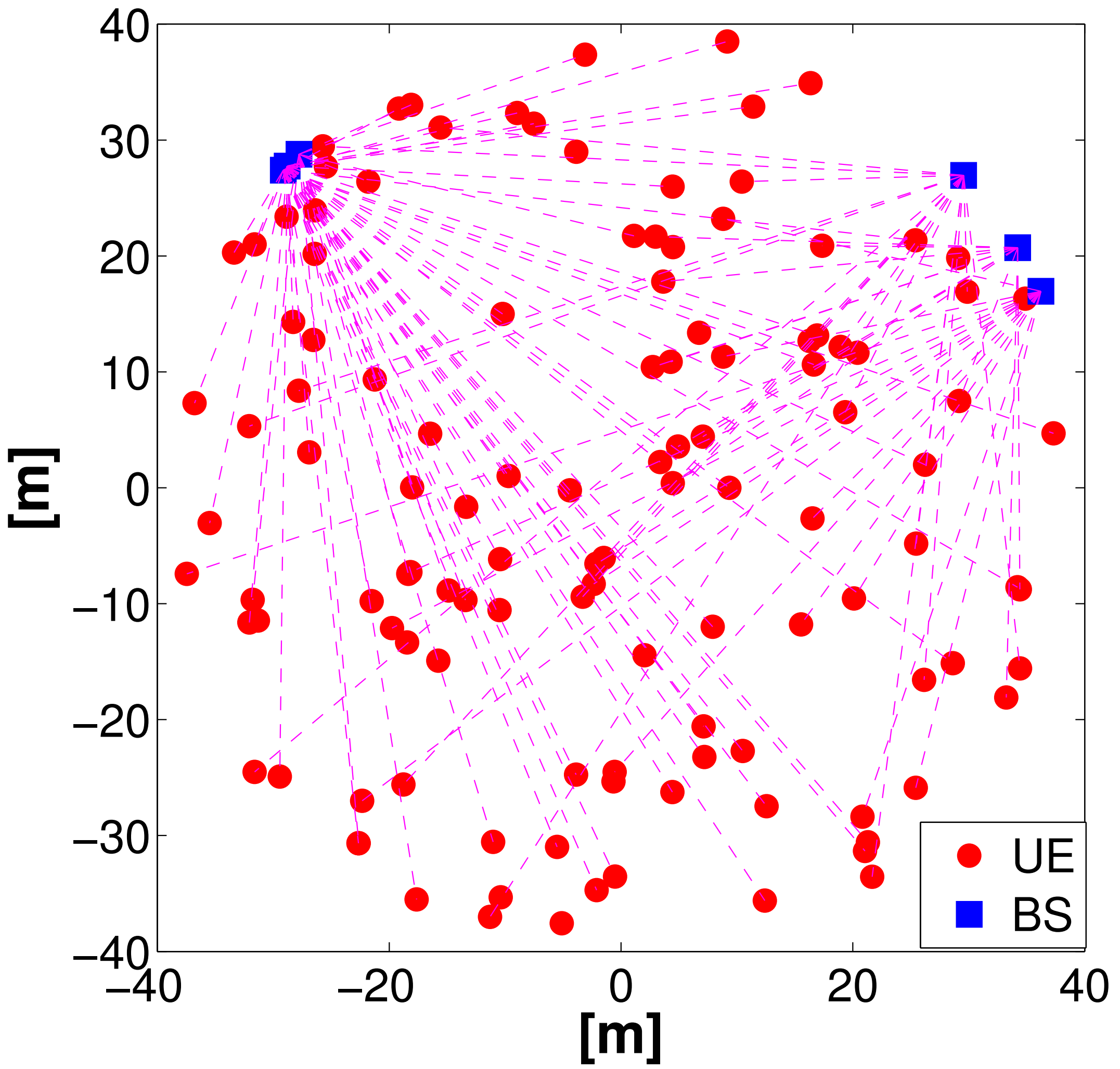

*Figure 24: Indicative BS-UE association example.*

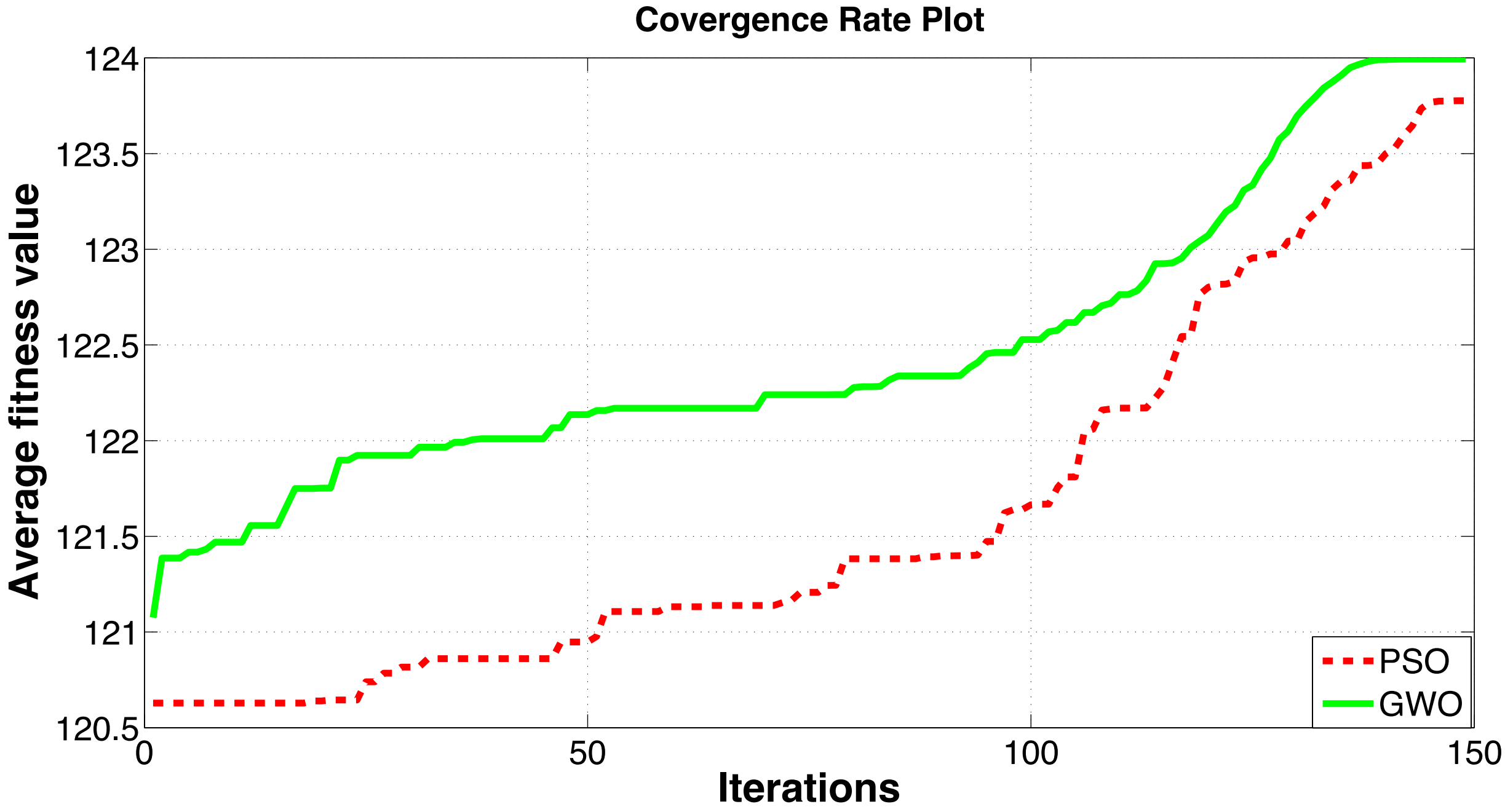

*Figure 25: Convergence rate plot of the average fitness per iteration obtained by GWO and PSO.*

Figure 18 demonstrates an indicative example of a BS-UE association scenario, whereas, Figure 19 illustrates the convergence rate measured as the average fitness per iteration, for both the GWO and PSO approaches. From this figure, it is evident that GWO-based algorithm was able to converge slightly faster than PSO and cover less at a higher objective function values. Additionally, the convergence rate for the first about 30 generations of GWO-based algorithm is quite faster than the PSO.

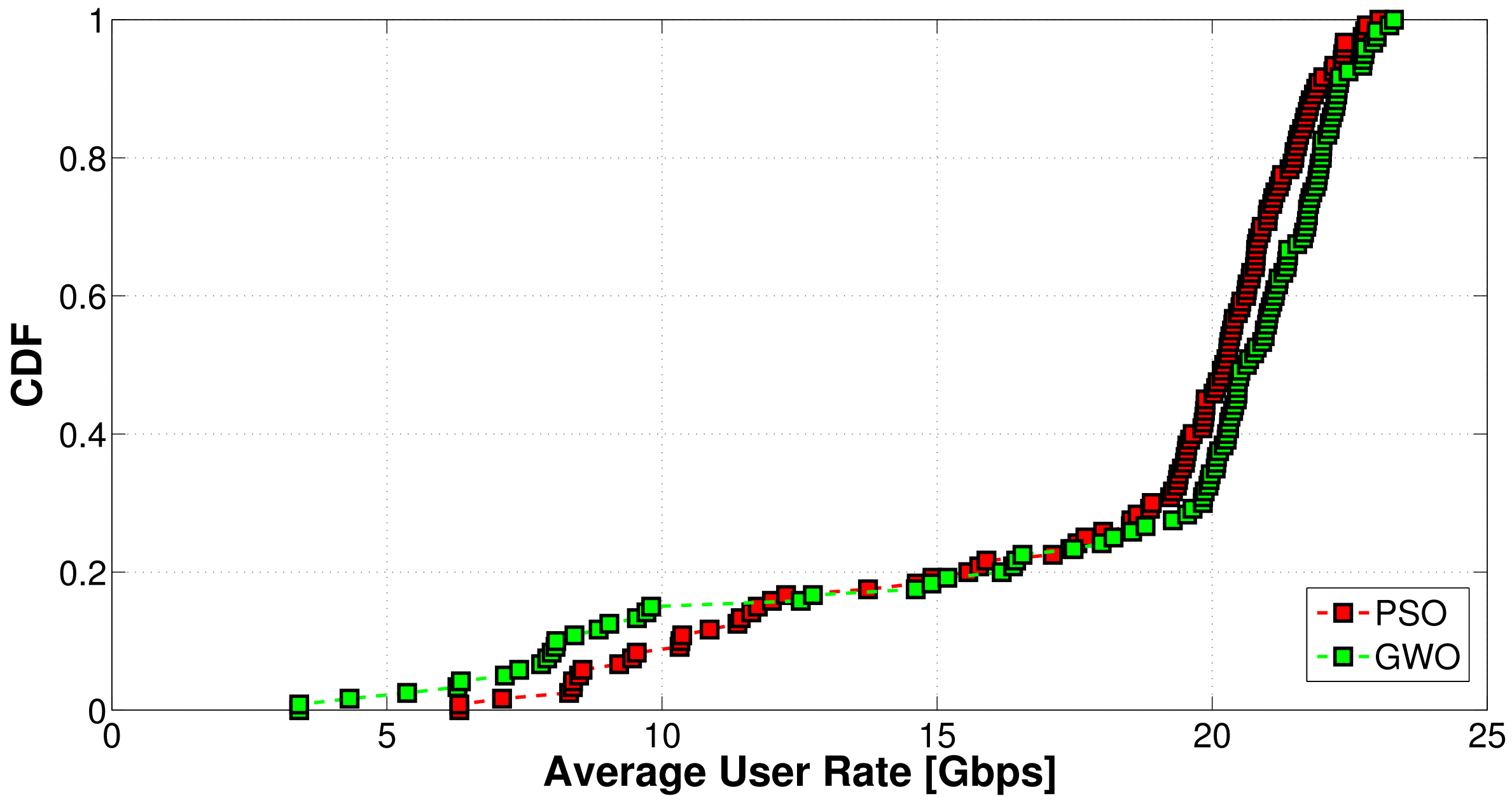

*Figure 26: CDF of the total data rates found by GWO and PSO.*

The cumulative distribution function (CDF) of the date rates obtained by both algorithms is depicted in Figure 20. It is observed that the PSO obtained values are higher for smaller data rates and probabilities less than 0.2. On the other hand, as the average user rate increases, the GWO-based algorithm outperforms the PSO. As a consequence, the GWO-based algorithm achieves faster and higher objective values compared to PSO. This indicates the effectiveness of the proposed solution framework. Moreover, Figure 21 shows the utility function value obtained by both algorithms in each algorithm run. We notice the GWO-based algorithm outperforms PSO in 17 out of the 20 cases. The percentage of users served by the network is depicted in Figure 22. Again, GWO-based algorithm clearly outperforms PSO in most of the cases.

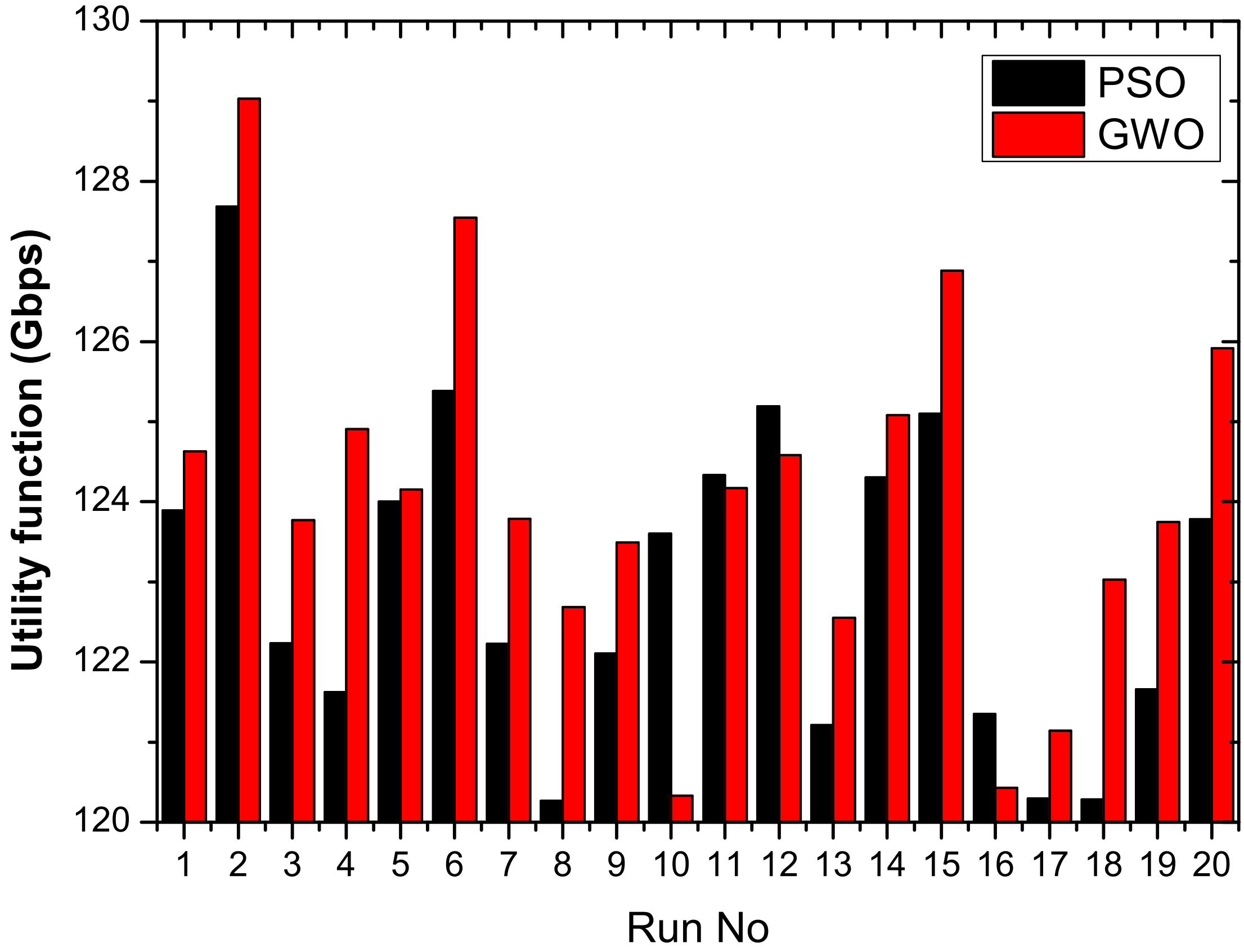

*Figure 27: Utility function value obtained by both algorithms in different executions.*

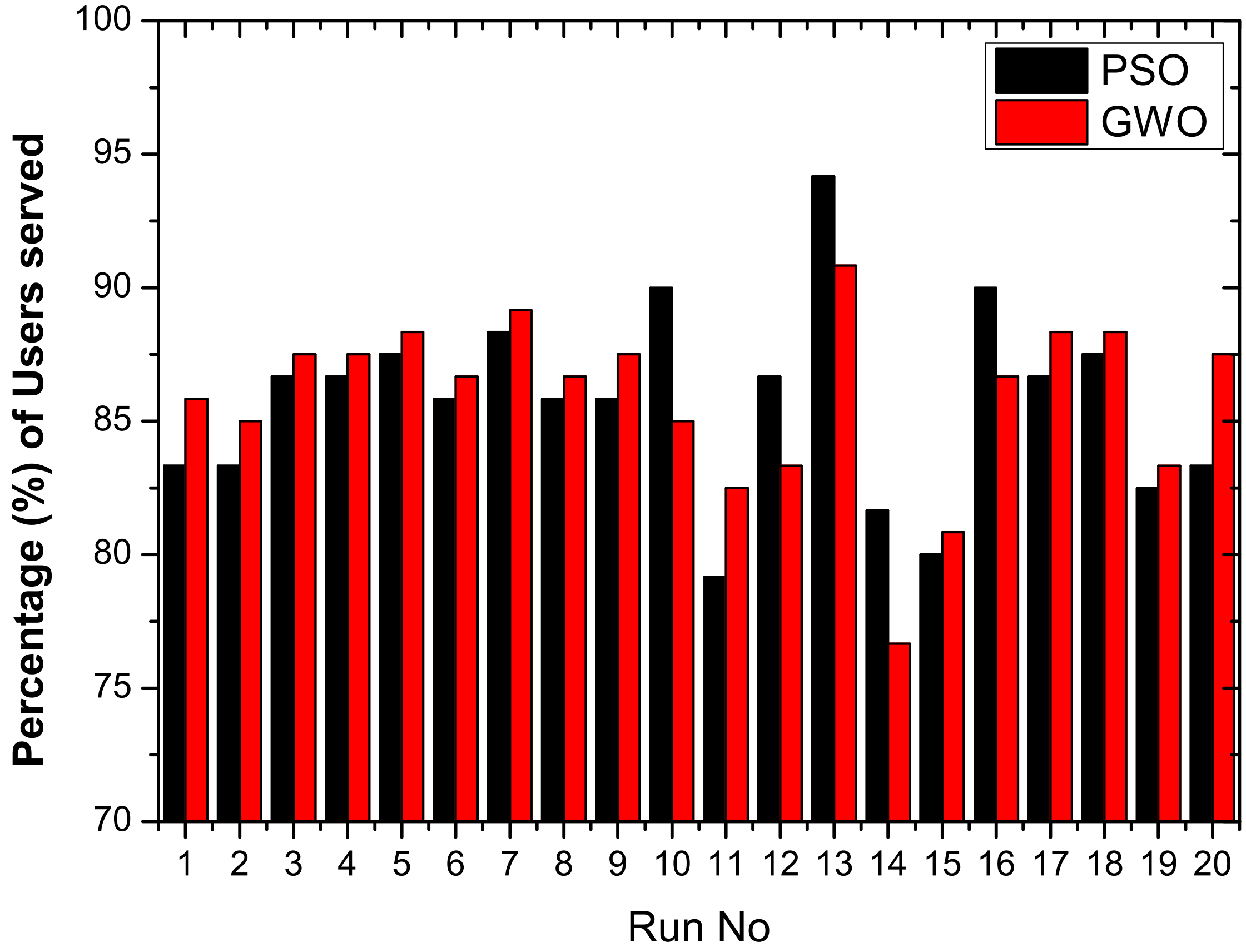

*Figure 28: Percentage of UEs served by the network obtained by both algorithms in each execution.*

Finally, we evaluate the algorithms performance for increasing UE number and we set the circle radius to 10 m and obtain results from 10 to 260 users with step 10 using both algorithms. In this sense, Figure 23 illustrates the results for this case. We observe that both algorithms obtain results that are very close. In general, PSO outperforms the GWO-based algorithm for small user number. On the other hand, GWO-based algorithm achieves better performances for larger number of UEs. This indicates that GWO-based algorithm is more suitable for higher dimensional problems. Moreover, we see that as the number of UEs increases, the utility function value tends to 160 Gbps, whereas the percentage of UEs served tends to 43%. Finally, from this figure, it is evident that both the algorithms are capable to serve all the UEs, when the user number is below 60.

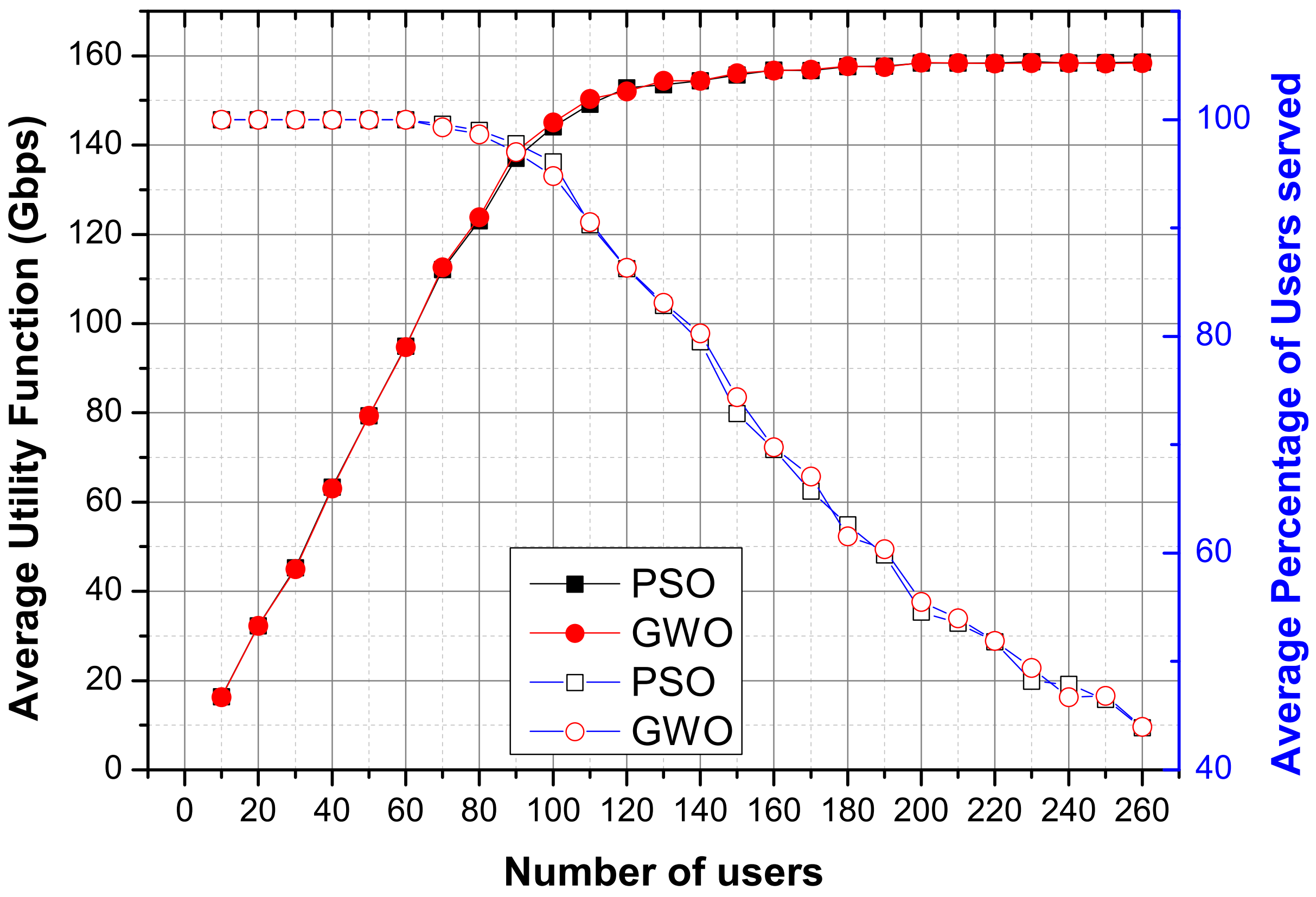

*Figure 29: Average utility function values and average parentage of UE covered as a function of the number of UEs, obtained by both algorithms for radius equals 10 m.*

### 6.5.3. Initial access

At THz frequencies, high antenna gains are needed to overcome the large path loss and other losses. In order to support this scenario, IA, where UE discovers THz access point and forms link-layer connection, needs to support BF/directionality at least in one end of the link (unless communication distances are very small or very long preambles/high coding gain are used to make it easier to discover users even with omnidirectional antenna modes). However, the use of low-complexity and low-power THz devices, along with the massive number of antennas, make traditional digital BF based on instantaneous channel state information (CSI) very expensive and in several cases infeasible. On the other hand, the use of analogue BF, based on predefined beam steering vectors (beam training codebook), each covering a certain direction with a certain beamwidth (BW), is considered as a feasible and effective alternative solution. However, one of the main drawbacks of analogue BF is the lack of multiplexing gain, which is addressed by the hybrid digital/analogue BF architecture.

Based on the assumed BF (i.e., fully digital, fully digital but low-resolution converters, hybrid, analogue, etc.) multiple beams could be transmitted/received at the same time, helping to reduce IA time. Otherwise, options include sequential scanning of all possible BF directions at the THz access point and UE (exhaustive search), using hierarchical search, etc. The reliance on the directional transmission in the IA in the mmWave communications renders the initial cell search procedures differ from the existing microwave systems. In the existing microwave systems, sophisticated BF is used only after the IA procedures have been completed. In order to reduce the time overhead, a hybrid approach has been proposed in several papers, where microwave bands (omnidirectional search) are used for the BS discovery and then reverting to mmWave bands for subsequent beam alignment and data transmission. As a result, the required functions for UE detection depend on the chosen approach for IA. For example, for hierarchical search, it is needed to support successively narrower beams.

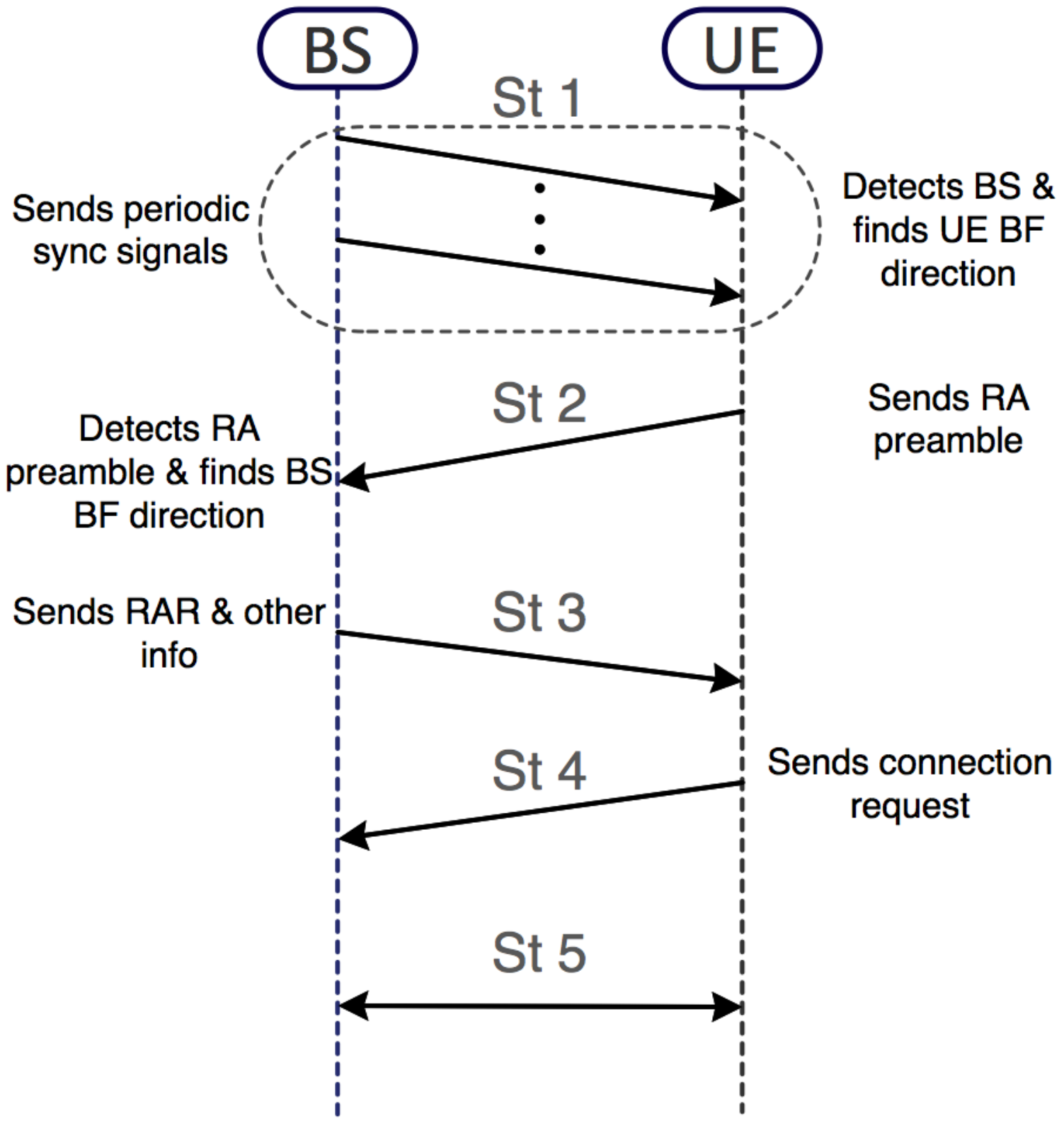

*Figure 30: IA procedure.*

In general, all the design options considered for IA follow the same basic steps shown in Figure. The main steps of the IA procedure are identical to the ones used in the third generation partnership project (3GPP) LTE-A, which are described in [94]. However, the depicted procedure needs major modifications for THz to enable both the UE and the BS to determine the initial BF directions in addition to detecting the presence of the BS and the access request from the UE. In more detail, the IA steps are:

1. Synchronization and signal detection: During this step, each BS periodically transmits synchronization signals that the UE can scan to detect the presence of the BS and obtain the downlink frame timing. In LTE-A, the first synchronization signal to detect is the Primary Synchronization Signal (PSS). On the other hand, for THz systems, the synchronization signal will additionally be used to identify the UE's BF direction, which (for analogue BF) is related to the angles of arrival of the signal paths from the BS.
2. Random access (RA) preamble transmission: Motivated by the LTE-A uplink frame structure, we assume that the uplink contained dedicated slots exclusively for the purpose of RA messages. After the synchronization and signal detection is completed, the location of these slots is known to the UE. The UE transmits a RA preamble in one of the RA slots. Since the UE BF direction is known after step 1, the RA preamble can be transmitted directionally. The BS scans for the presence of the RA preamble and learns the BF direction at the BS side.

3. RA response (RAR): After detecting a RA preamble, the BS transmits a RAR to the UE, which indicates the index of the detected response. At this point, both the BS and the UE know the BF directions so all transmissions can obtain the full BF gain. The UE receiving the RAR knows its preamble was detected.
4. Connection request: After receiving RAR, the UE desiring initial access will send some sort of connection request message on the resources scheduled in the uplink grant in the RAR.
5. Scheduled communication: All subsequent communication can occur on scheduled channels with the full BF gain on both sides.

Note that this IA procedure was initially presented in [94].

### 6.5.4. UE tracking

Once the THz access point (AP) and the associated UE have discovered the proper BF directions, the UE should be tracked in order to support nomadic mobility of UE or small movements at the transceivers of a fixed P2P link. If we adopt the conventional real-time channel estimation schemes, the pilot overhead will be unaffordable [98], [99]. As a consequence, the MAC and RRM layers should support beam-tracking method that consumes fewer resources compared with the full exhaustive beam discovery. For example, we can search just the beam direction adjacent to the current direction; this assumes that changes are small which may not always be the case.

Searching the technical literature, there exist several papers that provide different beam UE tracking approaches (see e.g., [100], [101], [102], [103], [104]). In particular, in [100], the authors proposed a beam tracking method, which is based on detecting the signal strength via training sequences appended to data packets. This approach has been experimentally validated by monitoring adjacent beams (in the angular domain) in [102]. The con of this approach is that it requires multiple beam pairs training. Moreover, in [103], the authors exploited the scarcity of the channel and reduces the problem in estimating the AoAs/AoDs in order to estimate the virtual channel matrix. This tracking solution relies on the assumption that the previous estimate has small difference to the current AoA/AoD and does not exploit any dynamics of the channels. Likewise, in [101], the authors presented a beam tracking approach that is based on fully scanning all possible beam combinations in order to create a measurement matrix to which the extended Kalman filter (EKF) is applied. The main drawback in [101] is the full scan requirement, which makes it difficult to track fast-changing environments, due to the long measurement time. In contrast to [101] , in [104], the beam tracking approach relies on a single measurement; therefore, it is considered to be more reliable for tracking fast-changing channels.

In practice, more efficient UE tracking schemes exploiting the temporal correlation of the time-varying channels are preferred. In general, excising UE tracking schemes can be classified into two categories. The key idea behind the schemes that belong in the first category is to model the time-varying channels in adjacent time-slot by a one-order Markov chain and apply classical Kalman filter in order to track the time-varying channels with low pilot overhead [105], [106]. However, the special sparse structure of the THz beamspace channels cannot be modelled as a one-order Markov process. As a consequence, these approaches are not capable of directly be extended to wireless THz systems with multiple antennas [107].

### 6.5.5. Interference management

Directional communications with pencil-beam operation drastically reduces multiuser interference in wireless THz networks [3]. An interesting question is whether in this case a THz network is noise-limited, as opposed to interference-limited networks. This fundamental question affects the design principles of almost all MAC layer functions. For instance, as the system moves to the noise-limited regime, the required complexity for proper resource allocation and interference avoidance functions at the MAC layer is substantially reduced. On the other hand, pencil-beam operation complicates negotiation among different devices in a network, as control message exchange may require a time consuming beam training procedure between transmitter and receiver [108]. In [109], the authors confirmed the feasibility of a pseudowired (noise-limited) abstraction in outdoor mesh networks, operating in high frequencies. However, as shown in [110], activating all links may cause an important performance drop compared to the optimal resource allocation in dense deployment scenarios due to non-negligible multiuser (MU) interference.

### 6.5.6. Handover

The suppression of interference in THz systems with pencil-beam operation comes at the expense of more complicated mobility management and handover strategies. In contrast with the long term evolution advanced (LTE-A), frequent handover, even for fixed UEs, is a potential drawback of THz systems, due to their vulnerability to random obstacles.

As a consequence, the MAC/RRM should be able to recover from user mobility outside the coverage of the current AP or from sudden blockage such as human blockage preventing to use current AP, preferably without full IA after current connection has been lost. Due to cost reasons, it is not expected that wireless THz networks will cover entire areas. Therefore, after losing current THz connection it may be necessary to move to lower frequency communication networks such as WLAN/LTE. Multi-connectivity may be required.

In order to avoid frequent handovers and reduce the overhead and the latency of re-association, the network should associate several BSs to each UE, by employing CoMP schemes [111], [112], [113], [114], [115], [116]. The cooperation among a UE, the associated BSs, and the macro-cell BS can provide smooth seamless handover through efficient beam-tracking and finding alternative directed spatial channels in case of blockage [117]. Here, two scenarios are foreseeable. A UE may receive data from several directions at the same time, but with a corresponding SNR loss for each beam, if we consider a fixed total power budget. Alternatively, the UE may be associated to several base (relay) stations, but only one of them is the serving BS, whereas the others are used as backup.

This scenario demands joint scheduling. Moreover, backup connections enable switching without extra latency, if the alignment and association to backup BSs are done periodically. In the light of a user-centric design, the macro BS can record all connections of a UE, predict its mobility, give neighbouring BSs some side information indicating when the UE is about to make a handover, so they can better calibrate the directed channel and be ready for the handover. If the UE can be served from more than one BS, it will be connected to at least two of them in order to be able to conduct fast switch operation.

### 6.5.7. Multiple access schemes

Novel MAC protocols are required for THz band communication networks, since classical solutions are unable to accommodate the following characteristics:

- The THz band provides devices with a very large bandwidth. As a consequence, the THz devices has not need to aggressively contend for the channel. In addition, this very large bandwidth results in extremely high bit-rates; hence, very low transmission durations are achieved. This causes a very low collision probability.

- The use of very large arrays and very narrow directional beams can also clearly reduce the multi-user interference. This comes with the cost that high-bit-rates and pencil-beams significantly increase the synchronization requirements [118].

As a result, the new MAC protocol need to be designed and developed that accommodates the following objectives:

- To support both random and scheduled access;
- To countermeasure deafness, by guaranteeing alignment between the TX and the RX, through the development of RX-initiated transmission schemes.
- To exploit fast-steerable pencil beams for network-wide objectives. The fast steering and pattern control capabilities of very large antenna arrays enable new functionalities that can be exploited to control interference. For example, in the DL, a TX can simultaneously send messages to multiple receivers, by transmitting time-interleaved pulses in different directions so that the pulses intended for each user add coherently at the desired RX, with no interference. There exists a trade-off between user acquisition complexity and communication robustness, in terms of BW choice. A possible scheme to fully harness the huge antenna array is to adapt wider BW during scanning phase for fast user acquisition, and intelligently steer focused thin beam for the subsequent data communication.

Motivated by the above requirement, the medium access can be organized in beacon intervals (BIs), similar to IEEE 802.11ay [119]. Each BI should support BF association, management frame exchange between the access point (AP) and the beam-trained stations, and data transmission. As a consequence, as illustrated in Figure 22, the BI consists of three phases, namely:

- Beacon header interval (BHI);
- Announcement transmission interval (ATI); and
- Data transmission interval.

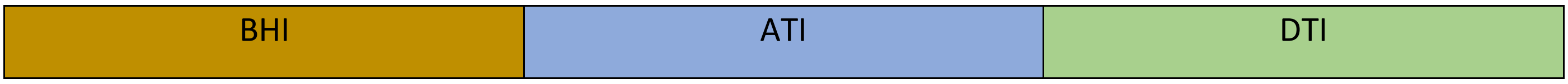

*Figure 31: The components of the BI.*

The BHI is responsible for the spatial synchronization, during the IA. Therefore, as demonstrated in Figure 23, it consists of two periods, namely beacon transmission interval (BTI) and associate BF training (A-BFT). During the BTI, the AP performs transmit sector sweep (TXSS), by sending synchronization signal (beacons), i.e., a sequence of packets, with each of them using a different antenna weight vector (AWV), in such a way that the transmissions include all directions, i.e., the sectors, of coverage of the AP. In the meanwhile, the STA continuously tries to detect a packet, by performing receive sector sweep (RXSS) and when it is able to lock to a beacon frame, it decodes its header and data field. The information contained in the payload enables the UE to associate to the AP. After the UE us associated with the AP, it randomly selects one of the A-BFT slots in order to perform TXSS. Note that since A-BFT is slotted, collisions may occur, during this phase, when two UEs selects the same A-BFT slot. Once collision occurs, the A-BFT slot may be unavailable for BF training, which will result in great waste of BF training opportunities. In order to further alleviate the problem of high collision in ultra-dense user scenarios, a secondary back-off mechanism (SBOM) is utilized[2].

[2] For more information about SBOM, the reader can refer to [128].

| BHI | |
|---|---|
| BTI | A-BFT |

*Figure 32: The structure of the BHI phase.*

During ATI, the AP exchanges management information with associated and beam-trained STAs, and only the AP can initiate a frame transmission. The frames transmitted in ATI are limited to the request and response frames such as management frame, acknowledgment (ACK) frame, etc. The details of specific request and response frames can be found in [120]. In particular, ATI is optional, and it is indicated by the ATI Present field that is set to 1 in the current DMG Beacon frame.

Transmissions within the DTI can use more than one channel or be performed over a bonded channel and supports channel access over:

- multiple channels through scheduling; and
- transmission opportunity (TXOP).

### 6.5.8. Caching

The benefits of device caching in the offloading and the throughput performance have been demonstrated in [121], [122], [123], [124], [125], [126]. In more detail, in [121] the spectrum efficiency of a device to device (D2D) wireless network, in which the UEs cache and exchange content from a content library, is shown to scale linearly with the network size, provided that their content requests are sufficiently redundant. Additionally, in [122], the previous result is extended to the UE throughput, which, by allowing a small outage probability, is shown to scale proportionally with the UE cache size, provided that the aggregate memory of the UE caches is larger than the library size. In order to achieve these scaling laws, the impact of the D2D interference must be addressed by optimally adjusting the D2D transmission range to the UE density. In this context, in [123], the authors proposed a cluster-based approach in order to countermeasure the D2D interference, where the D2D links inside a cluster are scheduled with time division multiple access (TDMA). The results corroborate the scaling of the spectrum efficiency that was derived in [121]. In [124], a mathematical framework based on stochastic geometry is proposed to analyse the cluster-based TDMA scheme, and the trade-off between the cluster density, the local offloading from inside the cluster, and the global offloading from the whole network is demonstrated through extensive simulations. Finally, in [125], the system throughput is maximized by jointly optimizing the D2D link scheduling and the power allocation, while, in [126], the offloading is maximized by an interference aware reactive caching mechanism. Based on the above, it is expected that device caching can significantly enhance the offloading and the delay performance of the THz cellular network, especially when the UEs exchange cached content through D2D communication.

In the P2MP scenario, the mobile user equipment (MUE) can be nomadic or mobile. In either case, the MUE can usually connect either simultaneously or sequentially to several access points (APs) using several radio access technologies (RATs), such as THz, mmWave, microwave, etc. When the MUE initiates an application, this application may include several sessions, with different QoS requirements. As an example, a voice communication will require low latency and no disconnection, while video download will benefit from high capacity, with lower constraints on disconnection or latency. Therefore, it is particularly important to select the appropriate RAT (or MAC domain) for each session, depending also on traffic load in each network.

Such coordination can be made usually by layer 4 protocols and above, by employing a central control topology. However, the case will often happen that local non-coordinated radio networks resources are also available. In that case network resources usage can be coordinated locally, taking into account layer 1 and 2 information from these different local RATs.

In such heterogeneous wireless environments, the MAC layer is expected to contribute in selecting the suitable RATs, during session initiation and communication to respect the QoS requirement of a session [127]. Additionally, different caching schemes, will mitigate the impact of specific channel impairments, such as disconnections or transmission capacity variations, and disconnection during handover. Anticipating mobile UE (MUE) handover by caching content in MUE, and prefetching content to the next cell will minimize disconnection effects. Motivated by this, a metaMAC protocol need to be introduced that ensures the dynamic coordination between the conventional MAC protocol and the caching schemes in order to optimize the MUE perceived QoS and to improve the bandwidth usage in the different RATs.

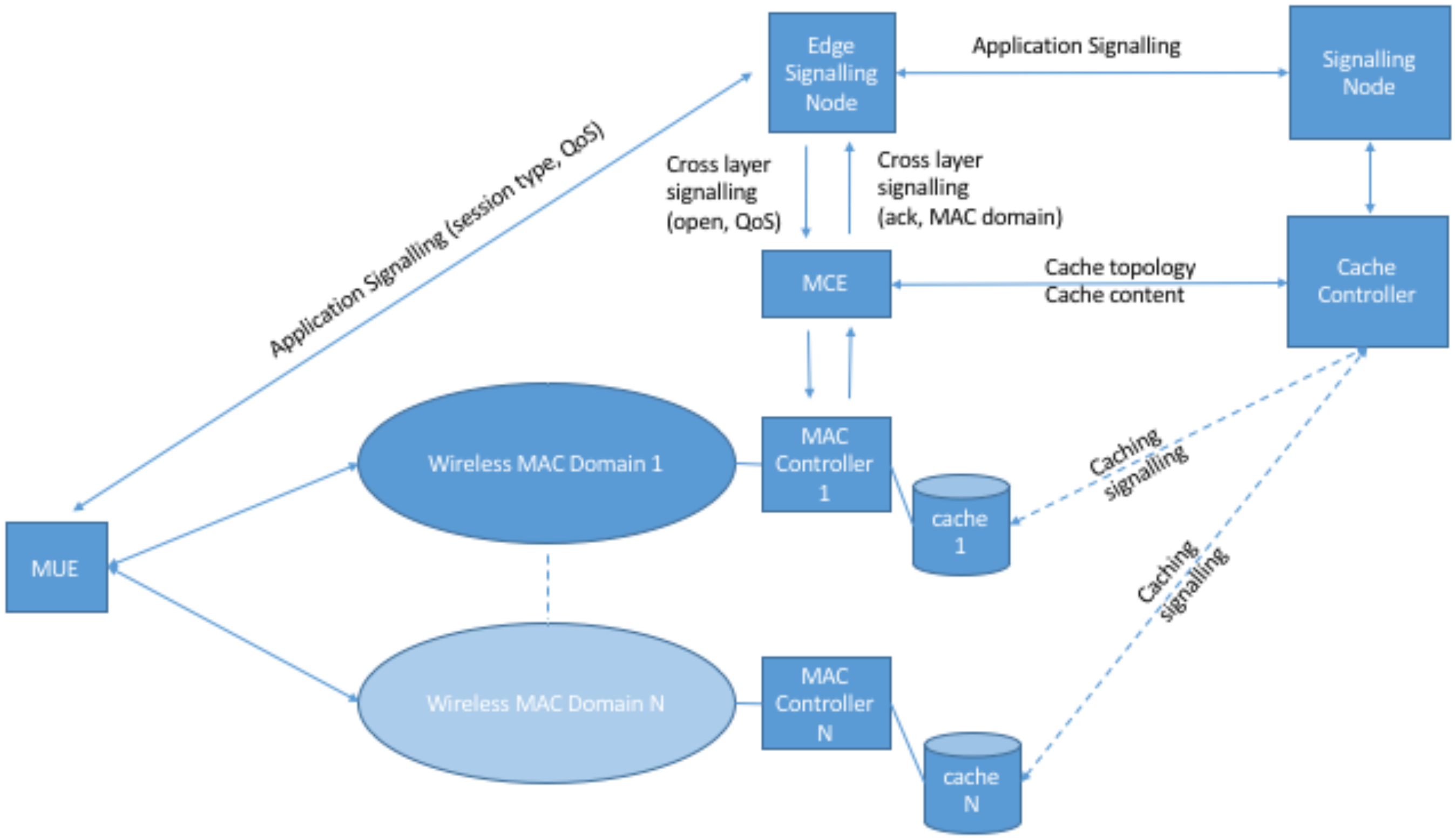

*Figure 33: System view of the different MAC and caching building blocks.*

Next, we provide a system view of the different metaMAC and caching building blocks. As illustrated in Figure 24, these blocks are:

- The MAC coordination entity (MCE) communicates with the edge signalling entity and MAC controllers to select the appropriate network once a session is initiated by the MUE. Additionally, the MCE communicates with the Cache Controller to take into account caching strategies and decisions. An MCE controls a defined set of RAT. The number of RATs controlled by a unique MCE is just a question of implementation.
- The cache controller controls the different cache nodes in the network, including definition of caching strategy, direct commands to cache content or transfer content from one cache to another caches. The cache controller has knowledge of topology of cache nodes. The Cache Controller exchanges with the MAC Control Entity.
- The MAC controllers terminate and manage the MAC layer of each network connected to the MUE.

- The edge signalling entities are the edge terminating points for QoS signalling. They communicate with the MCE to transmit information of the QoS required by a session.

The metaMAC is performed by the MCE. The authors, in [127], clarified that one option of MetaMAC is to coordinate several MAC domains (or RAT more precisely in our case) at the time slot level. This approach has several drawbacks, such as:

- Requirement to have fine grain access to all MAC controllers; and
- Difficulty to consider a multi-operator case.

These drawbacks make this kind of implementation hardly applicable to current networks.

The preferred approach in the project is rather to define cross layer signaling between MAC controllers and MCE allowing to allocate and dynamically change sessions to different MAC domains. According to Figure 24, when a MUE opens a session:

- The MUE can use an application layer protocol containing session requirements on QoS (in the form of a session description or QoS parameters); additionally, in the case the MAC layer is QoS aware, the MUE can send the same request at the MAC level.
- The edge entity then sends a cross-layer request to the MCE indicating the QoS requirement.
- The MCE communicates with the different MAC controllers and takes the decision on the MAC layer to use. It sends back an acknowledgement to the Edge entity and the MAC controller
- Once the MCE has selected the preferred MAC domains, it communicates with the Cache controller as some path could be preferred if content related to the session is already cached.

The same exchange can occur during the session if the MUE moves, or if changes occur in the network. Also note that the user could be connected to several MAC domains for the same session in the case the Layer 4 transport protocol allows it (MP-TCP, etc.)

# 7. Candidate architectures

In this section, we present the candidate architectures for the utilization of the corresponding TERRANOVA technical scenarios. In more detail, in Section 6.1, the corresponding architecture for the P2P links is demonstrated, while in Sections 6.2 and 6.3, the candidate architectures for P2MP and indoor quasi-omnidirectional THz systems are provided.

## 7.1. Technical scenario 1: P2P

Figure 28 shows two candidate architectures for the implementation of the point-to-point link of scenario 1, while the specifications are listed in Table 4 . For both architectures, a full-duplex symmetric link type is assumed, as well as the use of a single frequency window from 220 GHz – 300 GHz. Candidate (a) follows the transparent optical link architecture, while candidate (b) follows the digital optical link architecture with analogue media converter.

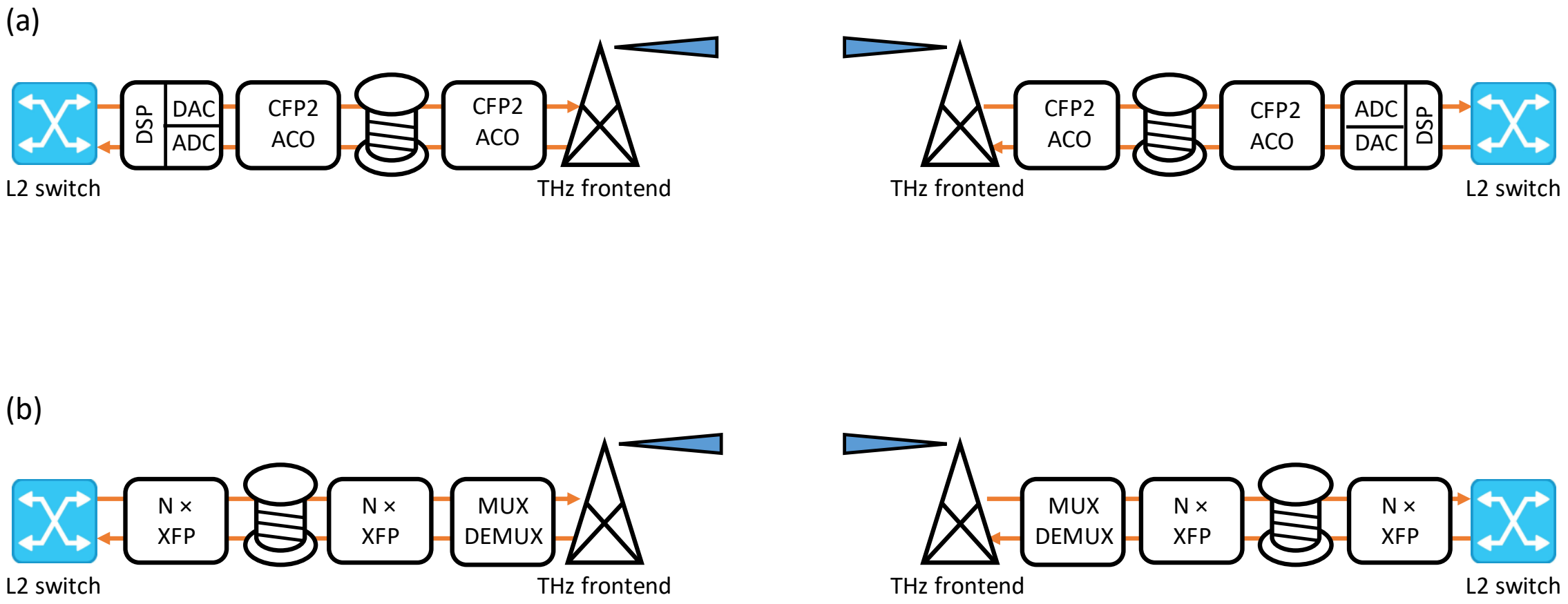

*Figure 34: Candidate architectures for the implementation of the technical scenario 1.*

*Table 4: Specifications of candidate architectures for the utilization of the technical scenario 1.*

| | **Candidate architecture (a)** | **Candidate architecture (b)** |
|---|---|---|
| **Link type** | LOS, FD, symmetric | LOS, FD, symmetric |
| **Duplex implementation** | Frequency | Polarization or frequency |
| **Expected throughput** | Up to 1 Tb/s full duplex | Up to 400 Gb/s full duplex in case of polarization duplex; up to 200 Gb/s full duplex in case of frequency duplex |
| **Used frequency window** | 220 GHz – 300 GHz | 220 GHz – 300 GHz |
| | | |
| **Optical channel** | SSMF | SSMF |
| **Optical transceiver** | CFP2-ACO | XFP |
| **Optical modulation** | Single-carrier PDM-QAM | N x NRZ |
| **THz modulation** | Single-carrier PDM-QAM and multi-carrier schemes (OFDM) | 4PAM |
| **Media converter type** | - | Analog MUX:<br>1:N/2 rate conversion<br>NRZ-to-PAM4 conversion |
| | | |
| **THz frontend type** | Double I/Q | SSB |
| **THz frontend bandwidth** | 40 GHz | 80 GHz in case of polarization duplexing;<br>40 GHz in case of frequency duplexing |
| **THz antenna** | 2 Polarizations (for polarization multiplexing) | 2 Polarization in case of polarization duplexing;<br>1 Polarization in case of frequency duplexing |
| **Beam forming type** | High gain | High gain |
| | Small angle | Small angle |
| **Dynamic beam steering** | Due to the fixed position, there is low beam steering requirements. | |
| **Dynamic UE detection** | Not required | |

| | |
|---|---|
| **Spatial synchronization** | Required |
| **Network Caching** | Required to reduce latency |
| **PHY Caching** | Not required |
| **CoMP** | Not required |

## 7.2. Technical scenario 2: P2MP

The candidate architecture for scenario 2 (point-to-multipoint) is shown in Figure 29, with the specifications listed in Table 5. The link between the single (macro) site to multiple (micro) sites is realized by space- and – time-division multiplexing (downlink) and multiple access (uplink), as the macro site beam is steering sequentially to each of the micro sites. Therefore, the macro site beam must be steered over large angles, while micro site beam does not require this functionality.

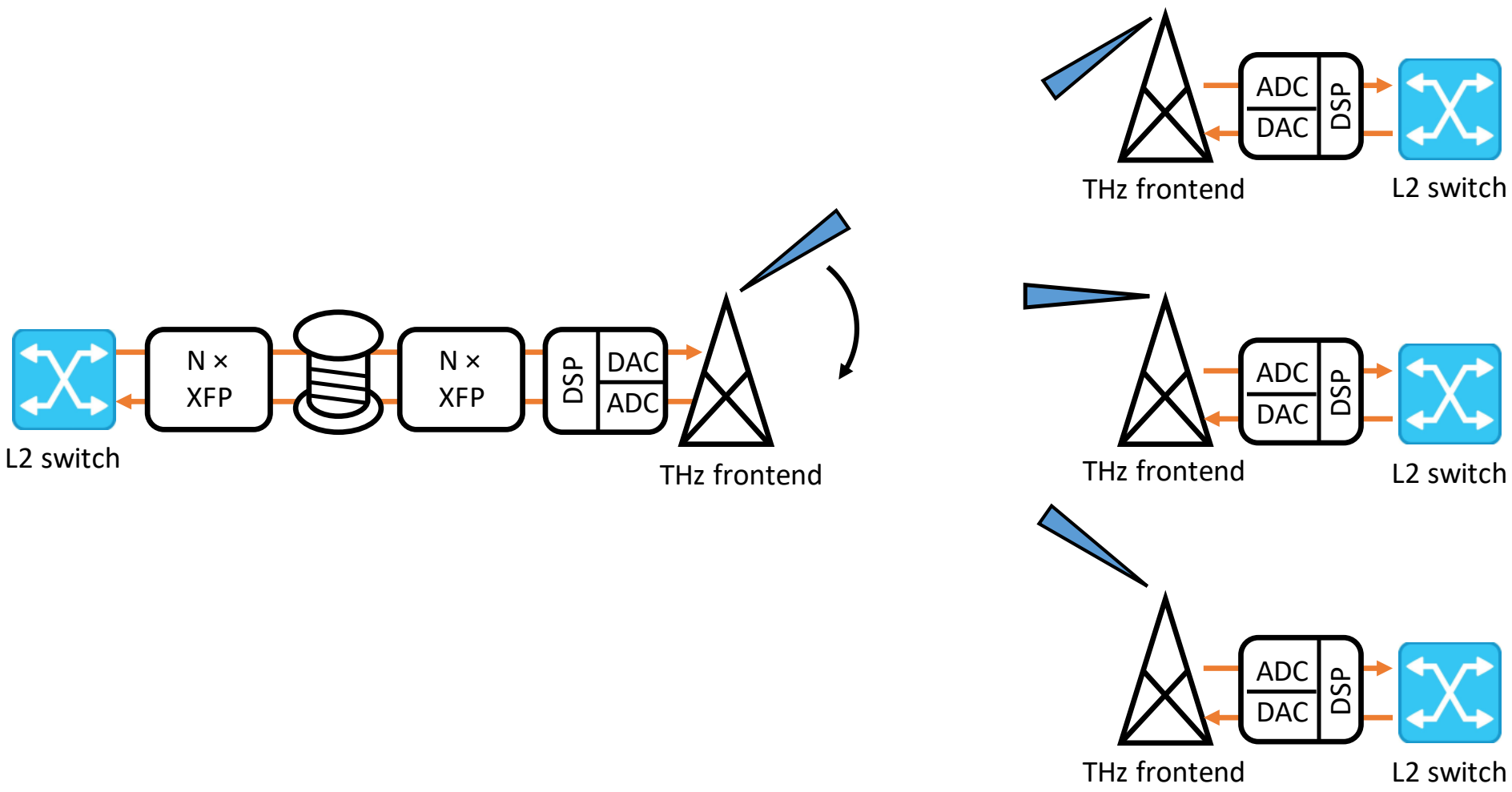

*Figure 35: Candidate architecture 1 for the implementation of technical scenario 2.*

*Table 5: Specifications of candidate architecture 1 for the implementation of technical scenario 2.*

| | **Candidate architecture 1** | |
|---|---|---|
| **Link type** | LOS, symmetric FD (per time slot) | |
| **Duplex implementation** | Frequency | |
| **Used frequency window** | 220 GHz – 300 GHz | |
| **Multi-user access** | Space and time-division multiple access | |
| | | |
| **Optical channel** | SSMF | |
| **Optical transceiver** | XFP | |
| **Optical modulation** | N x NRZ | |
| **Media converter type** | Digital (DSP + DAC/ADC) | |
| | | |
| | **Downlink** | **Uplink** |
| **Expected throughput** | Up to 500 Gb/s (e.g. 38-GBd PDM-256-QAM single carrier) | Up to 500 Gb/s shared by N users (e.g. 38-GBd PDM-256-QAM single carrier) |
| **THz frontend type** | Double I/Q | Double I/Q |
| **THz frontend bandwidth** | 40 GHz | 40 GHz |
| **THz antenna** | 2 Polarizations (for polarization multiplexing) | 2 Polarizations (for polarization multiplexing) |
| **Beam forming type** | High gain, space-division multiple access | High gain, space-division multiple access |
| **Dynamic beam steering** | Large angle | Small angle |
| **Dynamic UE discovery** | Fast and accurate with low discovery overhead. | - |
| **Spatial synchronization** | Required | |
| **Network Caching** | Required to reduce latency | |
| **PHY Caching** | Required to manage mobility and handovers | |
| **CoMP** | Required to manage mobility and handovers | |

Alternatively, the downlink can be realized by using spatial multiplexing only (candidate architecture b). That is, multiple fixed beams are employed to simultaneously transmit traffic to every micro site (Figure 30, Table x). This option increases significantly the throughput per link (per user), but it also increases the hardware complexity (antennas, RF chains, etc.), in order to support multiple beams that are associated to multiple data streams.

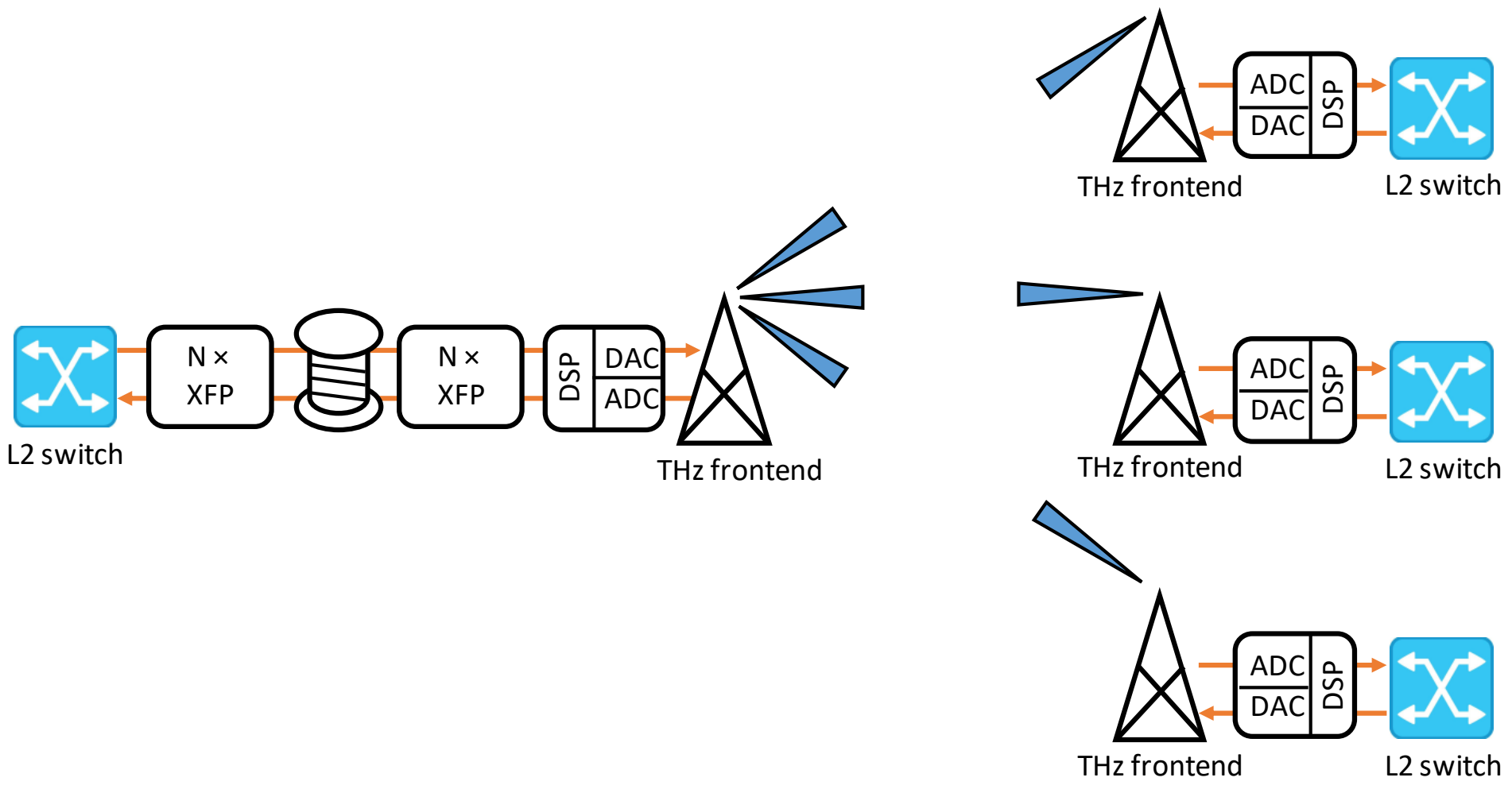

*Figure 36: Candidate architecture 2 for the implementation of technical scenario 2.*

*Table 6: Specifications of candidate architecture 2 for the implementation of technical scenario 2.*

| | Candidate architecture 2 |
|---|---|
| **Link type** | LOS, symmetric (in terms of throughput per link) |
| **Duplex implementation** | Frequency |
| **Used frequency window** | 220 GHz – 300 GHz |
| **Multi-user access** | STMA |
| | |
| **Optical channel** | SSMF |
| **Optical transceiver** | XFP |
| **Optical modulation** | N x NRZ |
| **Media converter type** | Digital (DSP + DAC/ADC) |
| | |

| | Downlink | Uplink |
|---|---|---|
| **Expected throughput** | Up to 500 Gb/s per link (e.g. 38-GBd PDM-256-QAM single carrier) | Up to 500 Gb/s per link (e.g. 38-GBd PDM-256-QAM single carrier) |
| **THz frontend type** | Double I/Q | Double I/Q |
| **THz frontend bandwidth** | 40 GHz | 40 GHz |
| **THz antenna** | 2 Polarizations (for polarization multiplex) | 2 Polarizations (for polarization multiplex) |
| **Beam forming type** | High gain, SDMA | High gain, SDMA |
| **Dynamic beam steering** | Large angle | Small angle |
| **Dynamic UE discovery** | Fast and accurate with low discovery overhead. | - |
| **Spatial synchronization** | Required | |
| **Network Caching** | Required to reduce latency | |
| **PHY Caching** | Required to manage mobility and handovers | |
| **Dynamic UE discovery** | Fast and accurate with low discovery overhead. | |
| **Spatial synchronization** | Required | |
| **Network Caching** | Required to reduce latency | |
| **PHY Caching** | Required to manage mobility and handovers | |
| **CoMP** | Required to manage mobility and handovers | |

## 7.3. Technical scenario 3: Indoor quasi-omnidirectional

The candidate architecture for the technical scenario 3, namely indoor quasi-omnidirectional, is shown in Figure 31, with its list of specifications summarized in Table 7. It is assumed that only the downlink to multiple UEs is using THz, while the uplink might be realized with other technologies. The downlink uses M beams, which broadcast the same data to different parts of the room. The indoor scenario as well as the presence of multiple beams requires to consider the downlink as nLoS link with multipath fading.

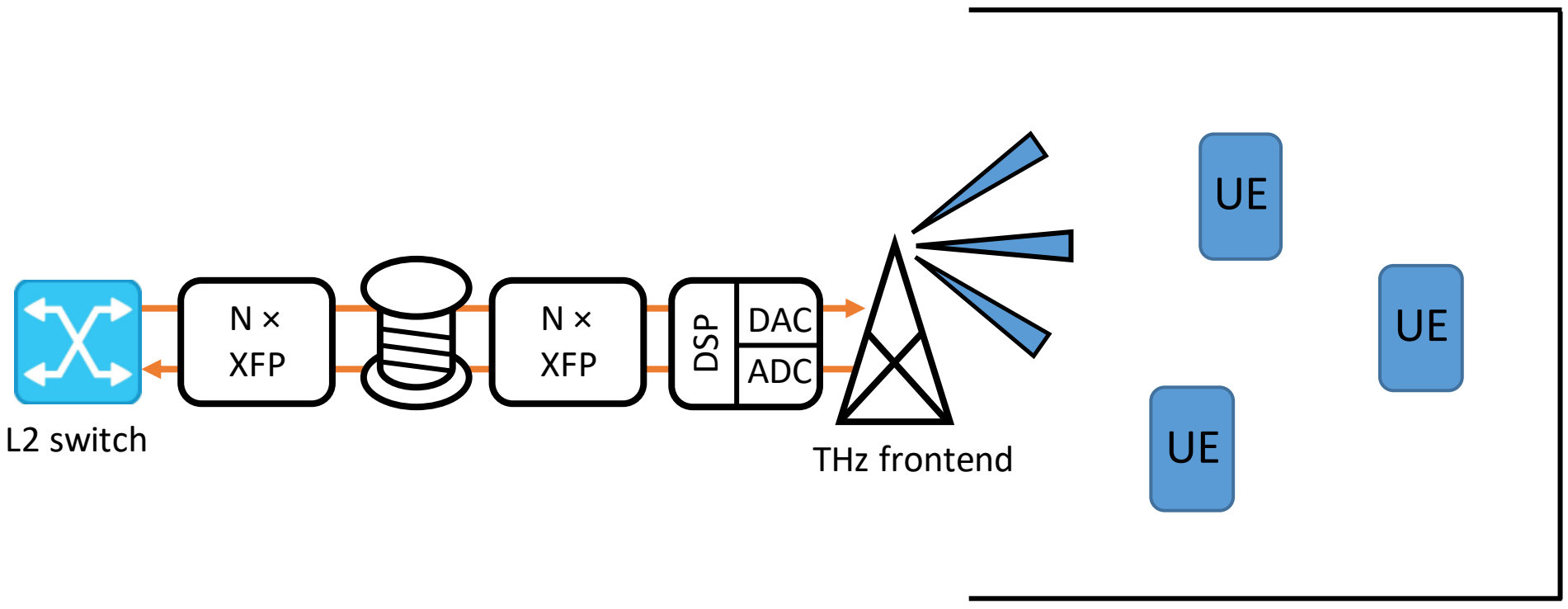

*Figure 37: Candidate architecture for the implementation of the technical scenario 3.*

*Table 7: Specifications of the candidate architecture for the implementation of technical scenario 3.*

| | Candidate architecture |
|---|---|
| **Link type** | nLoS, half duplex |
| **Duplex implementation** | THz downlink (uplink with other technologies) |
| **Used frequency window** | 220 GHz – 300 GHz |
| **Multi-user access** | TDMA |
| | |
| **Optical channel** | SSMF |
| **Optical transceiver** | XFP |
| **Optical modulation** | N x NRZ |
| **Media converter type** | Digital (DSP + DAC/ADC) |
| | |
| | **Downlink** |
| **Expected throughput** | Up to 400 Gb/s shared by multiple users (e.g. 64-GBd 256QAM single carrier) |
| **THz frontend type** | I/Q |
| **THz frontend bandwidth** | 80 GHz |
| **THz antenna** | M antennas for multiple indoor beams |
| **Beam forming type** | High opening angle for high coverage |
| **Dynamic beam steering** | Not required |
| **Network Caching** | Required to reduce latency |
| **PHY Caching** | Required to manage mobility and handovers |
| **Dynamic UE discovery** | Fast and accurate with low discovery overhead. |
| **Spatial synchronization** | Not required |
| **CoMP** | Required to ensure coverage |

# 8. Conclusions

This white paper presented the key system requirements of TERRANOVA. The development of the THz wireless system and baseband interface is in its early stage; hence, it is difficult to state precisely what the final system will be. However, it is already possible to identify the critical technology gaps and the appropriate enablers, which will boost the system's utilization. It is also possible to identify key use cases for TERRANOVA and further refine the design of the system so that it can take into account both the challenges and technology gaps as well as the expected usage scenarios.

Currently, themes that pervade multiple features are channel and noise modelling, which will allow the theoretical analysis of the THz link and network as well as the development of PHY and MAC schemes. The transceiver RF frontend and baseband design are important for matching the data rates of the optical and THz wireless link and it will be crucial to increase the spectral efficiency and link distance to meet the performance targets of many applications.

Given the system requirements defined in this white paper, TERRANOVA is expected to use novel technology concepts, including the joint design of baseband signal processing for the complete optical and wireless link, the development of broadband and spectrally efficient RF-frontends for frequencies >275 GHz, as well as channel modelling, waveforms, antenna array, physical layer and multiple-access schemes design.

From the physical layer perspective, fundamental inherent characteristic of transmission in the THz regime were discussed, namely the extreme bandwidth, the extremely small wavelength and the associated high attenuation as well as the material absorption, molecular absorption and the resulting frequency and distance dependent pathloss. It was hence recognised that the employment of pencil beamforming techniques is required in order to cope with distance-dependent performance (w.r.t. SNR, available bandwidth, spectral efficiency, etc) along with adaptive modulation and coding techniques and transmission through multiple frequency windows and multiple spatial modes.

The directional, 'beamspace', channel system model, resulting from the employment of pencil beamforming, calls for a suitable ('directional') MAC and RRM design, capable to address the challenges blockage, deafness, spectrum heterogeneity, detection and tracking agility and interference management. Both random and scheduled access need to be considered in order to strike the right balance between complexity and adaptability and at the same time support the directional ('beamspace') and multi-user (point to multi-point) scenarios.

With respect to the hybrid optical/THz system co-design, a brief review of wireless transmission reference systems and relevant standards was first presented (associated with the identified three basic topological application scenarios) and then optical transmission reference systems and relevant standards were discussed. Three interface implementation options were then described for the realization of the hybrid optical/ THz. A comparative analysis of the three presented architectures for the hybrid optical/THz link was performed, where it became apparent that the final selection among them is to a large extent application-dependent.

With regards to THz frontend, reference components and technologies were analysed, namely ISM-60 GHz, E-band and first generation THz frontend research prototypes (non commercially available) as well as beamforming frontends for 5G, 60 GHz WiGig and 300 GHz THz links. The analysis of typical standard transceiver architectures employed in millimetre-wave frequencies and comparison to the currently available THz solutions revealed that a quite significant functional gap has to be closed between the millimetre-wave and THz frontends. In that respect, the TERRANOVA media-converter imposes additional boundary conditions and specifications. The basic function of the TERRANOVA frontend and TERRANOVA media-converters is the mapping between a number of modulated optical carriers and a number of modulated electrical THz carriers. A generic solution needs to be devised, which is applicable to different fibre optical standards and compatible with the selected options for the architectures of the hybrid optical/THz link.

With regards to the beamforming subsystem, a classification of different beamforming options was presented (relevant to the TERRANOVA system concept and application scenarios) and appropriate THz antenna options were discussed.

From MAC/RRM protocol perspective, relevant reference MAC protocols were analysed, namely the IEEE 802.15.3c, IEEE 802.11ad, and IEEE 802.11ay MAC protocols along with the assisted beamforming MAC protocol for THz communication networks (TAB-MAC). The most critical functionalities and their respective implementations options were then discussed, i.e. physical control channel implementation, UE detection and tracking, random access, scheduled access and handover between beams.
In terms of RRM, the importance of dynamic cell formation and UE-BS association was emphasized and the corresponding optimization problem formulation was briefly discussed. Moreover, the benefits of device caching in terms of offloading and throughput performance need also to be considered in the overall system architecture.

Finally, specific candidate system architectures were presented for the three main technical scenarios in TERRANOVA, namely outdoor fixed P2P, outdoor/indoor individual P2MP, and outdoor/indoor "quasi"-omnidirectional links. From today's perspective, these candidate architectures are designed in order to provide full support for the previously identified target KPIs in each scenario. However, there are still several aspects and issues that need careful theoretical analysis, algorithm and protocol design, practical implementation and experimental validation to confirm their suitability to meet the KPIs.